\documentclass[twocolumn]{aastex631}

\usepackage{amsmath}
\usepackage{amssymb}
\usepackage{graphicx}
\usepackage{xcolor}
\usepackage{gensymb}
\usepackage{textcomp}
\usepackage{enumerate}
\usepackage[inline]{enumitem} % Customize first-level enumerate
\setlist[enumerate, 1]{left=0pt, itemsep=0pt, leftmargin=10pt, labelindent=10pt, listparindent=10pt, labelsep=5pt}
\usepackage{scrextend}
\usepackage[normalem]{ulem}
\usepackage{CJKutf8}
\usepackage[flushleft]{threeparttable}
\usepackage{soul}

\newcommand{\setofall}[3]{\{{#1}\}_{{#2}}^{{#3}}}
\newcommand{\allq}{\setofall{q_i}{i=1}{N}}

\newcommand{\allI}{\setofall{I_{g,i}}{i=1}{N}}

\defcitealias{2012ApJ...744..129B}{BD12}

\providecommand{\sorthelp}[1]{}

\shorttitle{cirrus decomposition in low surface brightness imaging}
\shortauthors{Liu et al.}
\graphicspath{{./}{figures/}}
\begin{document}

\title{Fuzzy Galaxies or Cirrus? Decomposition of Galactic Cirrus in Deep Wide-Field Images}

\correspondingauthor{Qing Liu}
\email{qliu@astro.utoronto.ca}

\author[0000-0002-7490-5991]{Qing Liu \begin{CJK}{UTF8}{gbsn}(刘青)\end{CJK}}
\affil{David A. Dunlap Department of Astronomy \& Astrophysics, University of Toronto, 50 St. George St., Toronto, ON M5S 3H4, Canada}
\affil{Dunlap Institute for Astronomy and Astrophysics, University of Toronto, Toronto, ON M5S 3H4, Canada}
\affil{Leiden Observatory, Leiden University, P.O. Box 9513, 2300 RA Leiden, The Netherlands}

\author[0000-0002-4542-921X]{Roberto Abraham}
\affil{David A. Dunlap Department of Astronomy \& Astrophysics, University of Toronto, 50 St. George St., Toronto, ON M5S 3H4, Canada}
\affil{Dunlap Institute for Astronomy and Astrophysics, University of Toronto, Toronto, ON M5S 3H4, Canada}

\author[0000-0002-5236-3896]{Peter G. Martin}
\affil{Canadian Institute for Theoretical Astrophysics, University of Toronto, 60 St. George St., Toronto,
ON M5S 3H8, Canada}

\author[0000-0003-4381-5245]{William P. Bowman}
\author[0000-0002-8282-9888]{Pieter van Dokkum}
\affil{Department of Astronomy, Yale University, New Haven, CT 06520, USA}

\author[0000-0002-1841-2252]{Shany Danieli}
\thanks{NASA Hubble Fellow}
\affil{Department of Astrophysical Sciences, 4 Ivy Lane, Princeton University, Princeton, NJ 08544, USA}

\author[0000-0002-9820-1219]{Ekta~Patel}\thanks{NASA Hubble Fellow}
\affiliation{Department of Physics and Astronomy, University of Utah, 115 South 1400 East, Salt Lake City, UT 84112, USA}

\author[0000-0003-0327-3322]{Steven R. Janssens}
\affil{Centre for Astrophysics and Supercomputing, Swinburne University, Hawthorn, VIC 3122, Australia}

\author[0000-0002-5120-1684]{Zili Shen}
\affil{Department of Astronomy, Yale University, New Haven, CT 06520, USA}

\author[0000-0002-4175-3047]{Seery Chen}
\affil{David A. Dunlap Department of Astronomy \& Astrophysics, University of Toronto, 50 St. George St., Toronto, ON M5S 3H4, Canada}
\affil{Dunlap Institute for Astronomy and Astrophysics, University of Toronto, Toronto, ON M5S 3H4, Canada}

\author[0000-0001-8855-3635]{Ananthan Karunakaran}
\affil{David A. Dunlap Department of Astronomy \& Astrophysics, University of Toronto, 50 St. George St., Toronto, ON M5S 3H4, Canada}

\author[0000-0002-7743-2501]{Michael A. Keim}
\affil{Department of Astronomy, Yale University, New Haven, CT 06520, USA}

\author[0000-0002-2406-7344]{Deborah Lokhorst}
\affiliation{NRC Herzberg Astronomy \& Astrophysics Research Centre,
5071 West Saanich Road, Victoria, BC V9E 2E7, Canada}

\author[0000-0002-7075-9931]{Imad Pasha}
\affil{Department of Astronomy, Yale University, New Haven, CT 06520, USA}

\author[0000-0002-2350-0898]{Douglas L. Welch}
\affil{Department of Physics \& Astronomy, McMaster University, Hamilton, ON L8S 4M1, Canada}

%% Note that the \and command from previous versions of AASTeX is now
%% depreciated in this version as it is no longer necessary. AASTeX 
%% automatically takes care of all commas and "and"s between authors names.

%% AASTeX 6.31 has the new \collaboration and \nocollaboration commands to
%% provide the collaboration status of a group of authors. These commands 
%% can be used either before or after the list of corresponding authors. The
%% argument for \collaboration is the collaboration identifier. Authors are
%% encouraged to surround collaboration identifiers with ()s. The 
%% \nocollaboration command takes no argument and exists to indicate that
%% the nearby authors are not part of surrounding collaborations.

%% Mark off the abstract in the ``abstract'' environment. 
\begin{abstract}
Diffuse Galactic cirrus, or Diffuse Galactic Light (DGL), can be a prominent component in the background of deep wide-field imaging surveys. The DGL provides unique insights into the physical and radiative properties of dust grains in our Milky Way, and it also serves as a contaminant on deep images, obscuring the detection of background sources such as low surface brightness galaxies. However, it is challenging to disentangle the DGL from other components of the night sky. In this paper, we present a technique for the photometric characterization of Galactic cirrus, based on (1) extraction of its filamentary or patchy morphology and (2) incorporation of color constraints obtained from Planck thermal dust models. Our decomposition method is illustrated using a $\sim$10 ${\rm deg^2}$ imaging dataset obtained by the Dragonfly Telephoto Array, and its performance is explored using various metrics that characterize the flatness of the sky background. As a concrete application of the technique, we show how removal of cirrus allows low surface brightness galaxies to be identified on cirrus-rich images. We also show how modeling the cirrus in this way allows optical DGL intensities to be determined with high radiometric precision.
\end{abstract}

%% Keywords should appear after the \end{abstract} command. 
%% The AAS Journals now uses Unified Astronomy Thesaurus concepts:
%% https://astrothesaurus.org
%% You will be asked to selected these concepts during the submission process
%% but this old "keyword" functionality is maintained in case authors want
%% to include these concepts in their preprints.
\keywords{Astronomy data reduction (1861); Astronomy image processing (2306); Interstellar dust (836); Low surface brightness galaxies (940); Sky surveys (1464)}

%% From the front matter, we move on to the body of the paper.
%% Sections are demarcated by \section and \subsection, respectively.
%% Observe the use of the LaTeX \label
%% command after the \subsection to give a symbolic KEY to the
%% subsection for cross-referencing in a \ref command.
%% You can use LaTeX's \ref and \label commands to keep track of
%% cross-references to sections, equations, tables, and figures.
%% That way, if you change the order of any elements, LaTeX will
%% automatically renumber them.
%%
%% We recommend that authors also use the natbib \citep
%% and \citet commands to identify citations.  The citations are
%% tied to the reference list via symbolic KEYs. The KEY corresponds
%% to the KEY in the \bibitem in the reference list below. 

\section{Introduction} \label{sec:intro}

Dust is an important component of the interstellar medium (ISM) in our Milky Way (MW) galaxy. It plays a critical role in star formation and galaxy evolution by serving as the catalyst of molecular hydrogen formation, the site for the photoelectric effect heating the ISM, the coolant of warm ISM, and the transporter of momentum (\citealt{2011piim.book.....D}). Dust is involved in numerous radiative transfer processes including thermal emission, absorption and scattering, polarization, luminescence, and radio emission from rotating grains. Dust models have been developed to match the observations, which largely improve our knowledge about the physical and radiative properties of interstellar dust (e.g., \citealt{2004ApJS..152..211Z}, \citealt{2007ApJ...657..810D}, \citealt{2011A&A...525A.103C}). 

Dust scattering is probably one of the most ubiquitous radiative processes among those mechanisms, occurring throughout the MW (\citealt{2003ARA&A..41..241D}). Observations of dust scattering can be traced back to pioneering work done by \cite{1937ApJ....85..213E} and \cite{1941ApJ....93...70H}. More extensive studies in the 1970-1990s using photographic plates (e.g., \citealt{1976AJ.....81..954S}, \citealt{1979A&A....78..253M}, \citealt{1987A&A...184..269L}, \citealt{1989ApJ...346..773G}, \citealt{1991ApJ...376..335P}) revealed the prevalence of Galactic cirrus, or diffuse Galactic light (DGL)
\footnote{In the astronomical literature, DGL is often referred to as the unresolved faint diffuse component of the sky background with an origin from the MW, which extends from mid-infrared to ultraviolet (UV). At wavelengths longer than near-infrared (NIR), dust emission starts to dominate over scattering (\citealt{2015ApJ...811...77S}). Here we refer to the optical DGL, and use the term interchangeably with diffuse Galactic cirrus below.}. However, dust scattering has been poorly mapped with modern CCD detectors over the subsequent three decades. 

This was mainly because of two facts: (1) the small size of digital sensors has led to most large telescopes being optimized for point-source depth in relatively small field-of-views, while cirrus often extends over degree scales on the sky. (2) When illuminated by the interstellar radiation field (ISRF) of the MW (\citealt{1983A&A...128..212M}), light from dust scattering is very faint, and is typically only a few percent of the brightness of the night sky in optical bands. 
The faint diffuse nature of optical cirrus (and many other low surface brightness sources) makes it extraordinarily vulnerable to various kinds of systematics in wide-field imaging, such as scattered light in the extended wings of the point-spread function (PSF), improper sky background subtraction, and flat-fielding.

Existing barriers have been broken recently, thanks to two major advances: novel instrumental designs optimized for low surface brightness imaging (e.g., \citealt{2014PASP..126...55A}, \citealt{2023PASP..135a5002L}), and improvements in data analysis techniques dedicated to the preservation of low surface brightness emission and the reduction of systematics (e.g., \citealt{2009PASP..121.1267S}, \citealt{2015ApJ...800L...3W}, \citealt{2016MNRAS.456.1359F}, \citealt{2017ApJ...834...16M}, \citealt{2018ApJ...857..104G}, \citealt{2020ApJ...894..119D}, \citealt{2023MNRAS.520.2484K}, \citealt{2023ApJ...953....7L}, \citealt{2024arXiv240513496C}, \citealt{2024MNRAS.528.4289W}). This progress has led to reprocessing deep imaging surveys with modern observing and data reduction techniques optimized for imaging the diffuse optical cirrus (e.g., \citealt{2013ApJ...767...80I}, \citealt{2016A&A...593A...4M}, \citealt{2020A&A...644A..42R}, \citealt{2023MNRAS.524.2797M}, \citealt{2023MNRAS.519.4735S}, \citealt{2023ApJ...948....4Z}, \citealt{2024AJ....168...88Z}).

%For example, Ineneka et al. (2013) investigated the high-galactic latitude cirrus region MBM32. Miville-Deschênes et al. (2016) analyzed cirrus around NGC2592 and NGC2594 using imaging from CFHT/MegaCam. Roman et al. (2020) studied the optical properties of cirrus in the deep Sloan Digital Sky Survey (SDSS) Stripe82 region. Mattila et al. (2023) studied the photometry and extinction of regions of dark nebula LDN1642 with varying optical depths, and Zhang et al. (2023) presented observations of Spider, a diffuse cirrus region at high Galactic latitude, obtained by the Dragonfly Telephoto Array.

Analysis of optical cirrus has pointed to one important conclusion: it has a strong spatial correlation with its mid-to-far-infrared (FIR) counterparts. The latter has its origin primarily in the thermal emission from dust grains in equilibrium with the radiation field, and has been extensively characterized by IR missions such as the IR Astronomical Satellite (IRAS; e.g., \citealt{1984ApJ...278L..19L}, \citealt{1998ApJ...500..525S}, \citealt{2011ApJ...736..119M}), the Herschel Space Observatory (e.g., \citealt{2010A&A...518L.105M}, \citealt{2011MNRAS.412.1151B}), and the Planck Satellite (e.g., \citealt{planck2011-7.0}, \citealt{planck2011-7.12}, \citealt{planck2013-p06b}, \citealt{planck2013-XVII}). 
This correlation confirmed that optical Galactic cirrus, or DGL, is mainly contributed by scattering of starlight by large dust grains (\citealt{2012ApJ...744..129B}), and therefore, it is `clean' to be used to constrain dust and ISRF properties and models. 
For example, the changes in the correlations shed light on optical depth effects in dust scattering due to the increase of dust column densities (e.g., \citealt{2013ApJ...767...80I}, \citealt{2020A&A...644A..42R}, \citealt{2023MNRAS.524.2797M}, \citealt{2023ApJ...948....4Z}). \cite{2023ApJ...948....4Z} showed that observations of optical cirrus can be used to constrain size distributions and compositions of dust grains, and furthermore, the anisotropy of the scattering phase function and the incident ISRF. Cirrus is likewise able to provide useful information about turbulence in the ISM. \cite{2016A&A...593A...4M} used cirrus as a probe of the turbulent cascade of the ISM and found no energy dissipation at 0.01 pc scales. 
In summary, imaging the optical cirrus provides valuable datasets for ISM researchers.

One person's gain may lead to another's pain. 
%Someone's success may come at the expense of someone else's suffering.
Cirrus is unwanted foreground contamination for researchers interested in extragalactic low surface brightness sources, many of which rely on identification with visual inspection. In near-field cosmology, an abundance of ultra-diffuse galaxies and dwarf satellite galaxies is a strong prediction of the hierarchical galaxy formation predicted by $\Lambda$CDM cosmologies (e.g., \citealt{1999ApJ...522...82K}, \citealt{2016ApJ...827L..23W}, \citealt{2019ARA&A..57..375S}). However, their detection and measurement can be drastically affected by pollution from cirrus (e.g., \citealt{2021ApJS..257...60Z}). Around nearby large galaxies, confusion arises between cirrus and collisional debris such as shells and tidal tails (e.g., \citealt{2020MNRAS.498.2138B}). In galaxy clusters, cirrus serves as contamination to the characterization of intra-cluster light (e.g., \citealt{2017ApJ...834...16M}). 
In fact, systematics from cirrus in the sky background pose some of the major challenges in recent-day deep imaging surveys reaching {\textit{g}-band} surface brightness limits of 29 mag/arcsec$^2$ and fainter. Without a doubt, cirrus contamination will be similarly non-negligible (and likely even more critical and pervasive) in the sky background of next-generation deep imaging surveys, e.g., those to be carried out by the Vera C. Rubin {Observatory} (\citealt{2022MNRAS.513.1459M}, \citealt{2024MNRAS.528.4289W}) and the Euclid Space Telescope (\citealt{2022A&A...657A..92E, 2024arXiv240513491E}; \citealt{2024arXiv240513496C}).

It is interesting to consider ways to disentangle the cirrus emission from other sources of light in the images, which would benefit both ISM and extragalactic studies. However, this is challenging because of its faint diffuse nature, complex morphology, and lack of well-calibrated radiometrically unbiased imaging datasets.
% and correspondingly, the poor understanding of the photometric characteristics, of cirrus. 
Current investigations have involved two tracks: 
\begin{enumerate}
    \item Using morphology to filter out `cirrus-like' signals. Many approaches have been developed to characterize the filamentary and patchy morphology of the diffuse ISM. For example, the pioneering work by \cite{1993AJ....106.1664A} used morphological filters with varying structure elements (a technique called `sieving'), to remove extended emission in IRAS imaging of the M81/M82 group.
    \item Using colors to distinguish cirrus from extragalactic sources. In particular, \cite{2020A&A...644A..42R} investigated the optical colors of cirrus in the deep Sloan Digital Sky Survey (SDSS) Stripe82 region using \textit{g}, \textit{r}, \textit{i}, and \textit{z} bands, and showed that with two colors, cirrus can be well differentiated from extragalactic sources via multi-band photometry (also see discussions in \citealt{2023MNRAS.519.4735S} and \citealt{2023MNRAS.524.2797M}).
\end{enumerate}

In the present paper, we combine both strategies by presenting an approach that applies morphological filtering with color constraints on deep wide-field images for the decomposition of optical cirrus. We focus on the optically thin cirrus to avoid optical depth effects. The data we used is from the Dragonfly Telephoto Array (Dragonfly for short), a telescope optimized for low surface brightness imaging. Even with only two filters equipped on Dragonfly, this approach can differentiate cirrus from most low surface brightness galaxies (LSBGs). A similar approach is promising to be applied to next-generation deep imaging surveys from the ground-based Rubin {Observatory} and the spaceborne Euclid Telescope. 

This paper is structured as follows: Section~\ref{sec:telecope_and_data} describes the Dragonfly Telephoto Array and the datasets. Section~\ref{sec:mrf} describes the foreground and background source subtraction techniques used. Section~\ref{sec:morph} presents the extraction of `cirrus-like' emission using morphological information. Section~\ref{sec:cirrus_color} explains the principles of color modeling on cirrus and demonstrates cirrus removal using Dragonfly imaging. Furthermore, we use several metrics to quantitatively evaluate the performance of our algorithm. Section~\ref{sec:LSBG} illustrates how this approach can be applied to facilitate LSBG searches via integrated light. Section~\ref{sec:discussion} discusses cirrus imaging with multi-band photometric surveys and investigates the optical DGL in the dataset. Finally, Section~\ref{sec:conclusion} presents the conclusions.

\section{Telescope \& Datasets} \label{sec:telecope_and_data}

To illustrate the methodology of cirrus decomposition, deep wide-field imaging datasets with high sensitivity to diffuse extended emission are required. The example datasets we use here were obtained by the Dragonfly Telephoto Array, which is briefly introduced in Section~\ref{sec:telescope}. Section~\ref{sec:reduction} summarizes the observations and data reduction. Section~\ref{sec:field} introduces the example fields used for demonstrating the decomposition approach.

\subsection{The Dragonfly Telephoto Array} \label{sec:telescope}
The Dragonfly Telephoto Array is an array composed of 48 Canon 400 mm $f/2.8$ IS II USM-L telephoto lenses, which together constitute a mosaic aperture telescope equivalent to a 1.0 m $f/0.39$ refractor. A Santa Barbara Imaging Group (SBIG) CCD camera with a field of view of 2.6$^\circ$ $\times$ 1.9$^\circ$ and a pixel scale of 2.85\arcsec/pix is equipped on each lens. %The equivalent FoV of a single exposure is around 2$^\circ$ $\times$ 3$^\circ$. 

The core concept of the design of Dragonfly is the optimization of the performance in low surface brightness imaging. Scattered light in the optical path is minimized by several key instrumental elements, including (1) zero pupil obscuration, (2) sub-wavelength nanostructure coatings on optical surfaces, and (3) all-refractive optics with excellent baffling. The 48 cameras take images in the Sloan \textit{g}- and \textit{r}-band. Two strategies are adopted in the observations to reduce camera-by-camera systematics: first, the pointings of individual lenses are offset by small amounts relative to each other so that ghost images are removed in stacking, and second, large ($\sim 15 \arcmin$)  dithers are performed in each visit/iteration of the observation. Readers are referred to \cite{2014PASP..126...55A} for the general description of the telescope design. We refer the readers to \cite{2020ApJ...894..119D} for a description of the current configuration of the broadband array.

\subsection{Data Acquisition \& Reduction} \label{sec:reduction}

The general strategy of Dragonfly's data acquisition is described in \cite{2020ApJ...894..119D}. In brief, Dragonfly takes 10-minute exposures with the 48 lenses and performs quality checks on each exposure. {Because images are undersampled, the mean FHWM of Dragonfly's PSF at New Mexico Skies under good conditions is $\approx 5\arcsec$.} Observations are obtained with a large dither angle to reduce systematics. For datasets used in this work, the dither angle adopted was $15\arcmin$. {A dark exposure with the same integration time as used by the science exposure was taken after each observing sequence. Darks passing quality checks with the same exposure times are average-combined into master darks.} {Twilight} flats were taken at the start and the end of the observing night, {in company with darks with the same exposure times as flats}. High-quality flats passing quality checks were combined into master flats. If no good flat was acquired for a unit on a night, flats from the nearest night were used. {Details about the quality checks of calibration frames are referred to \cite{2020ApJ...894..119D}.}

Raw frames were bias subtracted, dark subtracted, and flat-fielded using the upgraded Dragonfly data reduction pipeline \texttt{DFReduce} (Bowman et~al.~in~prep). Astrometric solutions were derived using the \texttt{astrometry.net} module (\citealt{2010AJ....139.1782L}).

Sky subtraction requires careful treatment. In cirrus-rich fields, conventional algorithms (e.g., using a box-averaging sky estimator or spline fitting) would inevitably be biased by cirrus. This limitation arises because these methods are designed to produce artificially flat sky backgrounds. Such systematics in sky modeling could severely hamper the photometric characterization of Galactic cirrus. To avoid this, sky subtraction of the dataset presented in this work was done following the procedures described in \cite{2023ApJ...953....7L}. In brief, in order to preserve the cirrus signal of interest while removing the time-varying large-scale sky pattern (mostly contributed by the zodiacal light and airglows), we adopted a sky modeling method using FIR/sub-mm data from Planck as priors, which proved to be effective in producing unbiased sky background model. The method relies on the assumption that the dust is optically thin on large scales and is under thermal equilibrium, which applies well to the scenarios in the context of this work. Details about the principles and procedures of sky modeling are described in \cite{2023ApJ...953....7L}. 

Finally, the exposures were combined following procedures in \cite{2023ApJ...953....7L} using Gaussian process modeling, which are more robust than typical stacking methods for images with correlated signals extending on large scales (e.g., cirrus fields) at low surface brightness levels.

\subsection{Example Datasets} \label{sec:field}

The datasets used in this work for the demonstration of the approaches consist of two fields. They were obtained by Dragonfly as part of a larger observing campaign that aims to map the nearby sky of M33. Throughout the paper, we denote them as Field A and Field B. The observations were taken in October 2020. 

Table \ref{table1} lists the Equatorial and Galactic coordinates, the areas used for cirrus modeling, numbers of effective exposures that passed the quality checks, and the 1$\sigma$ surface brightness limits at [$60\arcsec \times 60\arcsec$] scales. The surface brightness limits in mag/arcsec$^2$ are calculated using \texttt{sbcontrast} (\citealt{{2022ApJ...935..160K}}), a robust method to determine surface brightness limits, \textit{after} removing the Galactic cirrus (see below)\footnote{Cirrus effectively acts as a large scale variation constraining the surface brightness limit in its calculation. For reference, the surface brightness limits $\mu_{{\rm lim,}1\sigma}{\footnotesize (60\arcsec\times60\arcsec)}$ calculated before cirrus removal is 29.5 mag/arcsec$^2$ in \textit{g} and 28.8 mag/arcsec$^2$ in \textit{r} for Field A, compared to values listed in Table~\ref{table1}.}. {The magnitudes and surface brightness reported in this work are before Galactic extinction and reddening correction.} 

The field-of-view of the Dragonfly coadd is $\sim10\,{\rm deg}^2$. We use the central cutouts of the coadds for cirrus modeling because of two considerations\footnote{The images are projected following TAN-SIP convention before trimming. Note that projection effects could occur given the large field of view. We have examined that such effects do not affect the source modeling in Section~\ref{sec:flux_model} (due to a mismatch in astrometry) and the filtering process applied in Section~\ref{sec:cirrus_df} (due to distortion). However, caution needs to be taken where such effects become non-negligible.}: first, the coadd is noisier at the edges due to the large dither angle, which results in fewer exposures covering these regions. Second, diffuse light from extended wings of bright sources outside the area of investigation may contribute to the diffuse light background in the area, and needs to be modeled as well (Section~\ref{sec:mrf}). 

\begin{figure*}[!htbp]
\centering
  \resizebox{0.82\hsize}{!}{\includegraphics{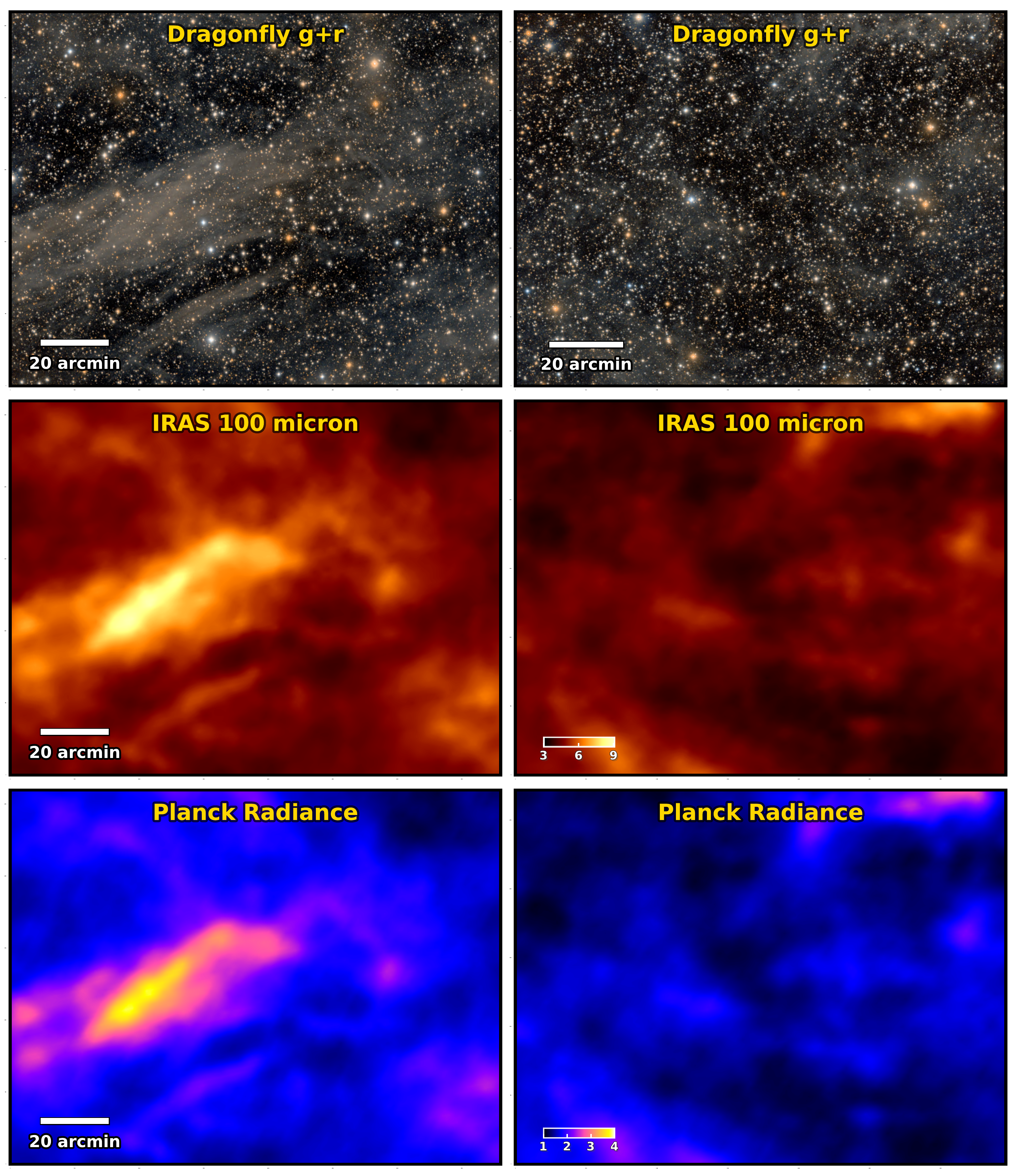}}
  \caption{Example dataset used in this work. Top row: Dragonfly \textit{g}+\textit{r} mosaic RGB image of the central $2.6\degree \times 1.8\degree$ area of Field A (left) and the central $2.4\degree \times 1.9\degree$ area of Field B (right). The green channel of the RGB is the average of \textit{g} and \textit{r} band data. Middle row: IRAS 100~$\mu m$ maps of the two fields {in the unit of MJy/sr}. Bottom row: Planck dust radiance maps of the two fields {in the unit of $\rm [10^{-7}\, W\,m^{-2}\,sr^{-1}]$}. The Dragonfly data, the IRAS data, and the Planck dust products show clear spatial correspondence.}
\label{fig:field}
\end{figure*}

The RGB images of the areas used for cirrus modeling in this work, created from the $g$ and $r$ band Dragonfly data (red channel: Dragonfly $r$, green channel: average {(in ADU)} of Dragonfly $g$ and $r$, blue channel: Dragonfly $g$), are displayed in the top panels of Figure~\ref{fig:field} (left: Field A, right: Field B). Both fields show the presence of cirrus. Field A has a wider dynamic range in the brightness of its cirrus. A bright cirrus patch extends over 2 degrees from the lower left portion of the image to the upper right portion. Compared with Field A, the cirrus in Field B is more diffuse and mostly occupies regions away from the field center. Below we use Field A as the main example field for demonstration of our cirrus decomposition approach. However, we also show results of Field B because Field B contains a confirmed M33 dwarf satellite galaxy, And~XXII, which is a perfect test case for demonstrating the application of cirrus decomposition to LSBG searches (see Section~\ref{sec:test_AndXXII}).

{The middle and bottom rows show the 100~${\rm \mu m}$ infrared maps from IRAS and the dust radiance maps from Planck used in the following sections, respectively. For IRAS data, we use products from the Improved Reprocessing of the IRAS Survey (IRIS; \citealt{2005ApJS..157..302M})\footnote{\url{https://www.cita.utoronto.ca/~mamd/IRIS}}. Dust radiance maps are products of the Planck all-sky thermal dust models (\citealt{planck2013-p06b}), which are retrieved from the Planck Legacy Archive\footnote{\url{https://pla.esac.esa.int}}. The 100~${\rm \mu m}$ maps and Planck dust radiance maps show good spatial correspondence with the optical cirrus maps obtained by Dragonfly. Further details about the infrared and thermal dust maps will be described in the sections below.}

\begin{deluxetable*}{cccccccccc}[!htbp] \label{table1}
\tablehead{
\colhead{Field} & \colhead{RA} & \colhead{Dec} & \colhead{$l$ } & \colhead{$b$} & Area & $I_{100}\,^a\tnote{a}$ &\colhead{band} & \colhead{$N_{\rm frame}\,^b\tnote{b}$} & \colhead{$\mu_{{\rm lim,}1\sigma}{\footnotesize (60\arcsec\times60\arcsec)}\,^c\tnote{c}$} \\
               \colhead {} & \colhead {J2000} & \colhead {J2000} & \colhead {[deg]} & \colhead {[deg]} & \colhead {[$\rm deg^2$]} & \colhead {[MJy/sr]} & \colhead {} & \colhead {} & \colhead {[mag/arcsec$^2$]}}
\caption{Summary of example datasets observed by Dragonfly}
\startdata 
Field A & 01h29m36s & +26d35m42s & 133.42 & -35.50 & 4.7 & 3.2--7.4 & g & 391 & 30.8 \\ 
  &   &   &   &  & & & r & 467 & 30.2\\ 
Field B & 01h28m35.52s & +28d30m00s & 132.74 & -33.66 & 4.6 & 3.0--5.4 & g & 252 & 30.5 \\
  &   &   &   &  & & & r & 227 & 30.0 \\ 
\enddata
\begin{threeparttable}
\begin{tablenotes}
    \small
      \item[a] Range of 100 $\mu m$ intensity from IRAS as the 1\%--99\% quantiles.
      \item[b] Number of effective frames that passed the quality control.
      \item[c] $1\sigma$ surface brightness limit on a spatial scale of $60\arcsec\times60\arcsec$ measured after diffuse light removal.
\end{tablenotes}
\end{threeparttable}
\end{deluxetable*}

\section{Foreground and background Source Subtraction}
\label{sec:mrf}

In Sections~\ref{sec:morph} and \ref{sec:cirrus_color}, the diffuse light in the entire field is modeled as an entity originating from dust scattering. Prior to this step, light other than cirrus emission should be modeled and subtracted. This is particularly important for low-resolution deep imaging, such as Dragonfly data, where unresolved stars and galaxies contribute to the sky background. This section introduces the modeling of light from foreground and background sources.

We use the \texttt{MRF} package, a software developed for modeling compact sources in low surface brightness imaging (\citealt{2020PASP..132g4503V}). In brief, \texttt{MRF} takes advantage of high-resolution imaging data and finds a matching kernel between the low-resolution image and high-resolution image using non-saturated isolated stars, and convolves the high-resolution image with the kernel to build the flux models, which can then be subtracted from the Dragonfly data to leave out diffuse emission in the image. To preserve any faint diffuse sources detected in the high-resolution image, extended sources below a given mean surface brightness threshold and above a given angular scale are excluded in the flux model.

\subsection{PSF Modeling} \label{sec:psf}

One major consideration is the incorporation of wide-angle PSF treatment in the PSF modeling. The wide-angle PSF characterizes the extended wing of the PSF on scales beyond tens of arcseconds, extending even to degree scales (\citealt{1971PASP...83..199K}). The wide-angle PSF can originate from a variety of processes, including propagation of the wavefront through the turbulent atmosphere, scattering from micro-roughness and micro-ripples of optical surfaces, and diffraction within detectors (\citealt{1971PASP...83..199K}, \citealt{1996PASP..108..699R}, \citealt{2009PASP..121.1267S}). Some studies also have proposed that it can arise from the scattering of aerosols or dust in the atmosphere (e.g., \citealt{2013JGRD..118.5679D}). At low surface brightness levels, modeling the wide-angle PSF can be challenging because of the degeneracy between the extended PSF wing, the diffuse light from various sources, and the sky background. The readers are referred to \cite{2014A&A...567A..97S} and \cite{2022ApJ...925..219L} for a review of the challenges and the importance of properly characterizing the wide-angle PSF for unbiased measurement in low surface brightness imaging.

The original \texttt{MRF} algorithm uses a static extended PSF wing model. As illustrated in \cite{2022ApJ...925..219L}, the wide-angle PSF may show temporal variation due to changes in observing conditions, such as atmospheric conditions and cleanliness of lens surfaces. As a result, an instantaneous characterization of the wide-angle PSF from the image is preferred over using static PSF models when such changes are non-negligible. To handle this, we have incorporated the wide-angle PSF modeling approach in \cite{2022ApJ...925..219L} into \texttt{MRF}, in which the scattered light in the background of the field is simultaneously fitted through forward modeling of the wide-angle PSF. The extended PSF wing is modeled by a combination of a Moffat function and a double broken power law.

A key difference from the examples shown in \cite{2022ApJ...925..219L} is that in the example datasets used in this work, Galactic cirrus covers the majority of the field, so it is now a major systematic in PSF modeling, making it much more challenging to model extended wings. Changes we have made to enable the wide-angle PSF to be estimated are as follows: 
\begin{enumerate}[label=(\roman*)]
    \item The core of the PSF (within $r_{\rm core} = 15\arcsec$) is built using isolated non-saturated bright stars. In the [$3\arcmin \times 3\arcmin$] cutout of each star, we mask the target star with a circular aperture with a radius of $20\arcsec$, and then run a SExtractor-like sky subtraction with a box size of $40\arcsec$ (16 pix) to subtract the diffuse light in the background. The fractional difference threshold is slightly increased compared to normal fields because of higher photon noise in the presence of cirrus.
    \item At intermediate radii ($r_{\rm core}$ to $r_{\rm halo}$ = $50\arcsec$), the PSF model is constructed iteratively. An initial stellar halo model out to $r_{\rm halo}$ is built by stacking isolated bright stars where no saturation occurs in this range, after subtracting the mean sky background in the [$5\arcmin \times 5\arcmin$] cutout, masking any nearby fainter source, and normalizing by the surface brightness at $r_{\rm core}$. The halo model is concatenated with the core model derived in (i) to build an intermediate PSF model. For each star in the stack, a radial profile is extracted from the cutout and fitted with the PSF model to determine its flux. The fitting range is between saturation and the radius at which the profile bends upward. The target star is then subtracted from the image using the flux and the PSF model, and a 2D local background is again evaluated by the SExtractor-like sky estimator with a box size of $40\arcsec$. This local background is subtracted to remove the diffuse light contamination proceeding to the next iteration of stacking and re-evaluation. We find a couple of iterations are sufficient to yield a stable result.
    \item At large radii ($> r_{\rm halo}$), we follow the Bayesian forward modeling approach in \cite{2022ApJ...925..219L} by assigning the parametric fitting results on the PSF model derived in (ii) as priors of the outer wings of the PSF model. The PSF is modeled out to $20\arcmin$; however, at large radii the outer wings might still suffer from the cirrus bias, where the power of the wing can be overestimated. Such bias is not significant here as we do not observe clear boundary effects in the fits, but caution is required when proceeding to a larger dataset.
\end{enumerate}

We note these specific treatments are `patches' to mitigate the systematics from cirrus on wide-angle PSF modeling. A more elegant, flexible, and self-consistent approach would be to incorporate cirrus in the modeling, which is challenging and will be explored in the future.

\subsection{Building Flux Models} \label{sec:flux_model}

\begin{figure*}[!htbp]
\centering
  \resizebox{\hsize}{!}{\includegraphics{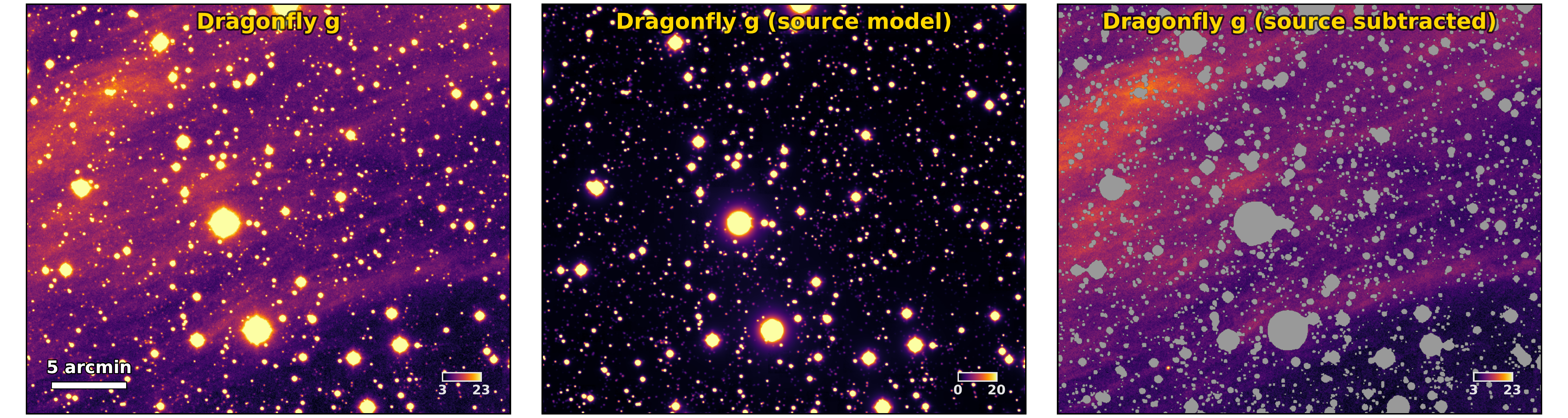}}
  \caption{Left: a [$0.6\degree\times0.5\degree$] zoom-in region of the Dragonfly g-band image of Field A. Middle: zoom-in of the source model constructed following procedures in Section~\ref{sec:mrf}. Right: the source model subtracted image. The central bright parts of stars and galaxies are masked out to 5$\sigma$. The image scale is in the unit of {kJy/sr in linear scale}.}
\label{fig:source_model_g}
\end{figure*}

With the constructed PSF model, we can proceed to flux model construction or source rendering. Similar to \cite{2022ApJ...925..219L}, the rendering is done with different treatments depending on the brightness of the source:

\begin{enumerate}[label=(\roman*)]
    \item For non-saturated sources fainter than a magnitude limit \texttt{bright\_mag\_lim}, the modeling follows \cite{2020PASP..132g4503V}. In brief, the \texttt{MRF} algorithm selects tens of isolated stars and creates a [$3\arcmin \times 3\arcmin$] cutout for each star in both low and high-resolution images. The low-resolution is upsampled by a factor of 3 using \texttt{IRAF}'s utility \texttt{magnify} and the high-resolution image is downsampled to the same pixel grid. It then computes the matching kernel for each star in the Fourier space and combines the kernels after clipping outliers. A source detection is run on the high-resolution image with a signal-to-noise ratio (S/N) of 2 and flux models are built by convolving the downsampled image with the matching kernel for the detected sources. Sources with {mean surface brightness above a limit \texttt{sb\_lim} $=24.5$ mag/arcsec$^2$ (before Galactic extinction correction)} and pixel area $>$ 40 pix$^2$ are removed from the flux models\footnote{That is to say, diffuse extended sources with mean surface brightness $>$ 24.5 mag/arcsec$^2$ and areas $>$ 40 pix$^2$ in each filter are preserved as `LSB' sources in the images. Further criteria will be required for a clean and complete detection of LSBGs, which is not the purpose of this work.}. Finally, the flux models are downsampled to the original pixel grid and subtracted from the image. We use imaging data from the Legacy survey DR9 (\citealt{2019AJ....157..168D}) as the high-resolution images. For the example fields in this work, \texttt{bright\_mag\_lim} is set to 16.
    \item For saturated bright stars ($g \lesssim 12.5$), they are rendered using normalization from profile fitting, similar to the procedures in PSF modeling above. In this case, the normalization from profile fitting is less affected by the presence of cirrus, given their high significance. The fit range is set as the range between the saturation and where the profile starts to deviate from the halo model by more than 1$\sigma$ {(the local standard deviation of the sky background using \texttt{photutils})}, where it indicates that the background systematics start to alter the profile shape.
    \item For non-saturated bright stars (brighter than \texttt{bright\_mag\_lim}), the normalization is measured from iterative PSF photometry. In each iteration, the local background in the [$5\arcmin \times 5\arcmin$] cutout is evaluated by a sky estimator with a box size of $40\arcsec$ and subtracted prior to PSF photometry. Faint stars and extended wings of bright stars contributing to the diffuse background are also subtracted during this step.
    \item Bright extended sources are currently not included in the flux models. Instead, we grow the mask from SExtractor to mask out the diffuse light from halos. More aggressive masking would be needed for nearby galaxies with large angular sizes and prominent extended disks/halos, although they are not present in the example datasets. Occasionally, some non-LSBGs fall below the `diffuse' limit and are retained in (i). These leftover extended sources and the diffuse light associated with them (e.g., halos) will also be picked out by the approach in {Section~\ref{sec:morph}} given their morphologies.
\end{enumerate}

This source model is subtracted from the image prior to cirrus modeling. The results of this process are demonstrated in Figure~\ref{fig:source_model_g}, where we show a [$0.6\degree\times0.5\degree$] region of the $g$-band image of Field A, the source model, and the source model subtracted image. Small-scale structures and the extended PSF wings are effectively removed, while large features are retained. It should be noted that the extended wings of bright sources outside the field-of-view might also contribute to the diffuse light background in the field. Therefore, we construct flux models on a larger sky area of the co-added image and use only the central area of the field in the subsequent cirrus modeling.

\section{Distinguishing Diffuse Structures Using Morphology}
\label{sec:morph}

The geometry of the diffuse ISM is largely molded by turbulence and magnetic fields (e.g., \citealt{2004ARA&A..42..211E}, \citealt{2010MNRAS.406.2713B}, \citealt{2014ApJ...789...82C}, \citealt{2023ASPC..534..153H}). As a consequence, dust emission as a tracer of the diffuse ISM has been observed to have 1D filamentary or 2D sheet-like structures both in state-of-the-art simulations (e.g., \citealt{2020MNRAS.497.4390C}) and observations in a variety of tracers (e.g., \citealt{1979ApJS...41...87S}, \citealt{2010MNRAS.406.2713B}, \citealt{2011A&A...529L...6A}, \citealt{2015A&A...579A..29B}, \citealt{2016A&A...593A...4M}, \citealt{2020MNRAS.492.5420S}). High-resolution observations of nearby galaxies have even revealed filamentary dust structures beyond the Milky Way (e.g., \citealt{2023ApJ...944L..13T}).

In contrast, LSBGs such as ultra-faint dwarf galaxies and ultra-diffuse galaxies (UDGs) have roundish (or `blobby') morphologies in their integrated light (e.g., \citealt{2015ApJ...798L..45V}, \citealt{2021ApJ...922..267C}). Therefore, a natural idea for distinguishing LSBGs from cirrus is to make use of their differences in morphologies.

% \blobby ISM may be a complex of filaments, clumps, and shells.

In recent years, various methods have been proposed to identify and extract filamentary structures from simulations/observations, including the density-based \texttt{DISPERSE} algorithm (\citealt{2011MNRAS.414..350S}) using critical manifolds, which was initially developed for identifying filaments in the Cosmic Web; the multi-scale filtering \texttt{GETFILAMENTS} algorithm (\citealt{2013A&A...560A..63M}) using wavelets; curvature-based approaches using the local Hessian matrix (\citealt{2014ApJ...791...27S}, \citealt{2015MNRAS.449.1782S}); algorithms based on mathematical morphology by \cite{2015MNRAS.452.3435K}; and more recently, approaches using machine learning (\citealt{2022arXiv220500683A}, \citealt{2023MNRAS.519.4735S}, \citealt{2023A&A...669A.120Z}).

In this section, we present a method for distinguishing blobby LSBG-like emission from patchy or filamentary `cirrus-like' emission using the Rolling Hough Transform (RHT), a widely adopted algorithm used for identifying ISM structures, for its simplicity and interpretability. Section~\ref{sec:RHT} introduces the RHT algorithm. Section~\ref{sec:D_w} discusses parameter choices. Section~\ref{sec:mask_infill} describes mask infilling. Section~\ref{sec:cirrus_df} presents results obtained by applying RHT on imaging data retrieved by Dragonfly.

\subsection{The Rolling Hough Transform}
\label{sec:RHT}

\begin{figure*}[!htbp]
\centering
  \resizebox{\hsize}{!}{\includegraphics{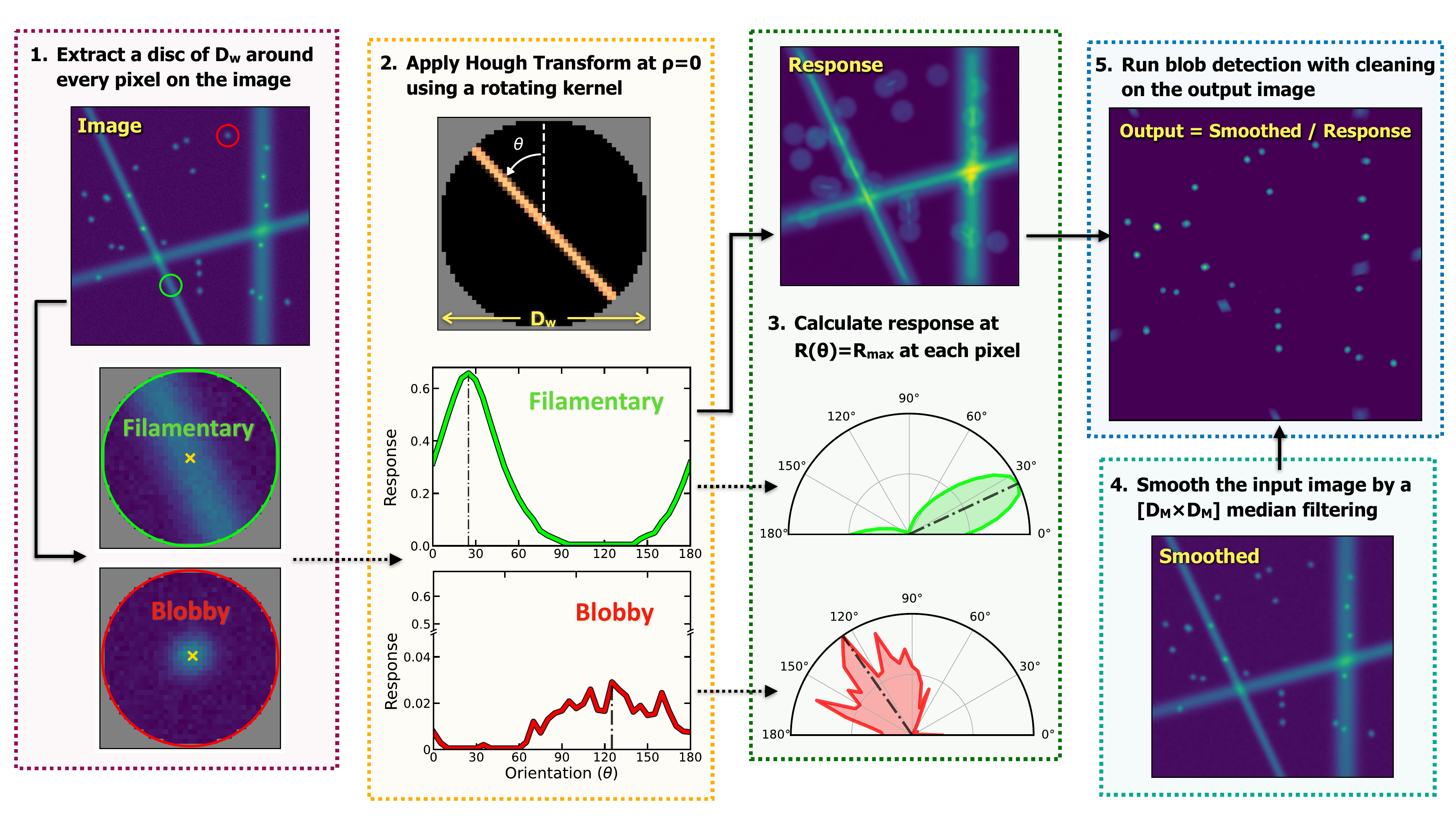}}
  \caption{Schematic of the RHT procedure applied to toy models. Step 1 extracts a local disc of diameter $D_w$ around each pixel in the input image. The disc is the window function rolling across the field. The insets show discs centering on a `filamentary' (upper) and a `blobby'  (lower) structure. Step 2 performs the Hough transform at $\rho=0$ to map the intensity in (x,y) space to response $R$ as a function of $\theta$. In this step, a filament would have significant peaks in $R(\theta)$ while $R(\theta)$ of a blob would be much flatter, as illustrated in the example insets. In each inset, a background response is subtracted for display to highlight the contrast. The $R(\theta, x, y)$ cube is used in step 3 to compute the peak response at each pixel, which is significant at a pixel belonging to filamentary structures. This step generates the response image $\widetilde{R}(x,y)$ {and (optionally) smooth it into $\widetilde{R}^s(x,y)$}. $R(\theta)$ is normalized to 0 to 1 for display. Step 4 smooths the {input image} by median filtering, {producing $\widetilde{I}^s(x,y)$}. Step 5 calculates the ratio of products in steps 3 and 4, and runs a blob detection with cleaning on the output image {$H(x,y)$}.}
\label{fig:rht_demo}
\end{figure*}

The Rolling Hough Transform is a machine vision technique developed for detecting and characterizing coherent ISM structures (\citealt{2014ApJ...789...82C}). It was initially applied to HI survey data by \cite{2014ApJ...789...82C} to quantify the alignment of HI fibers with the magnetic field. The algorithm was also successfully applied to Herschel IR data for the characterization of the ISM filaments in the Herschel Gould Belt Survey by \cite{2015MNRAS.452.3435K}. Below we provide a brief summary of the RHT algorithm and describe adaptations made to fit into our use case. The procedures are illustrated in Figure \ref{fig:rht_demo}. Detailed explanations and implementations of the original algorithm can be found in \cite{2014ApJ...789...82C}.

The RHT is a variant of the well-known Hough transform (\citealt{Hough1962}), which is a feature extraction technique, particularly for line detection, that has been widely applied in imaging analysis and computer vision. In the classical Hough transform, a straight line in Cartesian image space ($x$, $y$) is mapped into the polar parameter space ($\rho$, $\theta$) by:
\begin{equation}
    \rho = x \cos\theta + y \sin \theta
\end{equation}
where $\rho$ is the orthogonal distance to the origin and $\theta$ is the orientation (\citealt{Duda1972}). Any possible line segment in the ($x$, $y$) image space can be transformed into a single point in the ($\rho$, $\theta$) parameter space. In turn, a single point in the image space corresponds to a family of curves that overlap at the same point in the parameter space. As a result, collinear points in the image space `accumulate' and become significant in the parameter space. The Hough transform then selects local maxima in the parameter space that passes a specified threshold as candidates for linear features.

For the RHT, the key adaption from the Hough transform is to restrict $\rho=0$ and define the origin as the center of a circular domain (with a disc diameter $D_w$) placed on a given pixel on the image, and map the intensity distribution $I(x,y)$ in the image space within the domain to intensities in the RHT parameter space. At a given pixel ($x_i$, $y_i$), the transformed intensity (called the `response'):
\begin{equation}
R_i(\theta)=\iint_{disc} R(\theta, x_i,y_i)\,dxdy
\end{equation}
is a single variable function of $\theta$ (because $\rho=0$) summing over the disc, which measures the significance of linear structure at different orientations in the local neighborhood of that pixel. By `rolling' the circular disc across the image, one yields a distribution of $R(\theta, x, y)$.

We follow \cite{2014ApJ...789...82C} by defining $\theta=0 \degree$ to be the positive y-axis, while $\theta$ varies between $[0\degree, 180\degree)$ given the periodic behavior outside the domain. In practice, the response $R(\theta, x, y)$ is calculated at discretized $\theta$ values with a binning step. For larger $D_w$, the bin width needs to be smaller given the larger change in $R$ between steps. We adopt the rule of thumb binning used in \cite{2014ApJ...789...82C} for the number of $\theta$ bins in the domain $[0\degree, 180\degree)$:
\begin{equation} \label{Eq:theta_bin}
n_\theta = \pi\,\frac{\sqrt{2}}{2}\,(D_w-1),
\end{equation}
where $D_w$ is in pixel units and {$n_\theta$ is rounded up to an integer}. This mapping is done by convolving the image with a linear kernel with a rotating position angle relative to the image y-axis using the middle value of the bin at a step of $\delta\theta=\frac{\pi}{n_{\theta}}$.

The main purpose of the RHT approach in \cite{2014ApJ...789...82C} is to identify the ridge line of ISM structures (the `skeleton') as a probe of the interstellar magnetic field and to characterize its orientation. In \cite{2014ApJ...789...82C}, the original image was smoothed by a top-hat kernel and subtracted from the unsmoothed image, and the RHT was performed on this residual image to remove the underlying continuum of ISM. Our aim is to distinguish `blobs' from `cirrus-like' emission, and we therefore do not subtract the large-scale smooth component before doing the RHT. Instead, we compute the maximum of the response $R(\theta, x, y)$ at each pixel over $\theta$, $\widetilde{R}(x, y)$, as the peak response:
\begin{equation}
\widetilde{R}(x_i, y_i) = max\left\{ R(\theta\,|x=x_i,{\ }y=y_i)\right\}\,.
\end{equation}
This is based on the fact that $\widetilde{R}(x, y)$ is enhanced at pixels belonging to an extended `cirrus-like' structure tracing the filamentary or patchy morphology, while it represents the local mean intensity within the domain at pixels belonging to a blob without clear directional preference. 

When using $\widetilde{R}(x, y)$ to identify blobs based on their morphologies, two facts need to be taken into account: (1) a blob may have an elliptical morphology with elongation in one direction and hence non-negligible significance relative to the background response, and (2) nearby blobs (within the domain specified by $D_w$) at similar surface brightness or overlap with dense regions of cirrus patches may be misidentified as a directional preference. Note that for (2), it is not frequent in reality that two nearby blobs are real LSBG candidates, given their relative sparsity. However, contamination from the leftovers of star/galaxy subtraction may also contribute to the response computation. To mitigate these sources of confusion, $\widetilde{R}(x,y)$ is {optionally} smoothed by a {square} [$D_M \times D_M$] median filter, denoted by $\widetilde{R}^s(x,y)$\footnote{{The smoothing is applied on all the results of real and simulated data in this paper, except the toy model in Figure ~\ref{fig:rht_demo}. This is to better preserve one of the mock filaments with sharp boundaries and a small width relative to the smoothing filter size. The toy model is for demonstrative purposes only, and real cirrus structures would be more diffuse and extended.}}. We test that modifying $D_M$ does not dramatically affect the result, as long as it is sufficiently large to smear out the blobs. {Very small $D_M$ values should also be avoided to avoid pixelization effect.} We then look at the ratio between $\widetilde{R}^s(x,y)$ {(or $\widetilde{R}(x, y)$ for the toy model)}
 and the image smoothed by the median filter, denoted by $\widetilde{I}^s(x,y)$: 
\begin{equation}
H(x,y) = \widetilde{I}^s(x,y) \,/\, \widetilde{R}^s(x,y)\,.
\end{equation}
On $H(x,y)$, any structure extending on scales larger than $D_w$ in 1D/2D is suppressed. Effectively, this mapping from $I(x,y)$ to $H(x,y)$ subtracts a large-scale background along the `manifolds' based on local connectivity on scale $D_w$, in comparison with the box estimator used by conventional sky subtraction. 

Source detection is then run on the output image $H(x,y)$. Practically, this is performed using the \texttt{detect\_sources} and \texttt{deblend\_sources} utilities in the \texttt{photutils} package. Sources with very small axis ratios (\textit{b/a}) are cleaned from the list, which are likely contamination from dense regions of cirrus. A mask map is generated on the detections to mask out blobs, which are LSBG candidates. The masks are morphologically dilated with three iterations to include the outskirts of the blobs. This masked image is then infilled (see Section~\ref{sec:mask_infill}) proceeding to the color modeling of the cirrus in Section~\ref{sec:cirrus_color}. A demonstration of the algorithm on simulated galaxies and cirrus is presented in Appendix \ref{appendix:rht}.

\subsection{Choice of \texorpdfstring{$D_w$} {\,} in RHT} \label{sec:D_w}
The disc size, $D_w$, is an input parameter chosen by the user. In fact, it is one advantage of the RHT that $D_w$ can be changed to identify filamentary features on scales of interest (\citealt{2014ApJ...789...82C}). However, as \cite{2015MNRAS.452.3435K} pointed out, using a too-small $D_w$ will result in a pixelization bias where intensities along the x, y, and diagonal axes dominate, while using a too-large $D_w$ can potentially wipe out the structural information. \cite{2015MNRAS.452.3435K} adopted a $D_w$ of three times the beam width for ISM in the Herschel data. For our purpose, we would like to distinguish (relatively) small blobs, which are LSBG candidates or other sources, from the large-scale cirrus. The disc needs to be sufficiently large to fully contain the blob to differentiate it from cirrus structures extending on larger scales, but not so large as to include big cirrus patches. Based on the empirical rule for the disc size being at least three times larger than the scale of the phenomena of interest, we adopt a minimum of $D_w = 3\arcmin$, considering that very few LSBG candidates have angular extent larger than 30$\arcsec$ in their effective radii (\citealt{2016MNRAS.456.1359F}, \citealt{2022ApJS..261...11Z}). In practice, we have tweaked $D_w$ between 3$\arcmin$ to 6$\arcmin$ to optimally extract cirrus information while removing contamination, although the difference in performance is not dramatic. We note faint extended sources larger than this scale may be misidentified as cirrus patches. Meanwhile, at this stage cirrus at small scales with analogous morphologies to LSBG candidates (such as knots and clumps at high dust column density regions) may also cause confusion, which is one of the main motivations for using colors to further refine the discrimination in Section~\ref{sec:cirrus_color}.

\subsection{Infilling of Masked pixels}
\label{sec:mask_infill}

The core regions of bright stars and the blobs detected on the output image (including intrinsic LSBG candidates and the contamination from residuals of the star/galaxy subtraction) are masked in the image. This section describes the implications of this masking and describes the infilling of missing data.

%In general, this is attributed to the non-trivial problem of reconstructing missing pixels in image analysis/computer vision. 
Modern non-parametric machine learning techniques have been developed to tackle the problem of filling missing data in the images, e.g., using generative neural networks. Conventional statistical approaches, such as Gaussian process regression (GPR), are also popular and, in many cases, more robust and explainable. In astronomy, GPR has been widely used for interpolating missing/bad data (e.g., \citealt{2015ApJ...812..128C}). In particular, \cite{2022ApJ...933..155S} developed a method called Local Pixel-wise Infilling (LPI) that predicts the ISM background and its uncertainty behind foreground sources to improve source photometry. The LPI approach is similar to GPR, but does not need optimization over kernel parameters by using a non-parametric kernel estimated from local pixel covariance.

Here we employ an iterative mask infilling approach using the software \texttt{maskfill} (\citealt{2024PASP..136c4503V}). \texttt{maskfill} is a simple and robust method that performs inward extrapolation on the masked pixels using edges of unmasked pixels, leading to a smoothly varying spatial resolution in the filled regions and a seamless transition at the edges. Details about the algorithm can be found in \cite{2024PASP..136c4503V}.
This mask infilling approach avoids the deficiency of convolution-based interpolation using a fixed kernel, in which a too-small kernel cannot fill large `holes' and a too-big kernel produces over-smooth interpolation across the field. A comparison with results using a more time-consuming GPR approach is presented in Appendix \ref{appendix:infill}, which has similar outputs but is much slower in computational efficiency. However, it is promising to apply the LPI approach to infill the cirrus map as a more robust and physically driven solution, which we will explore in future work.

\subsection{Application on Dragonfly Imaging}
\label{sec:cirrus_df}

\begin{figure*}[!htbp]
\centering
  \resizebox{\hsize}{!}{\includegraphics{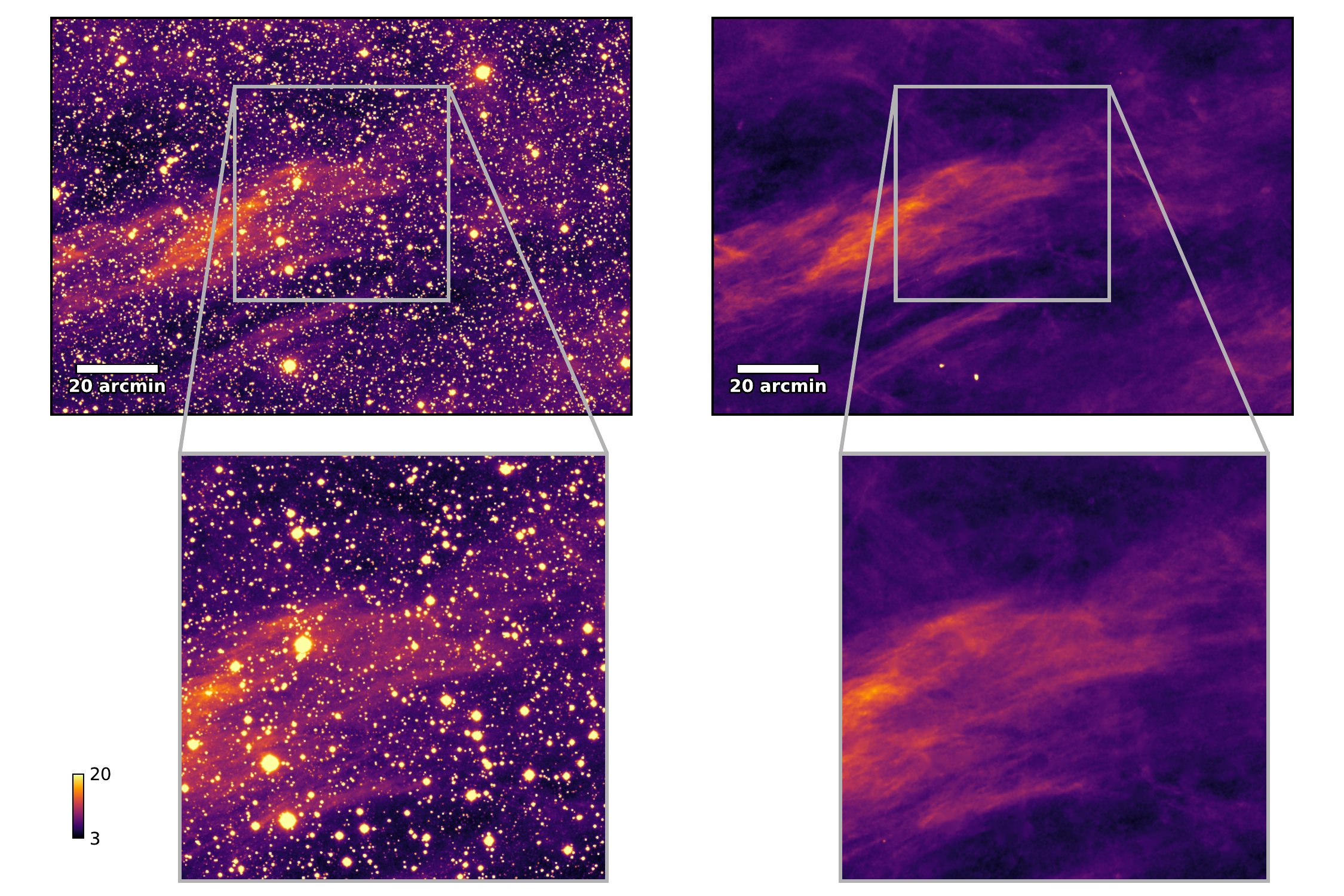}}
  \caption{Cirrus decomposition on Field A based on morphological information using RHT. Left: The central [$2.4\degree\times1.8\degree$] of the Dragonfly g-band observation of Field A. Right: The decomposed `cirrus-like' emission with patchy or filamentary structures extending on scales above $D_w=3\arcmin$. Masked pixels are filled by interpolation from nearby pixels. The insets display the zoom-out of a [$1\arcdeg \times 1\arcdeg$] cutout in the high column density regions of the image. The image scale is in the unit of {kJy/sr in linear scale}.}
\label{fig:cirrus_g}
\end{figure*}

Figure~\ref{fig:cirrus_g} presents the result of applying the above techniques to a deep image obtained by Dragonfly. The input image is a single-band image ($g$ or $r$ for Dragonfly), after subtracting the flux model (Section~\ref{sec:mrf}). On the input image, the pixels with values above 3 median absolute deviation (MAD) in the flux models are masked to exclude the poorly modeled and sampled cores, where MAD is the median absolute deviation of the image iteratively calculated after applying the mask. A preliminary mask infilling is done following as in Section~\ref{sec:mask_infill} to remove small `holes', which are mainly the central few pixels of fainter sources. The input image is then binned by [$4\times4$] using a median binning to increase the S/N.

We then applied RHT to the image using a disc size $D_w=3\arcmin$ and a smoothing size of $D_M$ = 5 pixels. The number of $\theta$ bins follows Equation \ref{Eq:theta_bin}. The output image contains signals with blobby morphologies within the scale of $D_w$. For blob detection, we adopt a detection threshold of 3 {times of standard deviation of the output image}, a deblending threshold of 0.001, and a number of deblending levels of 64. Detections with axis ratio $b/a<0.5$ are excluded to remove contamination from compact cirrus emission. 

The right panel of Figure \ref{fig:cirrus_g} shows the extracted `cirrus-like' emission in the central $2.6\degree \times 1.8\degree$ area of Field A in g-band. The original image is displayed to the left. The majority of light from stars (including the extended PSF wings) and galaxies in the field has been removed using approaches in Sec.~\ref{sec:flux_model}. Furthermore, blobby emissions are also removed, yielding a clean representation of `cirrus-like' emission in the field. {Two notable objects are visible near the middle bottom of the image though: the left object is a galaxy missed by the source modeling due to the presence of a nearby very bright star ($\rm V\sim7.2$ mag), and the right object is a galaxy improperly modeled by the source modeling. Future work will contribute to improving the source modeling to reduce such contaminations.}. 

Overall, the performance demonstrates the power of the approach in extracting `cirrus-like' emission in the image. However, as mentioned earlier, this approach works on a single filter and does not account for the physical correlation of cirrus emission between filters. We extend this method by building a color model as described in the following Section, which enables a full analysis of cirrus using multi-band photometry.

% Our approach using the RHT can be viewed as a `directional' low-pass filtering, which filters out signals without a clear directional preference on large scales.

% One is not recommended to look for candidates in the densest regions of cirrus given the contaminations there. 

% The original algorithm in Clark et al. (2014) includes a subtraction of the smooth component to suppress signals large-scale structures.

% Mask refill using regression with Gaussian kernel vs local covariance: Gaussian kernel is positive finite. Therefore it is not well performed in voids (flux overestimated). But given that we mostly mask cirrus knots, it is OK.

\section{Physical Constraints on Galactic Cirrus Based on Colors}
\label{sec:cirrus_color}

Another consideration for disentangling cirrus from extragalactic sources is to exploit the different origins of their emission, which should result in different colors. If cirrus is well-constrained in its spectral energy distribution (SED) locally, we can combine color information of cirrus with the decomposition method using morphological information (which is based on a single band). This section applies color constraints to the products (i.e., the `cirrus-like' emission maps) presented in Section~\ref{sec:morph}.

Cirrus emission at optical wavelengths primarily originates from the scattering of ISRF off dust grains\footnote{This does not take into account the possible luminescence of dust grains in NIR to optical bands, or the so-called `Extended Red Emission' (ERE). Studies around some reflection nebulae have shown evidence of excess light that cannot be explained by scattering alone (\citealt{2006ApJ...636..303W}). The ERE is suggested to have its origin in the interaction of far-UV photons with dust materials that yet have not been well understood (e.g., PAH$^{++}$). Note some studies favor the presence of ERE in optical DGL (\citealt{2008ApJ...679..497W}) while others suggest the opposite (\citealt{1999A&A...352..645Z}). Caution needs to be used in cases where the diffuse light comprises components different from dust scattering. ERE will be explored in further detail in a future work.}, whereas the integrated light of LSBGs is emitted by their own stellar populations. The SED of the cirrus at visible wavelengths is dependent on the properties of dust grains (which determine the absorption and scattering cross sections), the scattering phase function, and the illuminating ISRF. It is expected that the SED of the diffuse scattered light has spatial variation depending on the position in the Milky Way (\citealt{2017ApJ...849...31S}). However, many observations have shown that along different line-of-sights, the FIR and optical intensities of cirrus emission are well correlated (e.g., \citealt{2013ApJ...767...80I}, \citealt{2020A&A...644A..42R}, \citealt{2023MNRAS.524.2797M}, and reference therein). Therefore, it is worth investigating whether one can decompose the cirrus by assuming a fixed shape for its SED at a given line of sight, i.e., applying constraints in the optical colors of the cirrus.

Interstellar dust grains are primarily heated by starlight, and cool via re-radiation in mid-to-far infrared and submm. Here we make the following assumptions regarding the dust populations and the ISRF: (1) dust grains are in local thermal equilibrium (LTE), (2) physical properties (size distribution, composition, etc.) of the dust populations in the line-of-sight are similar, and (3) illumination from ISRF is homogenous. Very small grains can be heated far above equilibrium by hard-UV photons (\citealt{2001ApJ...551..807D}) and overshine in NIR. However, these are not the same population that contributes most to the scattered light in optical; at $\lambda \sim 0.6 \mu m$, dust scattering is dominated by large grains with sizes $a > (\lambda/2\pi) \approx 0.1 \mu m$ (\citealt{2011piim.book.....D}). These assumptions state that in optical bands (here for Dragonfly, $g$ and $r$) the scattered light should be correlated with the amount of light that is thermally emitted in FIR. As a result, the scattered light in different optical bands should be well correlated with each other. The decomposition of the cirrus with color constraints is done by identifying and extracting the corresponding amount of diffuse light in each band. 

In Section~\ref{sec:color_planck}, we correlate the Dragonfly observations with Planck products. In Section~\ref{sec:color_optical}, we correlate the Dragonfly $g$ and $r$-band data. In Section~\ref{sec:cirrus_removal}, we present the results of cirrus removal on the example dataset with color constraint based on the color model and the products in Sec.~\ref{sec:morph}. Section~\ref{sec:metric} shows metrics for evaluating the performance of the cirrus removal algorithm.

\subsection{Correlation with Planck Thermal Dust Model}
\label{sec:color_planck}

We first correlate the Dragonfly observations in $g$ and $r$ with the all-sky thermal dust model derived from Planck observations (\citealt{planck2013-p06b}). There are two major purposes: (1) to verify the correlation between optical scattered light and FIR dust emissions, and (2) to determine the zero-points to convert surface brightness in ADU/pixel into physical units (kJy/sr). The Planck dust model is retrieved from the Planck Legacy Archive. For the full description of the Planck thermal dust model, readers are referred to \cite{planck2013-p06b} (see also \citealt{planck2016-XLVIII}).

\subsubsection{Dust Tracer from Planck All-sky Thermal Dust Model} \label{sec:dust_tracer}

For dust in LTE, the optical depth is often used as a reliable tracer of the dust column density. The frequency-dependent optical depth $\tau_\nu$ is given by:
\begin{equation}
\label{eq:tau}
\tau_\nu = \sigma_{e,\nu}  N_H\,,
\end{equation}
where $\sigma_{e,\nu}$ is the dust emission opacity and $N_H$ is the gas column density. Alternatively, $\tau_\nu$ can be expressed in the form of:
\begin{equation}
\tau_\nu = \kappa_\nu  M_{{\mathrm d}}\,,
\end{equation}
where $\kappa_\nu$ is the dust emissivity and $M_{{\mathrm d}}$ is the dust mass column density. In the Rayleigh-Jeans limit, $\kappa_\nu$ is mostly described by a power law: $\kappa_\nu=\kappa_0 \cdot (\nu/\nu_0)^{\beta}$ (\citealt{1983QJRAS..24..267H}), which leads to the frequency dependence of $\tau_\nu$:
\begin{equation}
\label{eq:tau_freq}
\tau_\nu = \tau_0 \cdot (\frac{\nu}{\nu_0})^{\beta}\,,
\end{equation}
Below we adopt the reference optical depth at Planck {reference frequency $\nu_0=$ 353 GHz} (denoted by $\tau_{0}$ or $\tau_{353}$), which is derived from fitting the dust SED with the empirical modified black body approach (\citealt{planck2013-p06b}). Under LTE, the specific intensity of thermal emission $I_\nu$ is related to $\tau_\nu$ by:
\begin{equation}
\label{eq:I_nu}
I_\nu = \tau_\nu B_\nu(T)\,,
\end{equation}
where $B_\nu(T)$ is the Planck function for a black body at dust temperature $T$.

In \cite{2023ApJ...953....7L} we discuss the difference between using the radiance $\mathcal{R}$ and $\tau_{353}$ as the dust surrogate for their optical counterpart, i.e., the optical DGL. 
The radiance $\mathcal{R}$ is defined as the integral of thermal emission:
\begin{equation}
    \label{eq:radiance}
    \mathcal{R} = \int I_\nu d\nu\,.
\end{equation}
Assuming constant dust-to-gas ratio and other line-of-sight properties:
\begin{equation}
    \mathcal{R} \propto U \overline{\sigma_a} N_H\,,
\end{equation}
where U is the scaling factor of ISRF (U=1 is the local ISRF) depending on Galactic latitude, and $\overline{\sigma_a}$ is the absorption opacity, defined similarly to $\sigma_{e,\nu}$. At high galactic latitude, both $\mathcal{R}$ and $\tau_{353}$ are good tracers of dust column density given the relatively small variation in U and the dust opacity (\citealt{planck2013-p06b}). 

In Appendix \ref{appendix:dust}, we demonstrate that both tracers are expected to be well correlated with the optical scattered light under the aforementioned assumptions and several approximations. Here, we use $\mathcal{R}$ for the reason that the optical depth map presents larger scattering at small scales, which are smoothed out in the radiance map through integration. Furthermore, $\mathcal{R}$ is less affected by optical depth effects. In the optically thin regime, because of the large beam width of Planck {($\sim 5 \arcmin$)} compared to Dragonfly, the results using $\tau_{353}$ will be similar. The radiance maps of the example dataset are shown in the bottom panels of Fig. \ref{fig:field}.

\subsubsection{Linear models} \label{sec:model_planck}

To correlate Dragonfly data with the relatively low-resolution Planck dust map, we first subtract a median sky background value from the image in each band, and then convolve the PSFs of Dragonfly images to the beam width of Planck with apodization near the field edges. The Dragonfly data is downsampled to 10$\arcsec$ resolution to smooth out small structures and noise, and 0.1\% of data are clipped out as outliers. The median sky background value from the pipeline is likely to be biased by the presence of diffuse light in the image. To correct this bias, the pixel intensities in Dragonfly data are shifted by a constant sky value. We use the intercept pixel intensity, $a_{\lambda, p}$, as the background value to convert the intensities to physical units, assuming that the diffuse light from dust scattering should equal zero where the Planck dust tracer indicates there is no dust\footnote{This does not take into account other physical contributions to the optical diffuse light, including the Extragalactic background light (EBL) and the diffuse ionized medium, neither does contribution from dust in non-thermal equilibrium. Therefore the \textit{intrinsic} zero-point for scattered light from dust in Dragonfly observations should be slightly lower than $a_{\lambda}$. However, these are higher-order effects since EBL is much fainter than DGL in most sky areas involved here and diffuse ionized medium is typically faint at high Galactic latitudes. Furthermore, dust in the area of interest is mostly in LTE in the absence of ionizing sources.}. A linear correlation between Dragonfly data and the Planck thermal dust tracer $x_{p}$ (here dust radiance $\mathcal{R}$) is fit:
\begin{equation}
\label{eq:corr_df_plank}
    I_{\lambda} = a_{\lambda, p} + b_{\lambda, p} \cdot x_{p} \,,
\end{equation}
where $I_{\lambda}$ represents the surface brightness intensities of $g$ and $r$ data in [$kJy\cdot sr^{-1}$], respectively. 

In regions at high intensities, observations indicate that the correlation between optical and FIR data deviates from a single linear correlation (\citealt{2013ApJ...767...80I}, \citealt{2020A&A...644A..42R}, \citealt{2023MNRAS.524.2797M}, \citealt{2023ApJ...948....4Z}).  This non-linear part could be due to several factors, including optical depth effects (attenuation, multiple scattering, etc.), variations in the scattering cross-section, and changes in dust emissivity. To account for the possible break, alternatively, we fit a piecewise linear model between Dragonfly data and Planck:
\begin{equation}
\label{eq:corr_df_plank_cond}
    I_{\lambda} =
    \begin{cases}
    a_{\lambda, p} + b_{\lambda, p} \cdot x_{p}\,,& x_{p}\leq x_{\lambda}^c\\
    %c_{\lambda} \cdot (\tau_{353} - \tau_{353,c}) + b_{\lambda} + I_{\lambda,c},              & \tau_{353} > \tau_{353,c}
    c_{\lambda, p} + d_{\lambda, p} \cdot x_{p}\,.& x_{p} > x_{\lambda}^c
    \end{cases}
\end{equation}
The critical threshold $x_{p}^c$, at which the single linear correlation begins to break, is given by:
$x_{p}^c =  - (a_{\lambda, p} - c_{\lambda, p}) / (b_{\lambda, p} - d_{\lambda, p})$. The critical intensity in optical is: $I_{\lambda}^c = b_{\lambda, p} \cdot x_{p}^c + a_{\lambda, p}$. 
%$I_{p,c}$ and $I_{\lambda,c}$ represent the critical intenisties.

Equation \ref{eq:corr_df_plank_cond} could return a smaller residual because of the higher degree of freedom in the fitting. Therefore to do a model selection between Eq.~\ref{eq:corr_df_plank} and Eq.~\ref{eq:corr_df_plank_cond}, we calculate the Bayesian {Information} Criterion (BIC) of the best fit for each model:
    $\mathrm {BIC} =k\log(N)-2\log({\widehat {L}})\,,$
where $N$ is the sample size, $k$ is the number of free parameters, and $\widehat{L}$ is the likelihood function evaluated at the point of the maxima. The model with lower BIC is preferred. The best-fitted intercept at $x_{p}=0$, $a_{\lambda, p}$, is used as the new background value to convert $I_{\lambda}$ to physical units.

\begin{figure}[!htbp]
\centering
  \resizebox{\hsize}{!}{\includegraphics{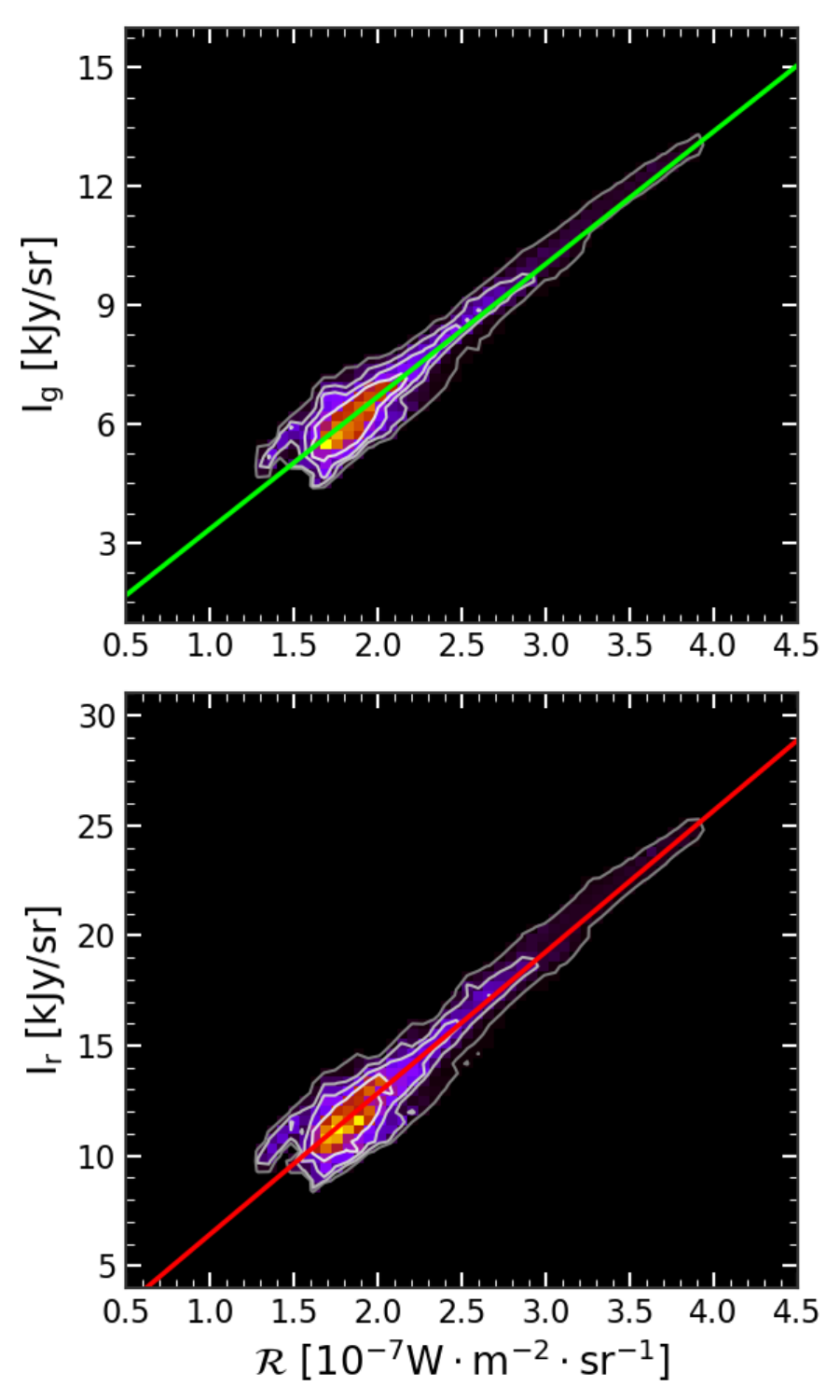}}
  \caption{Correlations of the Dragonfly $g$ and $r$ surface brightness intensities with Planck radiance $\mathcal{R}$ for the diffuse light in Field A. The Dragonfly data is convolved to Planck beam-width and binned 4x4. 
  %In both bands, a broken-linear model is preferred over a single linear model based on the BIC criterion, indicating a flattening at high dust column densities. 
  The linear fitting is indicated by the solid line.
  %The main linear component is indicated by the solid line and the flattened component is indicated by the dashed line. 
  The contours from outermost to innermost contain 99.5\%/90\%/50\% of the total data. The zero-points are shifted such that the intensity equals zero at the intercepts, assuming that there is no diffuse optical light from dust scattering where there is no dust.}
\label{fig:corr_planck}
\end{figure}

Figure \ref{fig:corr_planck} shows the correlations of the Dragonfly $g$ and $r$, after the zero-point shift, with Planck dust optical depth in Field A. At low intensities, both $g$ and $r$ data are well correlated with $x_{p}$. This is consistent with the correlation shown in Figure 3 of \cite{2023ApJ...948....4Z} using Herschel 250 $\mu m$ data. For this field, the model of Eq.~\ref{eq:corr_df_plank} has lower BIC. Therefore, no flattening, i.e., optical effects, is preferred in either band using radiance as the dust tracer. The ratio of the two fitted slopes, $b_{r,p}/b_{g,p}$, is $1.92\pm0.06$, which translates to $g-r = 0.70 \pm 0.03$. Note this color measurement is based on correlations with Planck, compared to that measured directly from Dragonfly data in the next section. The fitting results are summarized in Table \ref{table2}. The uncertainties include systematic errors in the photometric zero-points and fitting uncertainties estimated from bootstrap.

We apply the same model fitting on Field B. In this field, the cirrus is more diffuse than Field A with a smaller dynamical range in $x_p$. The model following Eq.~\ref{eq:corr_df_plank} is preferred with lower BIC, and therefore no clear flattening is detected. The results from the best fit are summarized in Table \ref{table2}, with a bluer color $g-r=0.56\pm0.03$ than Field A. Overall, the results show that there exists a good correlation between Dragonfly optical data and the dust tracer from Planck for the diffuse Galactic cirrus in both $g$ and $r$ bands.

We performed a similar analysis using the optical depth $\tau_{353}$ as the dust tracer $x_p$. The correlations show deviation at high intensities, similar to results in \cite{2023ApJ...948....4Z}. For Field A, the models according to Eq.~\ref{eq:corr_df_plank_cond} have lower BIC in both bands and therefore, they prefer a `bending' caused by optical depth effects. 
The transition occurs around $\tau_{353}^c\sim1.2\times10^{-5}$, or correspondingly, $I_g \sim 9$ kJy/sr and $I_r \sim 19$ kJy/sr\footnote{The detected transition in the example dataset occurs roughly at E(B-V)$\sim$0.16 based on the Planck dust model. Assuming an optical total-to-selective extinction ratio $R_V$ of 3.1, this corresponds to $A_V \sim 0.5$ or $\tau_V \sim 0.45$.}. The results are consistent with the statement that above a certain optical depth threshold, the dust is no longer diffuse or translucent to scattered light, and optical depth effects, including self-attenuation, reddening, and multiple scattering, become non-negligible.

\begin{deluxetable*}{ccccccc} \label{table2}
\tablehead{
\colhead{Field} & \colhead{Model} & $\left<\mathcal{R}\right>^\dagger$ & \colhead{$b_{r,p}$} & \colhead{$b_{g,p}$} & \colhead{$b_{r,p}/b_{g,p}$} & \colhead{$g-r$}
}
\caption{Field-averaged results from the correlation of Dragonfly $g$ and $r$ with Planck radiance.}
\startdata
Field A & Eq.~\ref{eq:corr_df_plank} & 1.9 & $6.58\pm0.15$ & $3.44\pm0.07$ & $1.92\pm0.06$ & $0.70\pm0.03$ \\ 
Field B & Eq.~\ref{eq:corr_df_plank} & 1.6 & $5.46\pm0.13$ & $3.27\pm0.07$ & $1.67\pm0.05$ & $0.56\pm0.03$ \\
\enddata
\begin{threeparttable}
\begin{tablenotes}
    \small
       \item $\dagger$ Field median radiance in the unit of $\rm [10^{-7}\, W\,m^{-2}\,sr^{-1}]$.
\end{tablenotes}
\end{threeparttable}
\end{deluxetable*}
% \begin{figure*}[!htbp]
% \centering
%   \resizebox{0.8\hsize}{!}{\includegraphics{corr_Dragonfly_Planck.png}}
%   \caption{.}
% \label{fig:corr_planck}
% \end{figure*}

\subsection{Correlation in Optical: The Color Model}
\label{sec:color_optical}

In this section, we correlate the Dragonfly observations in $g$ and $r$ bands and build a simple color model to explain the diffuse light emitted by dust scattering. This color model determines the amount of light in $g$ to be removed from the `cirrus-like' emission map in $r$ produced in Sec.~\ref{sec:morph}, and vice versa. This is supported by the result in Sec.~\ref{sec:color_planck} where both $g$ and $r$ imaging data show a good correlation with Planck at the resolution of Planck beam-width. The zero-points of pixel intensities are from the correlations with Planck data.

To do the correlation in optical bands, we reduce the pixel resolution to $10\arcsec$ by running a [$4\times4$] median binning on Dragonfly $g$ and $r$ images to increase S/N, and calculate the MAD of the images. Pixels with intensities 20 MAD higher or lower than the median sky are clipped as outliers, which accounts for $<0.3\%$ of the total. 

Similar to Sec.~\ref{sec:color_planck}, we build a linear model for the Dragonfly $g$ and $r$ data\footnote{Note that in general, the photometric data requires a PSF matching. We skip this because the difference of the Dragonfly PSF in the SDSS $g$ and $r$ bands is very small. For multi-band analysis across a wide range of wavelengths, e.g., using LSST, where PSF can vary in different bands, the images need to be convolved into the same PSF prior to the modeling.}. To prune the diffuse light from sources other than dust scattering, including LSBGs, stars, galaxy halo light, and possible contributions from EBL, we build a generative mixture model that includes an outlier population. The pixel intensity at pixel $i$ (e.g., in g-band) is:
\begin{equation} \label{eq:r_to_g}
    I_{g,i} =  q_i \cdot \left( A  + B \cdot I_{r,i} \right) + (1-q_i)\cdot I_{\rm bg, i}\,,
\end{equation}
where $q_i$ is a {0 or 1 binary integer} assigned to each pixel and $I_{\rm bg, i}$ belongs to a broader background (outlier) population: $I_{\rm bg, i}\sim\mathcal{N}(m_{\rm bg}, \sigma_{\rm bg}^2)$, described by its mean and variance, $m_{\rm bg}$ and $\sigma_{\rm bg}^2$\footnote{{The outlier population here, by its nature, should indeed be non-Gaussian. However, for the purpose of outlier pruning, the outlier model is not required to be accurate but rather, more importantly, to be included (\citealt{2010arXiv1008.4686H}). Given that $\sigma_{\rm bg}$ is much larger than $\sigma_i$, the difference of outliers superimposed on different underlying backgrounds would have negligible effects on the derivation of the key parameters (here A and B).}} Following \cite{2010arXiv1008.4686H}, the likelihood $\mathcal{L}$ is:
\begin{eqnarray}\displaystyle \label{eq:likelihood}
\mathcal{L} & = & p(\allI|A,B,\allq, m_{\rm bg},\sigma_{\rm bg}^2)
 \nonumber\\
  &=& \prod_{i=1}^N
 \left[p_{\rm fg}(I_{g,i}|A,B)\right]^{q_i}\cdot
 \left[p_{\rm bg}(I_{g,i}| m_{\rm bg},\sigma_{\rm bg}^2)\right]^{(1-q_i)}
 \nonumber\\
  &=& \prod_{i=1}^N \left[\frac{1}{\sqrt{2\,\pi\,\sigma_{i}^2}}
 \,\exp\left(-\frac{[I_{g,i}-A-B\cdot I_{r,i}]^2}{2\,\sigma_{i}^2}\right)\right]^{q_i}
 \nonumber \\ & \, & \,\times 
 \left[\frac{1}{\sqrt{2\,\pi\,[\sigma_{\rm bg}^2+\sigma_{i}^2]}}
 \,\exp\left(-\frac{[I_{g,i}-m_{\rm bg}]^2}{2\,[\sigma_{\rm bg}^2+\sigma_{i}^2]}\right)\right]^{(1-q_i)}\,,
\end{eqnarray}
where $\sigma_{i}$ stands for the uncertainties, $p_{fg}$ and $p_{\rm bg}$ are the {probability distribution functions} for the foreground (target) and background (outlier) points. {To marginalize over $q_i$, note that $q_i$ follows the binomial probability:}
\begin{equation} \label{eq:qi_binomial}
     p(\allq|f_{\rm bg}) = \prod_{i=1}^N (1-f_{\rm bg})^{q_i} f_{\rm bg}^{(1-q_i)}\,,
\end{equation}
{where $f_{\rm bg}$ is the probability that a pixel is drawn from the background (outlier) population. The likelihood therefore integrates to:}
\begin{eqnarray}\displaystyle \label{eq:likelihood_marginalized}
\mathcal{L} & = & p(\allI|A,B,f_{\rm bg}, m_{\rm bg},\sigma_{\rm bg}^2)
 \nonumber\\
 &=& \prod_{i=1}^N
 \left[ (1-f_{\rm bg})\,p_{\rm fg}(I_{g,i}|A,B)\right] + f_{\rm bg}
 \left[p_{\rm bg}(I_{g,i}| m_{\rm bg},\sigma_{\rm bg}^2)\right]
 \nonumber\\
 &\propto& \prod_{i=1}^N \biggl[\frac{1-f_{\rm bg}}{\sqrt{2\,\pi\,\sigma_{i}^2}}
 \,\exp\left(-\frac{[I_{g,i}-A-B\cdot I_{r,i}]^2}{2\,\sigma_{i}^2}\right)
 \nonumber \\ & \, & \, +
 \frac{f_{\rm bg}}{\sqrt{2\,\pi\,[\sigma_{\rm bg}^2+\sigma_{i}^2]}}
 \,\exp\left(-\frac{[I_{g,i}-m_{\rm bg}]^2}{2\,[\sigma_{\rm bg}^2+\sigma_{i}^2]}\right)\biggr]\,.
\end{eqnarray}

The best parameter set is given by maximizing the log-likelihood function on the data, {where free parameters include $A$, $B$, $f_{\rm bg}$, $m_{\rm bg}$ and $\sigma_{\rm bg}$.} {The uncertainty $\sigma_{i}$ is estimated with the local standard deviation of the sky background using \texttt{photutils}}. A similar model can be constructed mapping from g-band to r-band.

% \begin{deluxetable}{ccccc} \label{table3}
% \tablehead{
% \colhead{Field} & Model & \colhead{$r/g$} & \colhead{$A$ [kJy/sr]} & \colhead{$g-r$}
% }
% \caption{Field-average results from the best-fit color models for Dragonfly data.}
% \startdata 
% Field A & g to r & 1.82 & 0.18  & 0.65 \\ 
% & r to g & 1.95 & 0.24  & 0.72 \\ 
% & bisector & 1.89 & 0.03  & 0.69 \\ 
% Field B & g to r & 1.57 & 0.46  & 0.49 \\ 
% & r to g & 2.02 & 0.57  & 0.75 \\ 
% & bisector & 1.77 & 0.12  & 0.62 \\ 
% \enddata
% \end{deluxetable}

\begin{figure}[!htbp]
\centering
  \resizebox{\hsize}{!}{\includegraphics{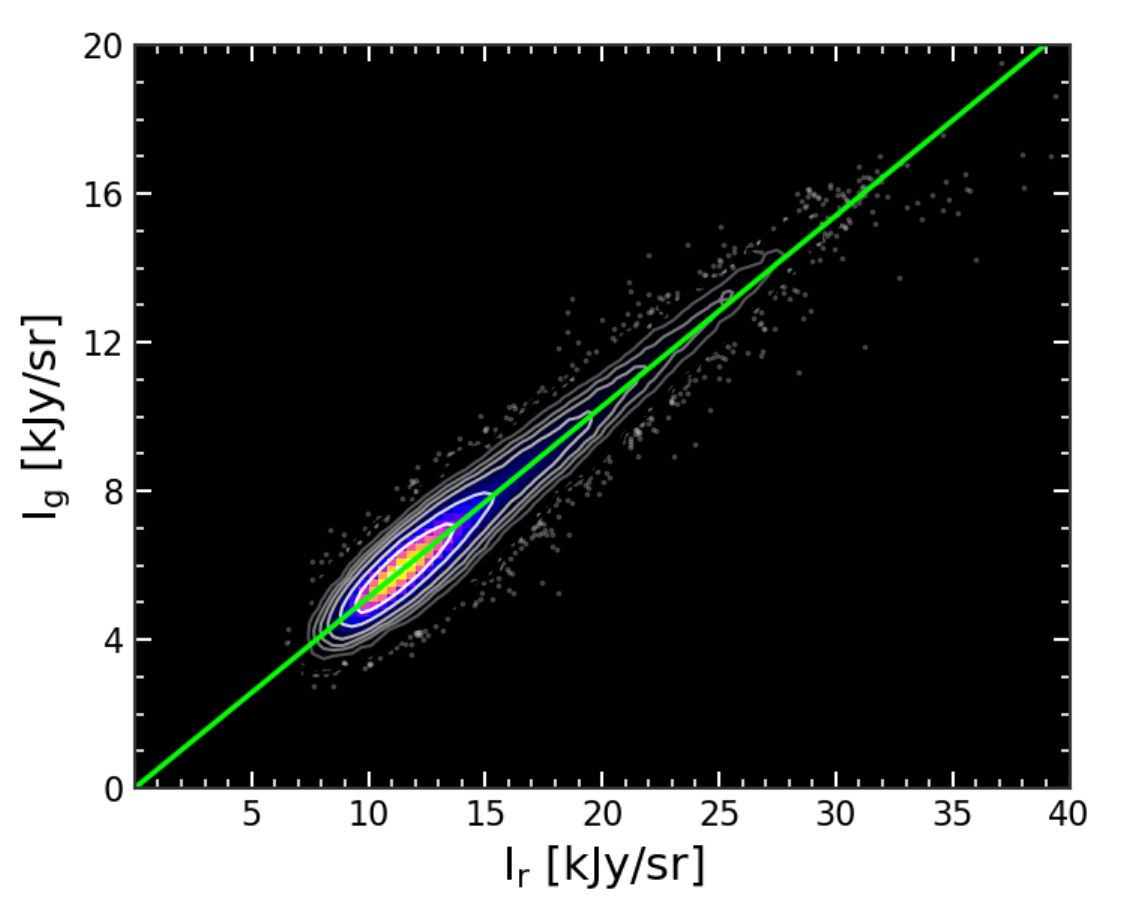}}
  \caption{Correlation between the Dragonfly $g$ and $r$-band surface brightness intensities for the diffuse light in Field A. The zero-points are calibrated based on correlations with Planck (Sec.~\ref{sec:color_planck}). The images are binned by a [$4\times4$] median binning (at a pixel resolution of 10$\arcsec$).  A mixture linear model accounting for outliers in terms of a single color model is fitted from r-band to g-band, indicated by the green line. There is no clear evidence of transition at high intensities, although higher-resolution data at different line-of-sights will be needed for further investigation. The contours from the outermost to innermost show 99.9/99/97.5/95/90/75/50\% of the total data points. The 0.1\% outliers are shown as small dots outside the outermost contour. A similar model is constructed mapping from g-band to r-band.}
\label{fig:corr_df}
\end{figure}

Figure \ref{fig:corr_df} shows the correlation between Dragonfly $g$ and $r$ data and the $r$-to-$g$ model. The green line is the best-fit linear model from the maximum likelihood estimation. A similar model is constructed mapping from $g$-band to $r$-band.
The parameters from the best fit are summarized in Table \ref{table3}, including the intercept at x-axis and $g-r$ color transformed from the slope of the linear model. 
Linear regression may potentially suffer from the low-intensity pixels because the data can be heteroscedastic. Therefore we compute the bisector of the two models following \cite{1990ApJ...364..104I} to correct the bias and list its slope, intercept, and the corresponding $g-r$ in Table \ref{table3}. The r to g ratios are $1.89\pm0.08$ and $1.78\pm0.08$, corresponding to $g-r$ = $0.69\pm0.05$ for Field A and $0.63\pm0.05$ for Field B. The fitted intercepts are close, but not equal to zero, indicating the amount of systematics in the zero-point calibration and possible contribution from other emissions such as EBL. No clear transition in the slope of the correlation is observed between the fitted range in the $g$ and $r$ correlation, indicating that a single color model is sufficiently good here to explain the dataset. However, it should be noted that this only applies to the dataset as presented, which is after binning (to a pixel resolution of $10\arcsec$) and clipping. More complex color models might be preferred at different line-of-sights or at higher resolutions where finer structures of cirrus are preserved. A higher-order color modeling can be implemented by generalizing Eq.~\ref{eq:likelihood_marginalized}. 

Overall, the results show that the optical diffuse light can be well explained by a single, simple color model. This color model is used for predicting data in one band from the other in Section~\ref{sec:cirrus_removal}.

\begin{deluxetable}{ccccc} \label{table3}
\tablehead{
\colhead{Field} & Model & \colhead{$r/g$} & \colhead{$A$ [kJy/sr]} & \colhead{$g-r$}
}
\caption{Field-averaged results from the best-fit color models for Dragonfly data.}
\startdata 
Field A & bisector$^\dagger$ & $1.89\pm0.08$ & $0.08\pm0.55$ & $0.69\pm0.05$ \\ 
& g to r & $1.84\pm0.12$ & $0.42\pm0.55$  & $0.66\pm0.07$ \\ 
& r to g & $1.95\pm0.12$ & $0.13\pm0.27$  & $0.72\pm0.07$ \\ 
Field B & bisector$^\dagger$ & $1.78\pm0.08$ & $-0.59\pm0.42$  & $0.63\pm0.05$ \\ 
& g to r & $1.56\pm0.10$ & $0.52\pm0.41$  & $0.48\pm0.07$ \\ 
& r to g & $2.06\pm0.13$& $0.96\pm0.22$  & $0.79\pm0.07$ \\ 
\enddata
\begin{threeparttable}
\begin{tablenotes}
    \small
       \item $\dagger$ Based on the ordinary least-squares bisector formula in Table 1 of \cite{1990ApJ...364..104I}. {The intercept (A) of the bisector model is calculated on $g$ to $r$.}
\end{tablenotes}
\end{threeparttable}
\end{deluxetable}

\subsection{Cirrus Removal with Color Constraint}
\label{sec:cirrus_removal}

\begin{figure*}[!htbp]
\centering
  \resizebox{\hsize}{!}{\includegraphics{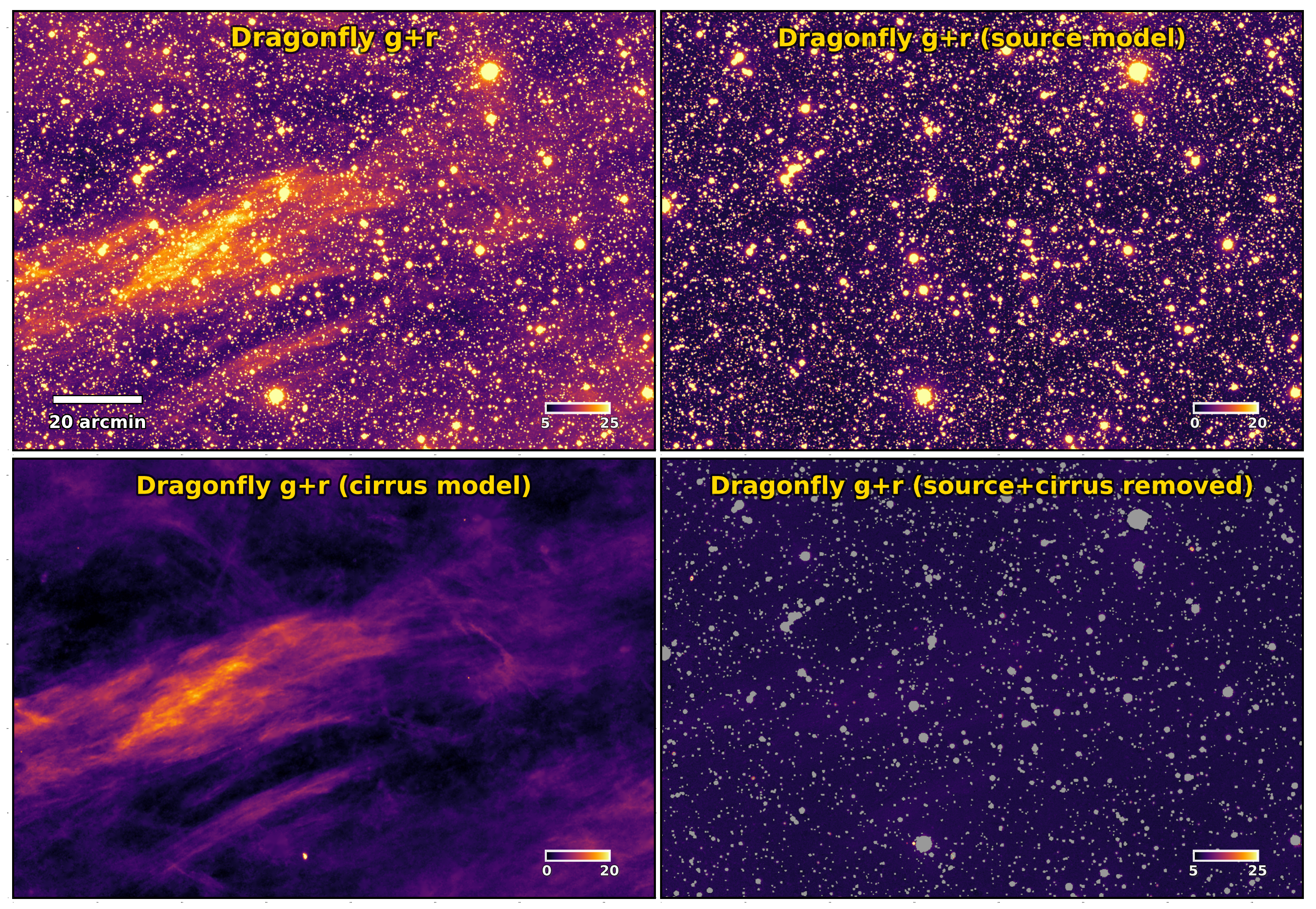}}
  \caption{Cirrus removal based on morphological information with color constraints. The top left panel shows the central [$2.4\degree\times1.8\degree$] of the original g+r image of Field A obtained by Dragonfly. The top right panel shows the constructed source model combining $g$ and $r$. The bottom left panel shows the constructed cirrus model combining $g$ and $r$. The bottom right panel shows the source and cirrus removed image of the same area. The images are in the same contrast. Compact sources (stars, galaxies, etc.) in the source models are removed (in $g$ and $r$, separately) and the cores are masked by $5\sigma$ above zero prior to the decomposition. The cirrus-removed image has a significantly flatter sky background. Unlike typical sky subtraction, this decomposition approach has a physical motivation based on cirrus characteristics.}
\label{fig:remove_cirrus}
\end{figure*}

In this section we apply the color model to the `cirrus-like' emission extracted based on morphologies in Sec.~\ref{sec:morph}, using Field A for demonstration. The r-band `cirrus-like' emission is used to predict the corresponding g-band emission according to Eq.~\ref{eq:r_to_g}, and similarly for the g-band. The predicted emission is subtracted from each band, and a $g+r$ residual image is constructed from the $g$ and $r$ band residual images, which is commensurate with the V band using the conversion derived from Table 3 of \cite{2006A&A...460..339J}: 
$V = 0.435\,g + 0.565\,r - 0.016 \,$\footnote{Note that this is Vega system V band, while the majority of this work adopts AB system. {Also note that there is a small ($<$0.01) difference in the coefficients between the ones derived from \cite{2006A&A...460..339J} and that of the equation used here, which follows the equations listed on \url{https://www.sdss3.org/dr8/algorithms/sdssUBVRITransform.php}. Practically, this difference is negligible for the purpose of this work}.}. \label{eq:g+r}
Below we use $g+r$ interchangeably with $V$. The $g+r$ image of the central [$2.6\degree\times1.8\degree$] region of the field created from the original Dragonfly data is shown in the top left panel of Figure \ref{fig:remove_cirrus}. The top right panel shows the $g+r$ source model constructed in Sec.~\ref{sec:mrf} and the bottom left panel shows the $g+r$ cirrus model constructed following Sec.~\ref{sec:morph} and Sec.~\ref{sec:color_optical}.

The bottom right panel of Fig.~\ref{fig:remove_cirrus} shows the $g+r$ residual image after removing foreground and background sources (Sec.~\ref{sec:mrf}) and removing cirrus using morphological information (Sec.~\ref{sec:morph}) with color constraints (Sec.~\ref{sec:cirrus_color}). In essence, the procedures remove the same amount of diffuse light with patchy or filamentary morphology that can be explained by a single color model in both optical bands. In the residual image, we apply a 5$\sigma$ mask to the centers of sources, and run a mask infilling (Sec.~\ref{sec:mask_infill}) for small holes with scales smaller than $10\arcsec$. After the cirrus removal, the sky background in the field is perceptually flat, which qualitatively proves the effectiveness of the approach. A quantitative evaluation is presented in the next section. 

Despite the removal of the majority of the diffuse light, there is a very faint large-scale diffuse light pattern in the residual image roughly spatially matching the high-intensity regions of the cirrus, which could result from changes in the dust properties and optical depth effects that cause changes in the cirrus color and/or the zero-point calibrated from FIR data. Notably, there are also some small-scale blobs in the residual images, which are composed of (i) sources missing in the foreground/background flux models (e.g., {faint sources around very bright stars, galaxies with improper segmentation}, variable stars) (ii) LSBG candidates (iii) bright cirrus blobs at field edges, and (iv) cirrus knots and clumps with abnormally red/blue colors relative to the {integrated} color based on the modeling. Sec.\ref{sec:LSBG} will investigate the blobs in the residual images. Other systematic contributions include imperfect flat-fielding, the diffuse ionized background (which should be low at high Galactic latitudes), and contribution from the EBL.

\subsection{Metrics of Performance}
\label{sec:metric}
To quantitatively evaluate the performance of the cirrus removal approach, we investigate several metrics of extended structures in the input image in the presence of cirrus ({with extended PSF wings of sources subtracted and centers masked}) and the output image with cirrus removed for the example shown in Fig.~\ref{fig:remove_cirrus}. Only pixels out of the $3\sigma$ source mask are included in the computation because the pixels within the source mask are from interpolation in the infilling procedure. The standard deviation of the distribution is calculated to derive the surface brightness limits in Table \ref{table1}.

\begin{enumerate}[label=(\roman*)]
    \item {Skewness}: We calculate the skewness of the {distribution function} of pixel intensities. Skewness is a measure of the asymmetry of the distribution. Skewness approaching 0 corresponds to higher symmetry. The input image has a positive skewness of 0.59, showing significant contribution from the diffuse light, mostly Galactic cirrus (with stars and galaxies subtracted). The output image has a skewness of 0.22, {which is around a factor of 3 lower than the input image}.
    
    \item {Gini coefficient}: We compute the Gini coefficient of the pixel intensity distribution. The Gini coefficient is a metric measuring the inequality in a given set of values, which was originally introduced in astronomy for quantitative galaxy morphology (\citealt{2003ApJ...588..218A}, \citealt{2004AJ....128..163L}):
    \begin{equation} \label{eq:gini}
        Gini = \frac{1}{\bar{I} n (n - 1)}
        \sum^{n}_{i=1} (2i - n - 1)\, I_i\,,
    \end{equation}
    where $n$ is the sample size, $I_i$ is the intensity of each pixel {sorted in ascending order}, and $\bar{I}$ is the mean intensity. A Gini coefficient of 0 represents a perfectly even distribution while 1 corresponds to an extreme inequality, {e.g.,} with all flux concentrating in one pixel. The pixel intensities {of the input and output images} are clipped with the lowest and highest 0.01\% of data masked {(8466311 pixels left)}, and then mapped into [0, 1] using the minimum and maximum intensities of the input image: $I=(I-I_{\rm min})/(I_{\rm max}-I_{\rm min})$. In the example field, $I_{\rm min}$ and $I_{\rm max}$ correspond to 4.0 kJy/sr and 21.3 kJy/sr. The Gini coefficient of the input and output images are 0.28 and 0.04, respectively, indicating a much flatter sky background in the output image.  
    
    One can further investigate the inequality of the intensity distribution of bright pixels by only including pixels brighter than a threshold in Eq.~\ref{eq:gini}. Figure~\ref{fig:gini} shows the Gini coefficient measured with different thresholds based on quantiles ({from 0 to 0.99}) of the intensity distributions in the input and output images. As the threshold increases, the pixel set shifts from being dominated by diffuse light in the background to being dominated by overdensities. The Gini coefficient of the {input} image quickly becomes closer to that of the output as the quantile approaches 1. The dashed line indicates the Gini coefficients of a flat sky with {a low-level (0.1\%) perturbation with the same normalization along different thresholds}, which are close to zero.

    \begin{figure}[!htbp]
    \centering
      \resizebox{\hsize}{!}{\includegraphics{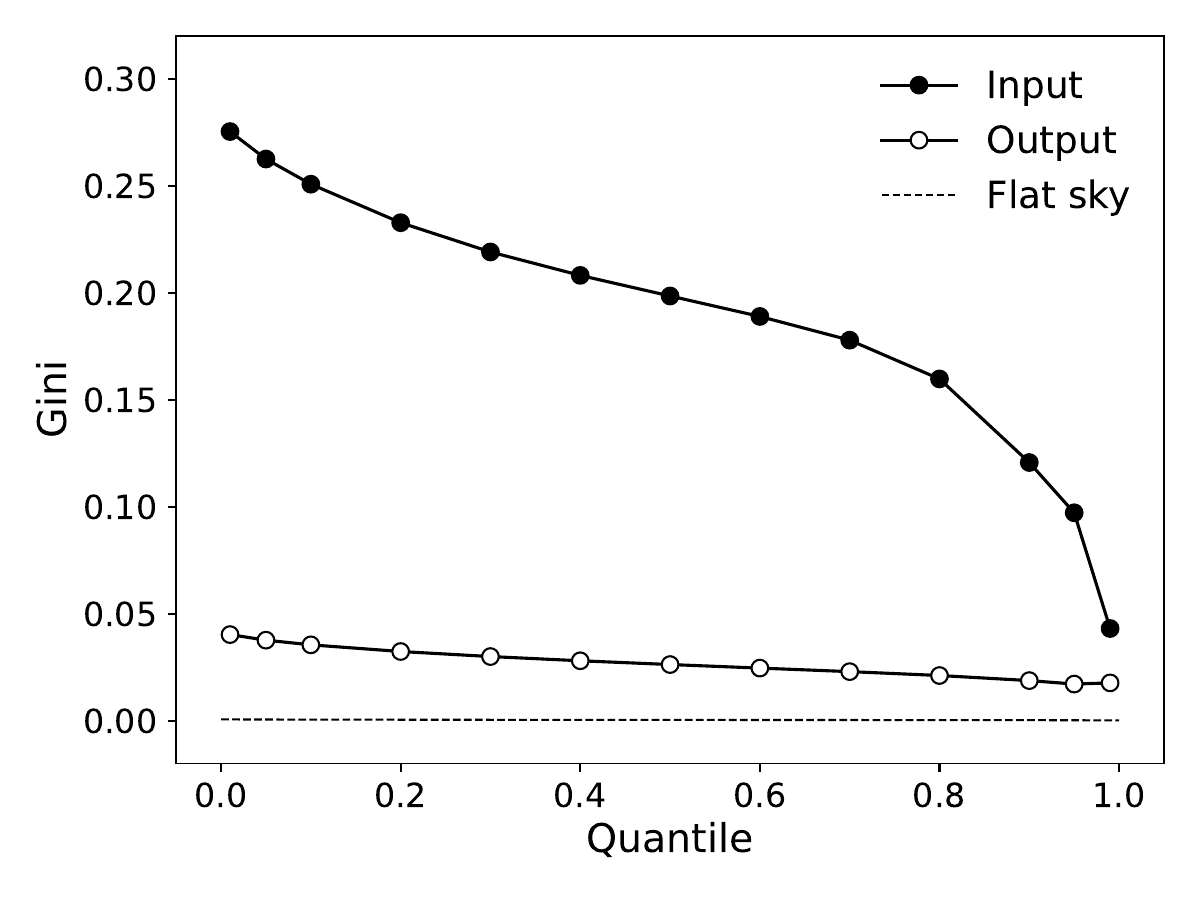}}
      \caption{Gini coefficient measured on the sky background before and after the cirrus decomposition. The metric is measured on the source subtracted Dragonfly g+r image (input) and the residual g+r image (output). The figure shows the variation of the metrics measured on the brighter subset of pixels above a given quantile. The dashed line shows the metric of a flat sky with {a low-level perturbation}, which indicates that the output image is close to a flat sky.}
    \label{fig:gini}
    \end{figure}

    \item {$\Delta$-variance}: {We compute the $\Delta$-variance spectrum (\citealt{1998A&A...336..697S}, \citealt{2001A&A...366..636B}, \citealt{2008A&A...485..917O}) for the input and the output image. In brief, the $\Delta$-variance method is a variant of the power spectra method that measures the power of structure on a range of spatial scales by convolving the image with a set of kernels with increasing kernel {width}. 
    % Following the formulation by \cite{2008A&A...485..917O}, to account for boundary and noise effects, the image is convolved by a Ricker kernel split into its core and outer annulus, and the annulus-convolved image is subtracted from the peak-convolved image:}
    % \begin{equation}
    %     F(\xi) = \frac{G_{\rm core}(\xi)}{W_{\rm core}(\xi)} - \frac{G_{\rm ann}(\xi)}{W_{\rm ann}(\xi)}\,,
    % \end{equation}
    % {where $\xi$ is the kernel size, $G$ is the convolved image, and $W$ is the convolved weight map used to down-weight noisy or missing data. The $\Delta$-variance is then computed as the weighted variance of F($\xi$):}
    %  \begin{equation}
    %     \sigma_\Delta^2 (\xi) = \frac{\Sigma\, \mathrm{Var}(F(\xi)) W_{\rm tot}(\xi)}{\Sigma W_{\rm tot}(\xi)}\,,
    % \end{equation}
    % {where $W_{\rm tot}(\xi) = W_{\rm core}(\xi) W_{\rm ann}(\xi)$, and the sum is performed over the image. 
    {We use the implementation in the \texttt{TurbuStat} package (\citealt{2019AJ....158....1K}) to calculate $\sigma_\Delta^2$, which adopts the formulation and kernel separation introduced by \cite{2008A&A...485..917O}. The implementation uses a Ricker kernel split into its core and outer annulus. The local standard deviation map of the local sky background was used as a weight map to down-weight noisy or missing data}.}

    {The top panel of Figure~\ref{fig:delta-var} shows the $\Delta$-variance spectra measured at different scales on the input and output images in magenta and blue, respectively. The kernel width is referred to as the `lag'. The bottom panel of Fig.~\ref{fig:delta-var} shows the ratio of $\Delta$-variance of the input ($\sigma_{\Delta,{\rm in}}^2$) and output ($\sigma_{\Delta,{\rm out}}^2$) as a function of spatial scales. The power of the cirrus structure is largely reduced in the output compared to the input on large scales, with a factor $\sim$10 on 1$\arcmin$ scales and a factor $\sim$200 on 5$\arcmin$ scales and larger. 
    
    In the input image, $\sigma_\Delta^2$ follows a power law over a wide range of scales. The fitted slope is $\gamma_\Delta=0.97\pm0.06$, as shown by the magenta dashed line. This corresponds to a power index of $\gamma=-\gamma_\Delta-2=-2.97$ for the power spectrum (\citealt{1998A&A...336..697S}), which is consistent with the expected value of $\gamma\sim-3$ from turbulence theories and observations (e.g., \citealt{1992AJ....103.1313G}, \citealt{2007A&A...469..595M}, \citealt{2016A&A...593A...4M}). On small scales, $\Delta$-variance is affected by beam effect, noise, and residuals of stars and galaxies, while on large scales, it is flattened by the limited size of the field relative to the filter size (\citealt{2008A&A...485..917O}). 
    {The $\sigma_\Delta^2$ spectrum of the output image reflects the residual pattern modulated by cirrus residuals, beam effect, and field edge effect.}
    A more detailed analysis of the coherence of the cirrus structures will be presented in our upcoming work, where here we focus on the result quantifying the amount of reduction of the cirrus structures with the decomposition algorithm.}

    \begin{figure}[!htpb]
    \centering
      \resizebox{\hsize}{!}{\includegraphics{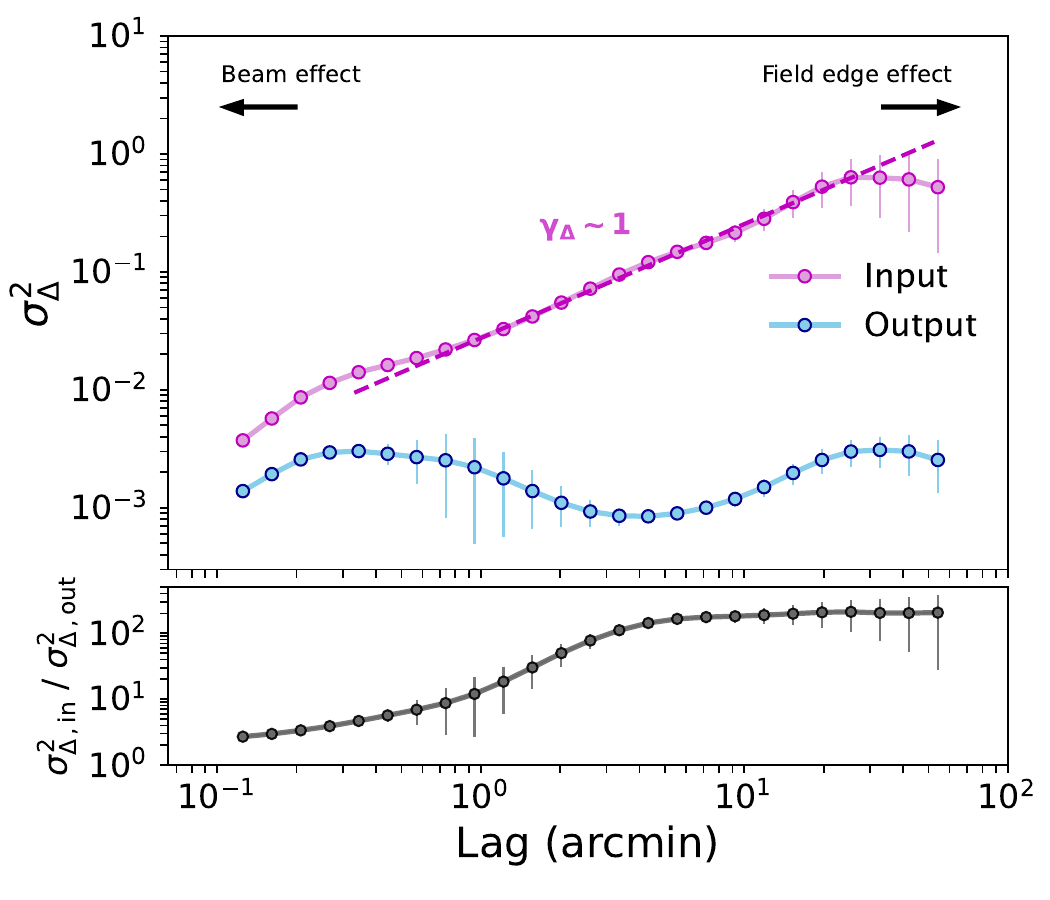
      }}
      \caption{\textit{Top}: $\Delta$-variance spectra measured on the source subtracted Dragonfly g+r image (input; magenta markers) and the residual (output; blue markers). $\Delta$-variance measures the amount of structure on different spatial scales. The power is largely reduced in the output image. {A fitted slope of $\gamma_\Delta=0.97\pm0.06$ is shown as the magenta dashed line}. \textit{Bottom}: The ratio of $\sigma_\Delta^2$ of the input and output images on different spatial scales. The powers of large-scale structures are largely reduced.}
    \label{fig:delta-var}
    \end{figure}
\end{enumerate}

Together these metrics quantitatively show that the cirrus-subtracted image is fairly {closer} to a flat sky, and therefore, the performance of the algorithm is acceptable. {It is worth noting that despite the simplicity of skewness (and likewise, $n-$th order moments) and Gini in their definition and calculation, unlike $\Delta$-variance, such metrics are pixel-wise and therefore, these do not encode 2D information. As a result, one should be careful about associating them with physical interpretations. In our follow-up work, we will present more metrics that further employ the spatial coherence of the cirrus.}

\section{Application to LSBG searches using Integrated Light} \label{sec:LSBG}

In this section, we demonstrate the application of the cirrus removal algorithm on LSBG searches using integrated light (e.g., \citealt{2018ApJ...856...69D}). This is done by recovering mock galaxies injected into the image with cirrus (Section~\ref{sec:test_inj}) and attempting to recover a known faint dwarf satellite galaxy, And~XXII (Section~\ref{sec:test_AndXXII}).

\subsection{Recovering Simulated Galaxies with Realistic Stellar Populations} \label{sec:test_inj}

We test whether the cirrus decomposition approach can facilitate LSBG searches. We inject mock galaxies into the cirrus-rich field and attempt to recover them from the residual image. Ideally, the algorithm should preserve light from the injected galaxies while removing the diffuse light from cirrus based on their differences in morphology and SED (i.e., colors).

\subsubsection{Injection-Recovery Test} \label{sec:test_single}

We used ArtPop to build mock UDGs with realistic stellar populations. \texttt{ArtPop} is a Python package for generating artificial images of stellar systems with synthetic stellar populations (\citealt{2022ApJ...941...26G}). {Details about the physical parameters used to generate the galaxy models and mock observations are described in Appendix~\ref{appendix:artpop}.} As a result, the integrated color of the mock UDG is $g-r$ = 0.54 (i.e., bluer than the field-averaged mean of cirrus), and the mean V-band surface brightness within the effective radius, $\mu_{{\rm eff}, V}$, is 26.0 mag/arcsec$^2$.

\begin{figure*}[!htbp]
\centering
  \resizebox{\hsize}{!}{\includegraphics{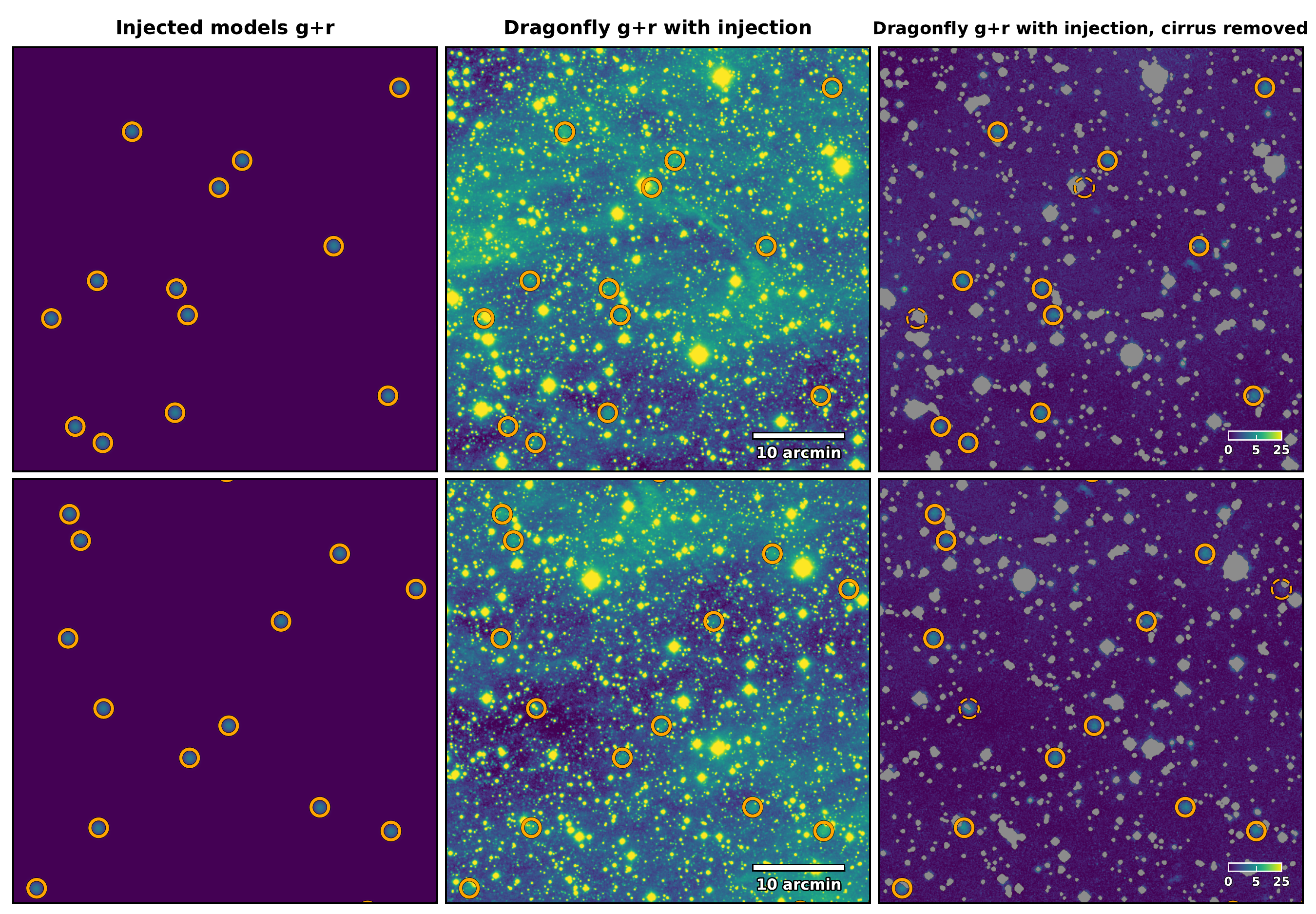}}
  \caption{Injection of mock UDGs in a cirrus-rich field and recovery after cirrus removal in Dragonfly imaging. In each row, the middle panel shows a [$50\arcmin \times 50\arcmin$] region of the g+r image with injections, and the injected galaxy models are shown in the panel to the left, indicated by orange circles. The right panel shows the cirrus removed image with sources subtracted and center masked. Injections that are detected as LSBG candidates are marked in solid orange circles, while those that fail to be recovered are marked in dashed orange circles. The two rows show different cutout regions in the field. The majority of the mock UDGs in the test are capable of being recovered after cirrus removal. {The image scale is in arcsinh stretch in units of kJy/sr.}}
\label{fig:test_inj}
\end{figure*}

The mock UDGs were randomly injected into the $g$ and $r$ band images of Field A. The middle panels of Fig.~\ref{fig:test_inj} show example [$50\arcmin \times 50\arcmin$] cutouts at two different positions in the field. The positions of the injected galaxies are indicated by orange circles. The $g$ and $r$ images with injections were then processed with the cirrus decomposition software after source subtraction. To reduce possible contamination and focus on the performance of the cirrus decomposition, we used the same flux models for the foreground/background sources constructed from the image without injection. However, the effects of injected mock sources on the construction of the flux models are generally small, and faint diffuse sources are, in principle, excluded from the flux models.

The right panels of Fig.~\ref{fig:test_inj} show the cirrus-removed residual image, combining $g$ and $r$, in the same region as the panel to the left. Sources are masked at $3\sigma$ level and small masks are infilled following the steps in Sec.~\ref{sec:mask_infill}. The injections are marked by orange circles. The mock UDGs become significant in the residual compared to the middle panels, in which they are flooded by the cirrus emission. Note, however, that there are some other diffuse signals remaining in the residual image, which are likely contaminations from cirrus knots/clumps with abnormally blue colors and/or compact morphologies. Our methodology is not perfect though, as some injections were unfortunately removed or simply blocked due to the blending with cirrus or bright stars/galaxies. {We now consider metrics used to quantify the effectiveness of our approach.}

\subsubsection{Performance Metrics} \label{sec:test_metrics}
To quantify the goodness of the recovery, we run a source detection on the residual image after a [$4\times4$] median binning using \texttt{SExtractor}. We apply tentative cuts on the detections based on the following criteria: (1) an S/N detection cut above 5 (2) a size cut of FLUX\_RADIUS $> 20\arcsec$ {with PHOT\_FLUXFRAC = 0.5} and (3) an axis ratio cut above 0.5. The detections are cross-matched with the injections with a maximum separation of 1.5 pixels. Injections that failed to be recovered by the source detection on the cirrus removed image are marked by dashed circles in the right panels of Fig.~\ref{fig:test_inj}. We calculate the recall and precision of the test, which are defined by:
\begin{equation}
    Precision = \frac{TP}{TP+FP}\,,\quad
    Recall = \frac{TP}{TP+FN}\,.
\end{equation}
TP, FP and FN stand for true positive, false positive, and false negative, respectively. The precision is a measure of the accuracy of the detection, and the recall represents the completeness of the recovery. The overall performance is evaluated by the F-score:
\begin{equation}
    F_1 = \frac{2}{{precision}^{-1}+{recall}^{-1}}\,.
\end{equation}

The precision is 0.77 and the recall is 0.73, yielding an F-score of 0.75
\footnote{Note that the reported measures did not include corrections in TP and FP to retain the bias from contaminations. Some $\sim$ 20 objects were detected with the above criteria, with only one overlapping with the injections within 3$\arcsec$. Among these objects, $<$6 of them are LSBG candidates with visual inspection.}.
Future work will explore reducing contamination (false positives) in the cirrus decomposition procedures and restoring the missed intrinsic candidates (false negatives). 

Overall, this injection-recovery experiment shows that the performance of the cirrus removal approach is encouraging and likely to be useful for facilitating LSBG searches via integrated light.

\subsubsection{Tests on a Grid of Models} \label{sec:test_grid}

To evaluate the variation of performance on LSBG candidate properties, we explored the parameter space of the model galaxy by building a grid of UDG models with varying physical parameters that change the effective surface brightness $\mu_{\rm eff, V}$ and the $g-r$ color of the galaxy models. {Details about the physical parameters used to generate the grid of galaxy models are described in Appendix~\ref{appendix:artpop}.}

\begin{figure*}[!htbp]
\centering
  \resizebox{\hsize}{!}{\includegraphics{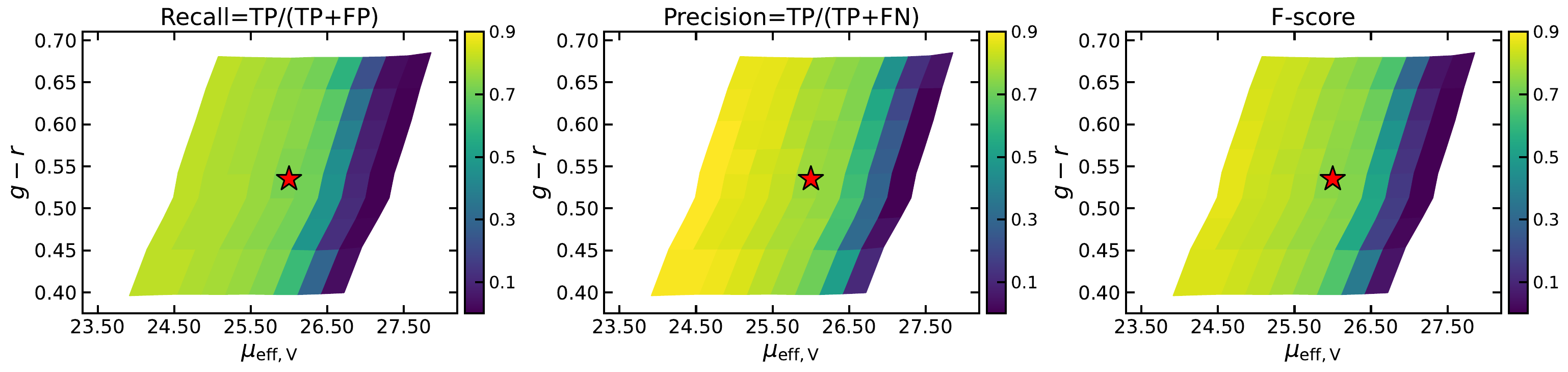}}
  \caption{Performance metrics for recovering injected galaxies in the cirrus removed image at varying effective surface brightness $\mu_{\rm eff, V}$ and $g-r$ color, including the recall, precision, and the F-score. Details of the model grid are described in Appendix~\ref{appendix:artpop}. Recall represents the completeness of the recovery. Precision measures the contamination rate. F-score indicates the overall performance. The red star indicates the fiducial model referred to in the main text. In general, fainter and redder galaixes are more challenging to be recovered with cirrus removal. {Note that $\mu_{\rm eff, V}$ and $g-r$ are apparent values without extinction and reddening correction. The galaxies overshadowed by cirrus would be intrinsically bluer and brighter.}}
\label{fig:test_grid_metrics}
\end{figure*}

Following the same procedures in Sec~\ref{sec:test_inj}, each model in the grid was injected 100 times into the $g$ and $r$ images with cirrus, and then a source detection was run in an attempt to recover them after running cirrus removal. The performance metrics resulting from this exercise are displayed in Figure \ref{fig:test_grid_metrics}. {The metrics are evaluated on the model grid and reprojected to the apparent parameter space ($\mu_{\rm eff, V}$ and $g-r$).} As the galaxy becomes fainter, the recovery rate drops rapidly, which is a joint result of the LSBG being harder to detect and it being harder to distinguish from the cirrus, especially with morphology. The metallicity has a weaker power on the recovery for brighter models; however, for fainter LSBGs, higher metallicity would lead to degradation of the performance as the galaxy becomes redder and, therefore, harder to be distinguished from cirrus with a similar color. 

{It is noteworthy that $\mu_{\rm eff, V}$ and $g-r$ in Fig.~\ref{fig:test_grid_metrics} are apparent observables without extinction and reddening correction. Therefore, in real observations, galaxies overshadowed by cirrus would be intrinsically bluer and brighter. This correction will be important to evaluate the completeness function of LSBGs as a function of physical parameters, while the gist of this experiment is to showcase that this decomposition approach facilitates reducing the confusion in the detection of LSBGs in a cirrus-riddled sky area.}

% Fe/H = -1 => g-r ~0.63, closer to cirrus => lower performance 

\subsection{Recovering M33 Satellite And~XXII} \label{sec:test_AndXXII}

{As a final test of our approach, we applied it to the recovery of a dwarf satellite of M33 identified originally via star counts.} And~XXII is a dwarf satellite galaxy of M33 discovered by the CFHT Pan-Andromeda Archaeological Survey (PAndAS; \citealt{2009Natur.461...66M}), a comprehensive observational campaign aimed at mapping the vicinity of {M31} to the depths needed to reach ultra-faint dwarfs. PAndAS identified only one M33 satellite candidate, And~XXII (\citealt{2009ApJ...705..758M}), using color-magnitude diagram (CMD) analysis. Spectroscopic follow-up was done by \cite{2013MNRAS.430...37C} using the DEep Imaging Multi-Object Spectrograph (DEIMOS) on the Keck II Telescope, which confirmed its identity as a strong candidate for being an M33 satellite. 

\begin{figure*}[!htbp]
\centering
  \resizebox{0.95\hsize}{!}{\includegraphics{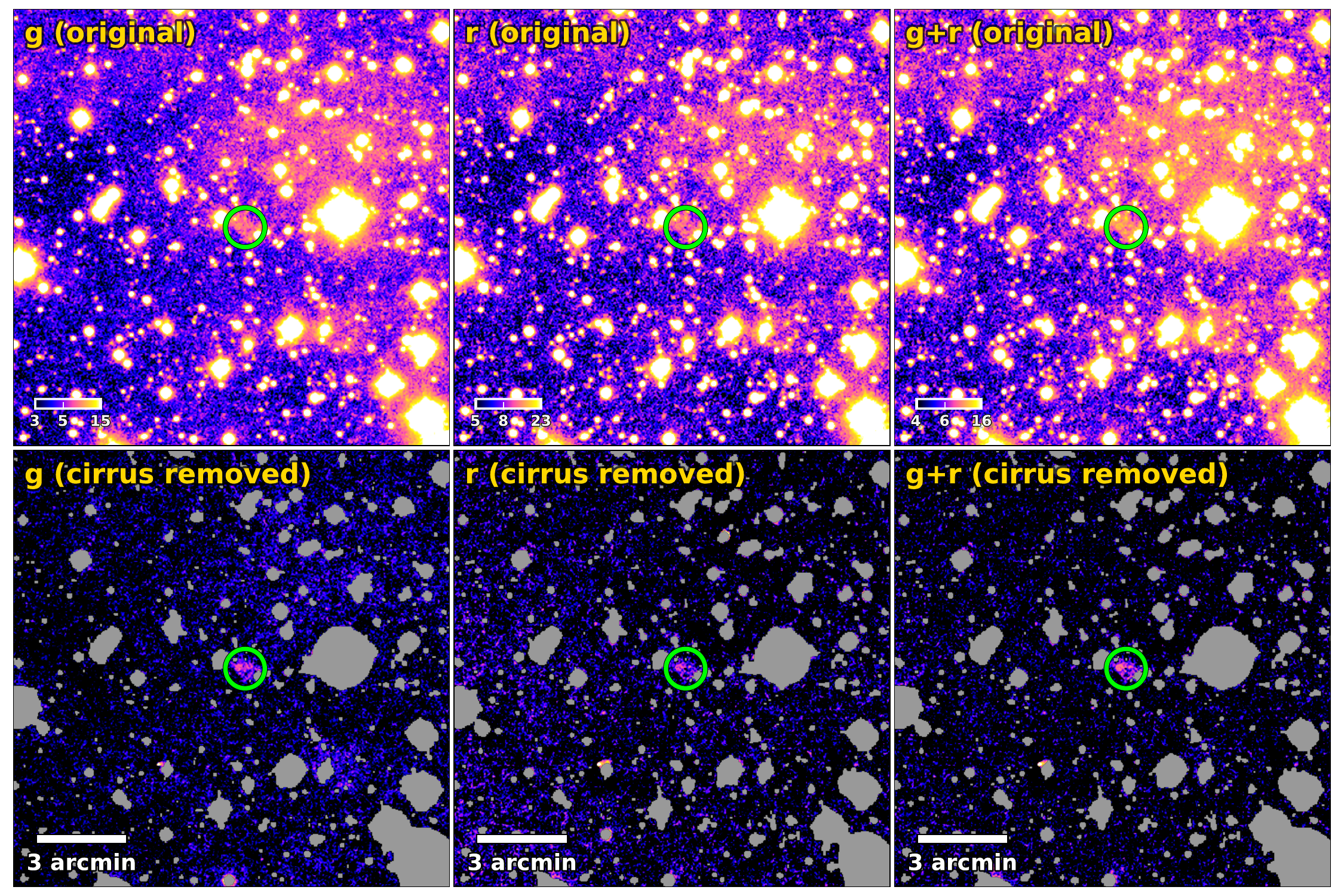}}
  \caption{Recovery of a confirmed dwarf satellite galaxy, And~XXII in the cirrus-riddled Field B. The left to right columns show [$16\arcmin \times 16 \arcmin$] cutouts around And~XXII (marked by the green circle) in g, r, and g+r bands in the original image (upper) and after cirrus removal (lower). In lower panels, sources are subtracted and masked at $3\sigma$ levels. {The image scale is in arcsinh stretch in units of kJy/sr. The g+r image is stretched to enhance the faint signal.}}
\label{fig:AndXXII}
\end{figure*}

The upper panels of Figure~\ref{fig:AndXXII} show [$16\arcmin \times 16 \arcmin$] cutouts around And~XXII in $g$, $r$, and $g$+$r$ data. 
%The $g$+$r$ data is created with Eq.~\ref{eq:g+r}. 
With Dragonfly imaging, the integrated light from And~XXII is clearly present in both $g$ and $r$ band data. And~XXII appears as a fuzzy blob, {with $R_{\rm eff}$ $\sim$ {15\arcsec} and an effective surface brightness $\mu_{{\rm eff}, g}\sim27$ mag/arcsec$^2$ in g-band.} However, there is an extended cirrus patch near And~XXII, adding confusion to the detection and identification of And~XXII with its integrated light alone. The extended PSF wing from a nearby bright star might also contribute to the diffuse light in the background.
The mean color of And~XXII in Dragonfly imaging is $g-r\sim0.35$ ($\sim$0.3 bluer than the field-mean color of cirrus), making it likely that this object is distinguishable from cirrus by using color constraints.

The cutouts of the residual images produced by procedures in Sec.~\ref{sec:morph} and Sec.~\ref{sec:cirrus_color} are shown in the lower panels of Figure \ref{fig:AndXXII}. Compact sources and their extended PSF wings have been subtracted following procedures in Sec.~\ref{sec:mrf}. The cirrus in the image is extracted and removed using morphological information (Sec.~\ref{sec:morph}) with color constraints (Sec.~\ref{sec:cirrus_color}). The majority of the extended cirrus emission is removed, and the diffuse light from And~XXII shows as an overdensity in the cirrus-removed image. Note that part of the light from the red giant branch in And~XXII was also subtracted because And~XXII is semi-resolved in the Legacy imaging. The remaining signal represents the diffuse integrated light from unresolved stellar populations. Therefore, LSBG searches using integrated light can be supplementary to the CMD analysis.

This result is encouraging, because cirrus is one of the major systematics that impedes LSBG searches in sky areas like these. Future deep imaging surveys can run cirrus decomposition software to facilitate LSBG searches and increase the significance of detections through control of systematics. Such techniques might be valuable additives to the pipeline of deep wide-field imaging surveys from, e.g., the Vera C. Rubin {Observatory}, the Euclid space telescope, and the Roman space telescope.

\section{Discussion} \label{sec:discussion}

\subsection{Differentiating Cirrus in Deep Imaging Surveys with Multi-band Photometry} \label{sec:discussion_surveys} 

The next 10 years will be a golden age for low surface brightness astronomy with the deployment of next-generation deep imaging surveys using state-of-the-art facilities: \textit{Rubin} will achieve $3\sigma$ surface brightness limits of 30.3 mag/arcsec$^2$ on 10$\arcsec$$\times$$10\arcsec$ scales in g and r-bands in a 10-year campaign, covering the entire Southern sky (\citealt{2024MNRAS.528.4289W}). The recently launched Euclid space telescope, which has delivered a wealth of its first science results (\citealt{2024arXiv240513491E}), has achieved a limiting surface brightness magnitude of 29.9 mag/arcsec$^2$ ($1\sigma$, 10$\arcsec$$\times$$10\arcsec$) with its Visible Imager (VIS) instrument (\citealt{2024arXiv240513496C}).
The superb sensitivity of these imaging surveys to faint diffuse emission calls for the essential need to characterize diffuse light from optical Galactic cirrus.

{The main driver of distinguishing the cirrus from extragalactic light using optical photometry is the difference in the physical mechanisms that mold the SED and the geometry of the optical cirrus.} The important work by \cite{2020A&A...644A..42R} showed that optical cirrus is well separated from extragalactic sources on the color-color diagrams, using observations of 16 cirrus patches in SDSS Stripe82 area, after carefully subtracting contaminating light. They proposed the following criterion based on the color of the cirrus:
\begin{equation}
    (r - i) < 0.43 \times (g - r) - 0.06\,,
\end{equation}
which is applied to the Hyper Suprime-Cam Subaru Strategic Program data (HSC-SSP; \citealt{2018PASJ...70S...4A}) and pixels dominated by cirrus emission can be well differentiated from extragalactic sources.

In this work, we show that by incorporating morphological information, diffuse light from dust scattering can be well differentiated from LSBG candidates, even with only two filters. This is promising -- although confusion between cirrus and LSBGs with similar colors may exist -- because deep imaging surveys with more filters should have the capacity to better characterize the cirrus by sampling the curvature of the dust-scattered light SED. 

It is worth noting that extragalactic sources with similar morphologies to that of cirrus, such as tidal tails, may not be well disentangled with this approach. In these cases, observations from multiple optical filters will be necessary. Data from other wavelengths or tracers of dust, including FIR (e.g., \citealt{2017ApJ...834...16M}), UV (e.g., \citealt{2015A&A...579A..29B}), atomic and molecular hydrogen (e.g., \citealt{2010MNRAS.406.2713B}), and polarization (\citealt{2023ApJ...959...40B}) will also be preferred. 

While Dragonfly resolution is poor, the large field-of-view of Dragonfly allows one to map Galactic cirrus in an unprecedentedly wide area of the sky (c.f., the Dragonfly Ultra Wide Survey from which the dataset in \citealt{2023ApJ...953....7L} was drawn), which will be presented in our following work. This dataset will be valuable as supplementary data for the \textit{Rubin} Observatory, the \textit{Nancy Roman} space telescope, and the \textit{Euclid} space telescope. It will also provide a training set for future deep learning-based techniques.

\subsection{The Optical DGL} \label{sec:DGL}

One major motivation for the decomposition of the optical cirrus from the other diffuse light is to obtain a `clean' representation of the optical DGL, which will benefit ISM studies by, e.g., constraining grain properties and parameters of scattering functions. The optical DGL is also an important source of foreground contaminant of the optical EBL, which is also known as the Cosmic Optical Background (COB; \citealt{2017NatCo...815003Z}, \citealt{2021ApJ...906...77L}). It is thus interesting to compare the DGL/cirrus measurements in the literature with measurements from Dragonfly. Below we show the preliminary optical DGL measured from the two example fields covering $\sim$9.3 $\rm deg^2$ sky area. This is a relatively small area compared to measurements from a wide area comparable to the entire sky (e.g., \citealt{2011ApJ...736..119M}, \citealt{2012ApJ...744..129B}). Measurements of the optical DGL from the Dragonfly Ultra Wide survey (Bowman et al. in prep) covering over 10,000 ${\rm deg^2}$ of the northern sky will be presented in the future.

\begin{figure*}[!htbp]
\centering
  \resizebox{0.9\hsize}{!}{\includegraphics{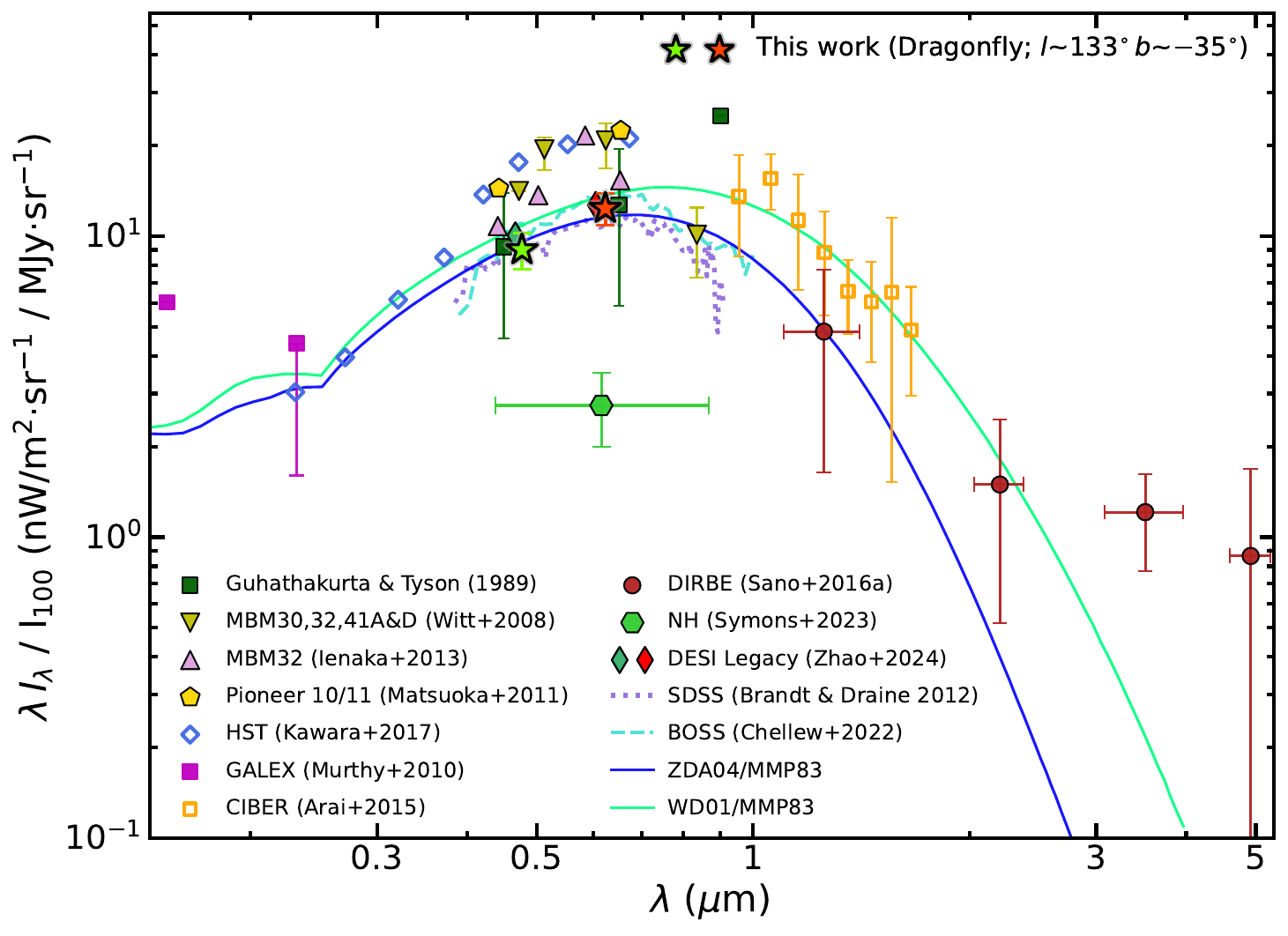}}
  \caption{The DGL SED from UV to NIR normalized by the $100\,{\rm \mu m}$ intensity. Dragonfly measurements of the diffuse background in the example datasets in $g$ and $r$-bands scaled by the mean $100\,{\rm \mu m}$ intensity, with CIB corrections included, are indicated by the green and red stars. Photometric/spectroscopic measurements in the literature are displayed as symbols/dashed lines. Model spectra of \cite{2012ApJ...744..129B} based on the MMP83 ISRF and dust models by WD01 and ZDA04 are indicated by green and blue curves, respectively. See the text for a summary of the compilation. For clarity, markers in optical bands are slightly shifted.}
\label{fig:DGL}
\end{figure*}

Dust regions with higher column densities have more scattering. Therefore it is typical to normalize the observed optical DGL by a tracer of column density. A widely adopted tracer is dust thermal emission at 100~${\rm \mu m}$. For the two example fields, we retrieve their 100~${\rm \mu m}$ intensity maps from IRIS (\citealt{2005ApJS..157..302M}), which includes improvements in the zodiacal light subtraction, calibrations, and artifacts of the original IRAS products. The 100~${\rm \mu m}$ maps ($I_{100}$) contain emission from extragalactic sources, known as the Cosmic Infrared Background (CIB). As pointed out by \cite{2017NatCo...815003Z}, it is important to subtract the CIB out. Therefore, we subtract a uniform value in each map by computing the mean offset between the observed intensity and the predicted intensity using the Planck thermal dust model of all pixels:
\begin{equation}
    \Delta I_{100} = I_{100} - I_{100}^{mod}\,,
\end{equation}
{where $I_{100}^{mod}$ is the predicted thermal dust emission at 100~${\rm \mu m}$ ($\nu=$ 3000 GHz) using the Planck thermal dust model given by Eq.~\ref{eq:I_nu}.} The mean offset, $\left< \Delta I_{100} \right>$, is $0.56\pm0.06$ and $0.59\pm0.05$ MJy/sr in Field A and Field B, respectively. 

The new intensity {after correction}: $I_{100}^{c}=I_{100}-\left< \Delta I_{100} \right>$ is used as the tracer of dust thermal emission. We fit a linear model between the optical and 100~${\rm \mu m}$ data in each band:
\begin{equation}
\label{eq:corr_df_100}
    I_{\lambda} = a_{\lambda} + b_{\lambda} \cdot I_{100}^{c} \,.
\end{equation}
The best-fit $b_{\lambda}$ in {$g$ and $r$-band} is 3.28 $\times10^{-3}$ {and} 1.73 $\times10^{-3}$ for field~A, and 2.56 $\times10^{-3}$ {and} 1.51 $\times10^{-3}$ for field~B. We use the brightness intensity corresponding to $\mu_{g, {\rm lim,}1\sigma}{(60\arcsec\times60\arcsec)}$, the $1\sigma$ surface brightness limit on scales of [$60\arcsec\times60\arcsec$] in g-band, as the representative diffuse light measurement across the field. Using the model of Eq.~\ref{eq:corr_df_100}, this corresponds to $\left<I_{100}\right>$ = $4.3\pm0.3$ and $3.8\pm0.3$ MJy/sr in field~A and field~B, respectively. The field-mean DGL intensity measurements of Dragonfly normalized by $\left<I_{100}\right>$ are shown in Figure \ref{fig:DGL}, where we overplot several classic measurements and models of the DGL/cirrus from UV to NIR in the literature. The error bars of the Dragonfly measurements include the difference between the example fields and uncertainty propagation including fitting errors estimated from bootstrap, systematic errors of photometric zero-points, uncertainties in intensity calibration from Planck dust models, and uncertainties of $\left<I_{100}\right>$ after CIB correction.

As noted earlier, the measurements presented thus far have been obtained from two Dragonfly fields, and in a future paper we will describe results obtained from about 250x this area. Nevertheless, it is interesting to place these measurements into the context of previous work in this area, and Figure~\ref{fig:DGL} also contains data points obtained from previous work. \cite{1989ApJ...346..773G}, \cite{2008ApJ...679..497W} and \cite{2013ApJ...767...80I} studied individual high-latitude clouds/fields with a typical $I_{100}$ ranging from 1 to 18 {MJy/sr} (see a compilation in Table 3 of \citealt{2013ApJ...767...80I}). \cite{2010ApJ...724.1389M} presented the diffuse UV background from GALEX.  \cite{2015ApJ...806...69A} and \cite{2016ApJ...821L..11S} reported the DGL in NIR measured by the Cosmic Infrared Background ExpeRiment (CIBER) and the Diffuse Infrared Background
Experiment (DIRBE), respectively. \cite{2012ApJ...744..129B}, \cite{2017PASJ...69...31K} and \cite{2022ApJ...932..112C} utilized blank-sky spectra obtained from the Sloan Digital Sky Survey (SDSS), the Hubble Space Telescope, and the Baryon Oscillation Spectroscopic Survey (BOSS), respectively. \cite{2011ApJ...736..119M} presented results from the Pioneer spacecraft over 1/4 of the sky. \cite{2023ApJ...945...45S} analyzed the reported the DGL measured by the broadband imager equipped on the New Horizons (NH) mission (see also analysis in \citealt{2021ApJ...906...77L}). \cite{2024AJ....168...88Z} analyzed the sky area in the Dark Energy Spectroscopic Instrument (DESI) Legacy imaging survey (\citealt{2019AJ....157..168D}) with a mean $\left<I_{100}\right>$ of 4.9 MJy/sr. Finally, the solid curves stand for model spectra of the observed DGL in \cite{2012ApJ...744..129B} using dust models from \cite{2001ApJ...548..296W} (WD01; green) and \cite{2004ApJS..152..211Z} (ZDA04; blue) with the \cite{1983A&A...128..212M} ISRF (MMP83). Compared with other measurements and models, our results match the dust models and do not indicate a clear presence of ERE in the $r$ band. Further investigation of the DGL and its constraints on grain properties over a larger sky area will be explored in our future work.

\section{Conclusions} \label{sec:conclusion}

Optical Galactic cirrus originates from the scattering of the interstellar radiation field by interstellar dust grains. It can provide unique insights into the physical and radiative properties of dust grains in our MW. An unbiased photometric characterization of optical cirrus depends on the preservation of large-scale low surface brightness emission in images. This has only been possible in deep imaging using CCD detectors recently, thanks to advances in observation and data reduction techniques. Radiometric calibration also requires a thorough understanding and exquisite control of the various systematics that may contribute to the diffuse light in images. 

In this work, we investigate the photometric characterization of optical Galactic cirrus in deep imaging of a $\sim$9.3 $\rm deg^2$ sky area obtained from the Dragonfly Telephoto Array, a telescope optimized for low surface brightness imaging. We have employed careful treatment to preserve cirrus in our data reduction pipeline, including sky background subtraction, flat-fielding, and removal of scattered light from the wide-angle PSF, to reduce systematics in the photometric characterization of optical cirrus.

We applied the Rolling Hough Transform, an algorithm developed for identifying and characterizing ISM structures, to extract `cirrus-like' emission based on morphology. The algorithm has good performance in distinguishing blobby emissions from patchy or filamentary emissions in both simulated data and observations from Dragonfly.

We add constraints in optical colors to the extracted `cirrus-like' emission by assuming a common SED for the optical diffuse light from dust scattering. This is based on the following assumptions: 1) dust grains are in local thermodynamic equilibrium, 2) the physical properties of dust populations are similar, and 3) the incident interstellar radiation field is homogenous. None of these assumptions necessarily hold in all cases, but when they do hold, the optical diffuse light from dust scattering should have a good correlation with dust thermal emission in FIR/sub-mm and, accordingly, be well correlated with each other in different bands. We verify the correlation in FIR/sub-mm by correlating Dragonfly $g$ and $r$-band data with the Planck dust thermal radiance from the all-sky Planck thermal dust model, which is used as a tracer of the dust content. The correlation corresponds to a $g-r$ = $0.70\pm0.03$ and $0.56\pm0.03$ for the two example fields. The zero-points are also calibrated using Planck. We found a good correlation of Dragonfly $g$ and $r$ data, yielding a {color of} $g-r$ = $0.69\pm0.05$ and $0.63\pm0.05$ for the {diffuse light in the} two example fields. Cirrus decomposition is performed by combining color modeling and morphological extraction, producing a fairly flat sky background. 

We present several metrics to quantitatively evaluate the performance of our proposed cirrus decomposition algorithm. Furthermore, we applied our approach to Dragonfly images with (1) injections of simulated galaxies with realistic stellar populations and (2) a known ultra-faint dwarf satellite galaxy to demonstrate its efficacy in distinguishing low surface brightness galaxies from cirrus. Galaxies with mean effective surface brightness and colors different from cirrus incur low-level confusion with cirrus and are able to be recovered. Our approach will facilitate the detection and identification of ultra-faint dwarf galaxies and ultra-diffuse galaxies on the exquisite imaging datasets expected from the forthcoming Vera C. Rubin {Observatory} and the recently commissioned Euclid space telescope.

Finally, we measured the intensity of the optical diffuse galactic light, which is the diffuse region of the optical cirrus, in the example fields. We used $\mu_{g, {\rm lim,}1\sigma}{(60\arcsec\times60\arcsec)}$,  the $1\sigma$ surface brightness limit on a [$60\arcsec\times60\arcsec$] spatial scale, as the representative surface brightness of diffuse galactic light. The measured intensities in $g$ and $r$ are normalized by $I_{100}$, the thermal emission intensity at 100~${\rm \mu m}$ from IRAS, with a cosmic infrared background correction derived using the Planck thermal dust model. We compared the Dragonfly measurements with observations and models reported in the literature. Our measurements match models and do not suggest a clear presence of extended red emission. This suggests that upcoming Dragonfly measurements of the diffuse galactic light, covering a much larger sky area, will usefully constrain dust properties in our Milky Way.

\vspace{10mm}

Q.L. is supported by an Ontario Graduate Scholarship. The research of R.G.A. and P.G.M. is supported by grants from the Natural Sciences and Engineering Research Council of Canada. The Dunlap Institute is funded through an endowment established by the David Dunlap family and the University of Toronto. S.D. is supported by NASA through Hubble Fellowship grant HST-HF2-51454.001-A awarded by the Space Telescope Science Institute, which is operated by the Association of Universities for Research in Astronomy, Incorporated, under NASA contract NAS5-26555. EP acknowledges financial support provided by a grant for \textit{HST} archival program AR-16628 through the Space Telescope Science Institute (STScI). EP also acknowledges financial support provided by NASA through the Hubble Fellowship grant \# HST-HF2-51540.001-A awarded by STScI. STScI is operated by the Association of Universities for Research in Astronomy, Incorporated, under NASA contract NAS5-26555. AK acknowledges support from NSERC, the University of Toronto Arts \& Science Postdoctoral Fellowship program, and the Dunlap Institute. 

This research has made use of data from the Planck Legacy Archive (PLA). The Planck Legacy Archive provides online access to all official data products generated by the Planck mission. 
This research has made use of the APASS database, located at the AAVSO website. Funding for APASS has been provided by the Robert Martin Ayers Sciences Fund. The authors thank the excellent technical staff at the New Mexico Skies Observatory. 

This research has made use of data from the Legacy Surveys. The Legacy Surveys consist of three individual and complementary projects: the Dark Energy Camera Legacy Survey (DECaLS; Proposal ID \#2014B-0404; PIs: David Schlegel and Arjun Dey), the Beijing-Arizona Sky Survey (BASS; NOAO Prop. ID \#2015A-0801; PIs: Zhou Xu and Xiaohui Fan), and the Mayall z-band Legacy Survey (MzLS; Prop. ID \#2016A-0453; PI: Arjun Dey). DECaLS, BASS and MzLS together include data obtained, respectively, at the Blanco telescope, Cerro Tololo Inter-American Observatory, NSF’s NOIRLab; the Bok telescope, Steward Observatory, University of Arizona; and the Mayall telescope, Kitt Peak National Observatory, NOIRLab. Pipeline processing and analyses of the data were supported by NOIRLab and the Lawrence Berkeley National Laboratory (LBNL). The Legacy Surveys project is honored to be permitted to conduct astronomical research on Iolkam Du’ag (Kitt Peak), a mountain with particular significance to the Tohono O’odham Nation.

NOIRLab is operated by the Association of Universities for Research in Astronomy (AURA) under a cooperative agreement with the National Science Foundation. LBNL is managed by the Regents of the University of California under contract to the U.S. Department of Energy.

This project used data obtained with the Dark Energy Camera (DECam), which was constructed by the Dark Energy Survey (DES) collaboration. Funding for the DES Projects has been provided by the U.S. Department of Energy, the U.S. National Science Foundation, the Ministry of Science and Education of Spain, the Science and Technology Facilities Council of the United Kingdom, the Higher Education Funding Council for England, the National Center for Supercomputing Applications at the University of Illinois at Urbana-Champaign, the Kavli Institute of Cosmological Physics at the University of Chicago, Center for Cosmology and Astro-Particle Physics at the Ohio State University, the Mitchell Institute for Fundamental Physics and Astronomy at Texas A\&M University, Financiadora de Estudos e Projetos, Fundacao Carlos Chagas Filho de Amparo, Financiadora de Estudos e Projetos, Fundacao Carlos Chagas Filho de Amparo a Pesquisa do Estado do Rio de Janeiro, Conselho Nacional de Desenvolvimento Cientifico e Tecnologico and the Ministerio da Ciencia, Tecnologia e Inovacao, the Deutsche Forschungsgemeinschaft and the Collaborating Institutions in the Dark Energy Survey. The Collaborating Institutions are Argonne National Laboratory, the University of California at Santa Cruz, the University of Cambridge, Centro de Investigaciones Energeticas, Medioambientales y Tecnologicas-Madrid, the University of Chicago, University College London, the DES-Brazil Consortium, the University of Edinburgh, the Eidgenossische Technische Hochschule (ETH) Zurich, Fermi National Accelerator Laboratory, the University of Illinois at Urbana-Champaign, the Institut de Ciencies de l’Espai (IEEC/CSIC), the Institut de Fisica d’Altes Energies, Lawrence Berkeley National Laboratory, the Ludwig Maximilians Universitat Munchen and the associated Excellence Cluster Universe, the University of Michigan, NSF’s NOIRLab, the University of Nottingham, the Ohio State University, the University of Pennsylvania, the University of Portsmouth, SLAC National Accelerator Laboratory, Stanford University, the University of Sussex, and Texas A\&M University.

BASS is a key project of the Telescope Access Program (TAP), which has been funded by the National Astronomical Observatories of China, the Chinese Academy of Sciences (the Strategic Priority Research Program “The Emergence of Cosmological Structures” Grant \# XDB09000000), and the Special Fund for Astronomy from the Ministry of Finance. The BASS is also supported by the External Cooperation Program of Chinese Academy of Sciences (Grant \# 114A11KYSB20160057), and Chinese National Natural Science Foundation (Grant \# 12120101003, \# 11433005).

The Legacy Survey team makes use of data products from the Near-Earth Object Wide-field Infrared Survey Explorer (NEOWISE), which is a project of the Jet Propulsion Laboratory/California Institute of Technology. NEOWISE is funded by the National Aeronautics and Space Administration.

The Legacy Surveys imaging of the DESI footprint is supported by the Director, Office of Science, Office of High Energy Physics of the U.S. Department of Energy under Contract No. DE-AC02-05CH1123, by the National Energy Research Scientific Computing Center, a DOE Office of Science User Facility under the same contract; and by the U.S. National Science Foundation, Division of Astronomical Sciences under Contract No. AST-0950945 to NOAO.

\vspace{5mm}

%% Similar to \facility{}, there is the optional \software command to allow 
%% authors a place to specify which programs were used during the creation of 
%% the manuscript. Authors should list each code and include either a
%% citation or url to the code inside ()s when available.

\software{astropy v5.3.4 \citep{2013A&A...558A..33A, 2018AJ....156..123A, 2022ApJ...935..167A}, numpy v1.26.4 \citep{harris2020array}, scipy v1.12.0 \citep{2020NatMe..17..261V}, matplotlib v3.8.0 \citep{Hunter:2007},
          Source Extractor v2.25 \citep{1996A&AS..117..393B}, reproject v0.11 \citep{2020ascl.soft11023R}, photutils v1.9.0 \citep{2016ascl.soft09011B}, TurbuStat v1.3 \citep{2019AJ....158....1K}, scikit-image v0.22.0 \citep{van2014scikit}, scikit-learn v1.2.2 \citep{2011JMLR...12.2825P}
          }

%% Appendix material should be preceded with a single \appendix command.
%% There should be a \section command for each appendix. Mark appendix
%% subsections with the same markup you use in the main body of the paper.

%% Each Appendix (indicated with \section) will be lettered A, B, C, etc.
%% The equation counter will reset when it encounters the \appendix
%% command and will number appendix equations (A1), (A2), etc. The
%% Figure and Table counter will not reset.

\newpage
\appendix

\section{Demonstration of Rolling Hough Transform on Simulated Images}
\label{appendix:rht}

In this section, we show the performance of the RHT algorithm on simulated images to decompose cirrus using morphological information described in Section~\ref{sec:morph}. 

\begin{figure*}[!htbp]
\centering
  \resizebox{\hsize}{!}{\includegraphics{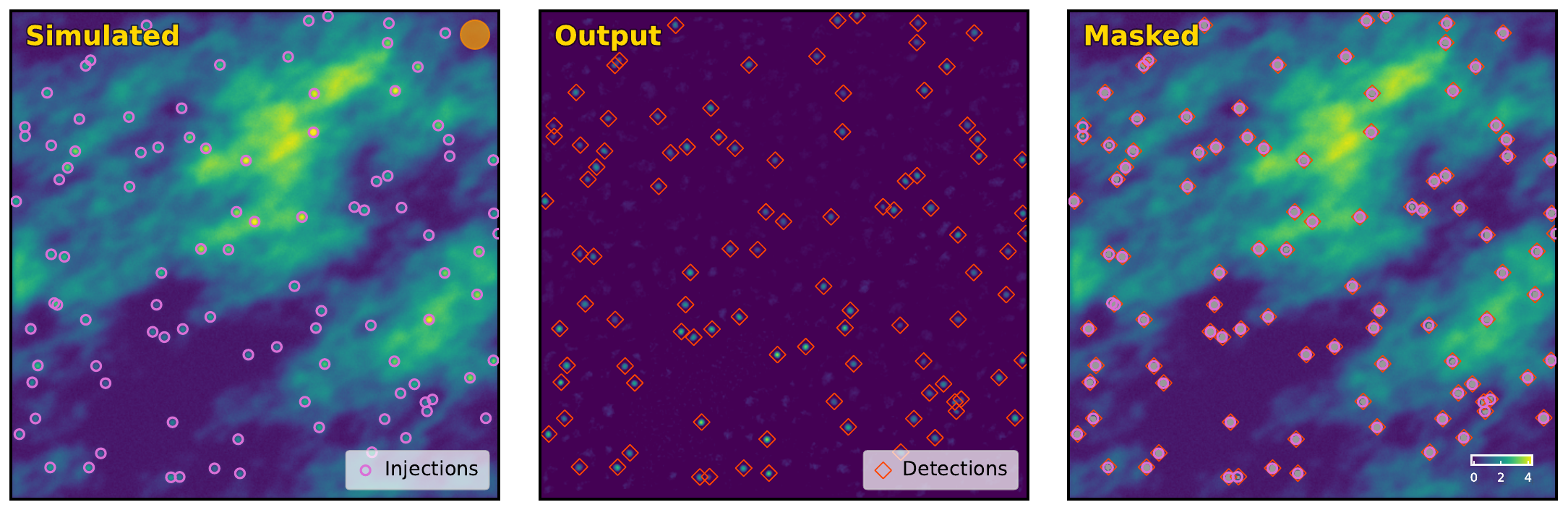}}
  \caption{Demonstration of disentangling `cirrus-like' emission from fuzzy blobs based on morphological information using RHT. Left: simulated image with injections of mock galaxies and cirrus. The mock galaxies are indicated by the magenta circles. The orange circle shows the disc size used in the RHT. Middle: output after applying RHT. The red circles indicate detections of blobs after cleaning. Right: simulated image after masking the sources in the middle panel.}
\label{fig:viz_RHT}
\end{figure*}

The simulated images are generated with toy galaxy models and mock ISM structures that mimic the cirrus emission. The image size is [800$\times$800] $\rm pix^2$. The mock galaxies are injected 100 times at random positions using Gaussian models with semi-major axis $a$ ranging from $8\arcsec$ to $15\arcsec$ and {ellipticity $\epsilon$ ranging from 0 to 0.3}. The amplitude {(peak value) of the Gaussian} ranges from 1 to 2 ADU. The mock `cirrus-like' emission is a 2D fractional Brownian Motion field (\citealt{2003ApJ...593..831M}) generated using the \texttt{simulator} module in the \texttt{TurbuStat} package (\citealt{2019AJ....158....1K}) based on a 2D power spectrum with a power index of $\gamma=-3$ expected from turbulence theories and observations (\citealt{1992AJ....103.1313G}, \citealt{2007A&A...469..595M}, \citealt{2016A&A...593A...4M}). An ellipticity of 0.5 with a position angle of $\frac{\pi}{3}$ ({defined as the angle clockwise relative to the positive y-axis}) is applied to generate the cirrus image from the 2D power spectra. The cirrus image is convolved with a Gaussian beam with FWHM of $5\arcsec$ to smooth out minor structures. The mean surface brightness of the mock cirrus is 1.5 ADU/pix$^2$. Finally, a low-level {Gaussian} noise ($\sim5\%$) is added to the simulated image to avoid computational singularities. A realization of the simulated image is shown in the left panel of Figure \ref{fig:viz_RHT}. The injected toy models are indicated by magenta circles.

We apply RHT on the simulated images with a disc size of $D_w=2\arcmin$, which is illustrated by the orange circle in Figure \ref{fig:viz_RHT}. The output image $H(x,y)$ after applying RHT is shown in the middle and right panels of Figure \ref{fig:viz_RHT}. The detections above a threshold of 3 after a cleaning requiring $b/a > 0.5$ are indicated by the red circles. In comparison with the injections, a small number of detections indeed result from mock cirrus emission with similar morphologies to blobs within the target scale. The masked map generated from the detections is shown in the right panel. Overall, the injections are well recovered by the algorithm though some confusion from cirrus exists.

\section{Comparison of \texttt{maskfill} and Gaussian Process Regression Approach}
\label{appendix:infill}
Sec.~\ref{sec:mask_infill} introduces the procedures of infilling the masked pixels, which include undersampled or saturated stellar cores, blobby LSBG candidates, leftover foreground and background sources missing in the flux modeling, and compact bright cirrus knots/clumps. Here we show the comparison of the \texttt{maskfill} approach and the approach using Gaussian process regression, or GPR.

The experiment was run on a [400x400] $\rm pix^2$ cutout of mock cirrus generated in the same procedure as Appendix \ref{appendix:rht} with a different random seed. We masked the bright part of the cirrus with a circular aperture with a radius of 5 pix, and randomly placed aperture masks with radius of 2 to 5 pix to mimic the masking from blobby emission and bright stars. Small masks were randomly placed and grown to mimic masking from undersampled cores of fainter stars and saturated cores of intermediate bright stars. The masked image was infilled separately by the \texttt{maskfill} approach in Sec.~\ref{sec:mask_infill} and by the GPR approach. For the GPR approach, we used the \texttt{GaussianProcessRegressor} utility in the \texttt{scikit-learn} package. We adopted a Radial Basis Function (RBF) kernel with a scale length fit range between 1 to 10 pix. The pixels were sparsely sampled by splitting into five folds (a `fold' is an instance of training sample) for computational efficiency and cross-validation. In each fold, the GPR model was trained with 80\% of measurements. The average of the cross-validation outputs was taken as the output.

\begin{figure}[!htbp]
\centering
  \resizebox{0.6\hsize}{!}{\includegraphics{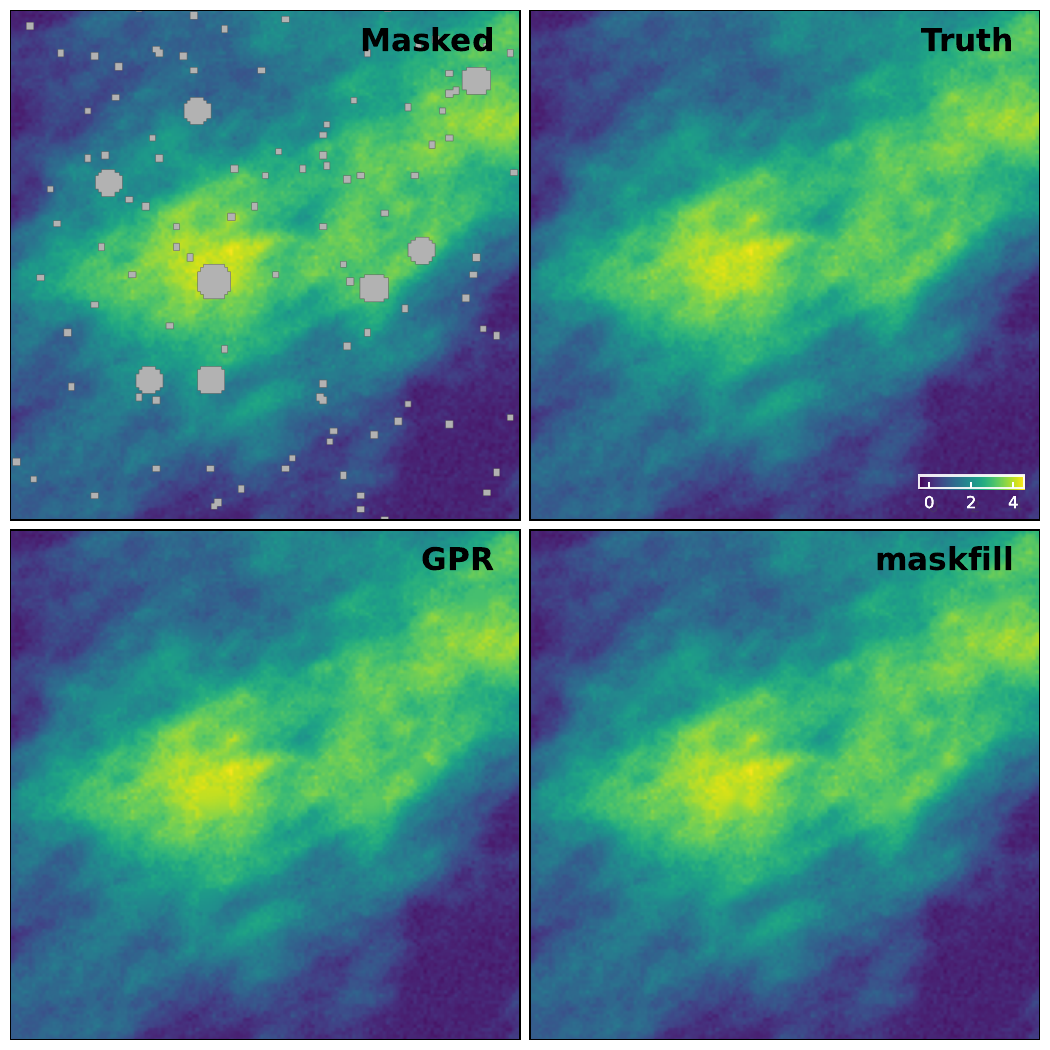}}
  \caption{Demonstration of mask infilling using the \text{maskfill} approach and the GPR approach. Upper left: [150x150] {$\rm pix^2$} cutout of the input image. Upper right: the same cutout of the ground truth {(mock cirrus in the unit of ADU)}. Lower left: cutout of the image infilled by the GPR approach. Lower right: cutout of the image infilled by the \text{maskfill} approach.}
\label{fig:compare_infill}
\end{figure}

Figure \ref{fig:compare_infill} shows the zoom-in comparison between the results, which displays [150x150] {$\rm pix^2$} cutouts in the high-intensity regions of the input image (upper left), the ground truth (upper right), the image infilled by the GPR approach (lower left) and the image infilled by the \texttt{maskfill} approach (lower right). {Overall, both approaches produce similar results, but in terms of computational time, the \texttt{maskfill} approach is more efficient than the GPR approach.} As noted in the text, future work will explore more data-driven approaches with fewer assumptions, such as the LPI approach (\citealt{2022ApJ...933..155S}).

\section{The Correlation of the DGL and Thermal Dust Tracers}
\label{appendix:dust}
Dust emission can be characterized by a number of different tracers (e.g. optical depth, radiance, FIR intensity, etc.). In this appendix, we briefly clarify the correlations seen when comparing optical dust emission from scattered light to various tracers of thermal dust emission.

Consider a simple model in which a plane-parallel dust slab is illuminated by ISRF at optical wavelength $\nu$ (\citealt{1937ApJ....85..107H}), the optical DGL can be written as:
\begin{equation}
    \label{eq:DGL_model}
    I_{\nu, {\rm sca}} = \frac{\omega_\nu}{1-\omega_\nu} I_{\nu, {\rm ISRF}} \left[1 - e^{-(1-\omega_\nu)\tau_\nu}\right]\,,
\end{equation}
where $\omega_\nu$ represents the albedo and $I_{\nu, {\rm ISRF}}$ is the incident strength of ISRF. This does not include the scattering anisotropy of dust grains, which has a Galactic latitude dependence\footnote{A more complicated model including scattering and ISRF anisotropies can be referred to \cite{2012ApJ...744..129B}, Eq. A7.} (e.g., \citealt{2017ApJ...849...31S}). The effect of multiple scattering is not considered either.

In the optically thin limit $\tau_\nu\ll 1$, Eq.~\ref{eq:DGL_model} approximates to:
\begin{equation}
    \label{eq:I_sca_tau}
    I_{\nu, {\rm sca}} = \omega_\nu \tau_\nu I_{\nu, {\rm ISRF}}  \propto \tau_\nu\,.
\end{equation}

From Eq.~\ref{eq:tau}, $\tau_\nu$ at optical wavelength is given by:
\begin{equation}
    \label{eq:tau_nu_353}
    \tau_\nu\, = \frac{\int n_{\rm d}(a) \cdot \pi a^2 \cdot \,Q_{\nu,{\rm ext}}(a)\,da}{\int n_{\rm d}(a) \cdot \pi a^2 \cdot \,Q_{\rm 353, ext}(a)\,da} \, \tau_{\rm 353}  =  \frac{Q_{\nu,{\rm ext}}}{Q_{\rm 353, ext}}\,\tau_{\rm 353}\,,
\end{equation}
{where $n_{\rm d}(a)$ is the dust column density as a function of grain size $a$, $Q_{\nu,{\rm ext}}$ is the extinction efficiency as a funtion of grain size}, and $Q_{\nu,{\rm ext}}$ is the equivalent extinction efficiency integrated over grain size distribution. $Q_{\nu,{\rm ext}}$ is determined by the dust model.
% and $\left<\cdot\right>$ stands for the integral over a grain size distribution. 
Combining Eq.~\ref{eq:I_sca_tau} and \ref{eq:tau_nu_353} and assuming constant grain size distribution and composition, $I_{\nu, {\rm sca}}$ has the correlation (when $\tau_\nu\ll 1$):
\begin{equation}
    I_{\nu, {\rm sca}} \propto \, \tau_{\rm 353}\,
\end{equation}
% \sout{Using Eq.\ref{eq:tau_freq}, this yields to the linear correlation between $I_{\nu, {\rm sca}}$ and $\tau_{\rm 353}$ at $\nu_{0}$ = 353 GHz}:
% \begin{equation}
%     \label{eq:I_sca_tau}
%     \sout{I_{\nu, {\rm sca}} \propto (\frac{\nu}{\nu_{0}})^\beta \, \tau_{\rm 353} \propto \nu^\beta \, \tau_{\rm 353}\,,}
% \end{equation}
% \sout{with a weak dependence on $\beta$ given its small spatial variation.}

On the other hand, the optical DGL can also be expressed as:

\begin{equation}
    \label{eq:DGL_ISRF}
    \nu I_{\nu, {\rm sca}} = 
    \left<\alpha_{\nu, {\rm sca}}\right> \int I_{\nu, {\rm sca}}\,d\nu\,,
\end{equation}
if defining $\left<\alpha_{\nu, {\rm sca}}\right>$ as a (dimensionless) scaling factor that scales DGL at frequency $\nu$ from the total scattered light:
\begin{equation}
    \label{eq:alpha_sca}
    \left<\alpha_{\nu, {\rm sca}}\right> =
    \frac{\nu\, Q_{\nu,{\rm sca}}I_{\nu, {\rm ISRF}}}{\int Q_{\nu,{\rm sca}}I_{\nu, {\rm ISRF}}d\nu}\,,
\end{equation}
where $Q_{\nu,{\rm sca}}$ is the scattering efficiency intergrated over grain size. $\left<\alpha_{\nu, {\rm sca}}\right>$ is dependent on the dust model and the incident ISRF. Based on the energy conservation law that the energy absorbed equals the energy thermally emitted, Eq.~\ref{eq:DGL_ISRF} leads to:

\begin{equation}
    \label{eq:DGL_IR}
    I_{\nu, {\rm sca}} = \left<\alpha_{\nu, {\rm sca}}\right> \frac{\overline{\omega}}{1-\overline{\omega}} \int I_{\nu, {\rm thermal}}\,d\nu\,,
\end{equation}
% \begin{equation}       % old equation 
%     \label{eq:DGL_IR}
%      I_{\nu, {\rm sca}} 
%      =\left<\alpha_{\nu, {\rm sca}}\right> \int I_{\nu, {\rm sca}}\,d\nu = \left<\alpha_{\nu, {\rm sca}}\right> \frac{\overline{\omega}}{1-\overline{\omega}} \int I_{\nu, {\rm thermal}}\,d\nu
% \end{equation}
% based on the energy conservation law that the energy absorbed equals the energy thermally emitted, 
where $\overline{\omega}$ is the spectrum-averaged albedo. %$\left<\alpha_{\nu, {\rm sca}}\right>$ is the scaling factor that scales the DGL intensity at frequency $\nu$ from the total scattered light, which depends on grain properties.
Therefore, by definition of $\mathcal{R}$ (Eq.~\ref{eq:radiance}), $I_{\nu, {\rm sca}}$ has the correlation:
\begin{equation}
    I_{\nu, {\rm sca}} \propto \mathcal{R}\,
\end{equation}
Note that radiance is less affected by optical effects because FIR is nearly optically thin with little extinction.

Similarly, by defining a scaling factor ${\left<\sigma_{100, {\rm abs}}\right>}$ that converts total thermal emission to IRAS 100~${\rm \mu m}$ bandpass power ($\sim 0.52$ for models from 0.5 to 1.5 times the local ISRF; \citealt{2012ApJ...744..129B}), Eq.~\ref{eq:DGL_IR} leads to:
\begin{equation}
I_{\nu, {\rm sca}} \propto \frac{\left<\alpha_{\nu, {\rm sca}}\right>}{\left<\alpha_{100, {\rm abs}}\right>} \frac{\overline{\omega}}{1-\overline{\omega}} \, I_{100}\,,
\end{equation}
which explains the observed correlation with IRAS 100~${\rm \mu m}$ intensities.

Finally, we can clarify the behavior of the correlation under LTE. Using Eq.~\ref{eq:I_nu}, $\mathcal{R}$ can be analytically expressed in terms of the Gamma ($\Gamma$) and Riemann zeta functions ($\zeta$):
\begin{equation}
\label{eq:rad}
\mathcal{R} = \tau_{353} \frac{\sigma_{{\mathrm SB}}}{\pi} T^4 \left(\frac{k_BT}{h\nu_0}\right)^\beta \frac{\Gamma(\beta+4)\zeta(\beta+4)}{\Gamma(4)\zeta(4)} \,,
\end{equation}
{where $\sigma_{{\mathrm SB}}$ is the Stefan-Boltzmann constant, $k_B$ is the Boltzmann constant, and $h$ is the Planck constant.} Therefore the optical DGL has the following dependency:
\begin{equation}
    \label{eq:Inu_beta_T}
    I_{\nu, {\rm sca}} \propto \tau_{\rm 353} \,T^{(\beta+4)} \,\Gamma(\beta+4)\zeta(\beta+4) = \Lambda(T,\beta)\,\tau_{353}\,.
\end{equation}
The $\Lambda(T,\beta)$ dependence on $T$ and $\beta$ has a small spatial variation at high latitudes.
In the example dataset, the fractional fluctuation $\frac{\delta\Lambda}{\Lambda}$ derived from the Planck thermal dust model is $\sim0.09$.

In summary, the two tracers (optical depth and radiance) converge when (1) in the optically thin limit and (2) $\beta$ and $T$ do not present large spatial variation in the field of interest, which typically holds at high Galactic latitudes (\citealt{planck2013-p06b}). It is worth noting that Eq.~\ref{eq:Inu_beta_T} does not hold at non-LTE regions. Therefore, caution needs to be taken in these cases to assume the same correlations, e.g., dust scattering illuminated by a nearby OB star where photodissociation, luminance, and scattering and thermal emission from ultra-small grains might occur.

\section{Generation of Mock UDGs in the Injection-Recovery Test} \label{appendix:artpop}
In Section~\ref{sec:test_inj}, we generated mock UDGs, injected them into the images, and recovered them after cirrus removal. Here we describe the details of the galaxy model construction. The mock UDGs were constructed using ArtPop, a Python package for generating artificial observations of stellar systems with synthetic stellar populations (\citealt{2022ApJ...941...26G}). To accomplish this goal, we adopted a set of realistic parameters as follows.

In the test presented in Section~\ref{sec:test_single}, a mock UDG with a stellar mass of $M_* = 10^8 M_{\odot}$ was created from the MIST isochrones (\citealt{2016ApJ...823..102C}, \citealt{2016ApJS..222....8D}) using a simple stellar population (SSP) following a Kroupa initial mass function (\citealt{2001MNRAS.322..231K}). The mock galaxy was set with a metallicity of $[Fe/H]=-1.5$ and an age of 9 $Gyr$. The galaxy was placed at $D=20\,Mpc$ and projected onto the image plane {at the pixel resolution of Dragonfly (2.85\arcsec/pix)}, where stars were sampled following the distribution of a 2D S{\'e}rsic profile with a S{\'e}rsic index $n_{\rm sersic}=0.8$, an ellipticity $\epsilon=0.1$, and an apparent effective radius $R_{\rm eff}=20{\arcsec}$ ($\sim2\,kpc$ at $D=20\,Mpc$). For computational efficiency, only stars brighter than a magnitude limit of 32 mag {in \textit{g} band} were sampled and fainter stars were combined into an integrated component. To convert the model into mock observations, we used the observing configuration of Dragonfly (equivalent to a 1m telescope) with equivalent exposure time. {We adopted the PSF model retrieved in Sec.~\ref{sec:psf} with a} seeing of FWHM $=5\arcsec$. The observed flux was transformed into the SDSSugriz photometric system given that Dragonfly's $g$ and $r$ filters match that of SDSS. The integrated color of the mock UDG from the stellar population synthesis is $g-r$ = 0.54, and the effective V-band surface brightness, $\mu_{{\rm eff}, V}$, is 26.0 mag/arcsec$^2$. {It should be noted that the model does not account for Galactic extinction, and therefore real LSBGs observed in the cirrus fields {with similar physical parameters} would be redder and fainter.}

\begin{figure}[!htbp]
\centering
  \resizebox{0.36\hsize}{!}{\includegraphics{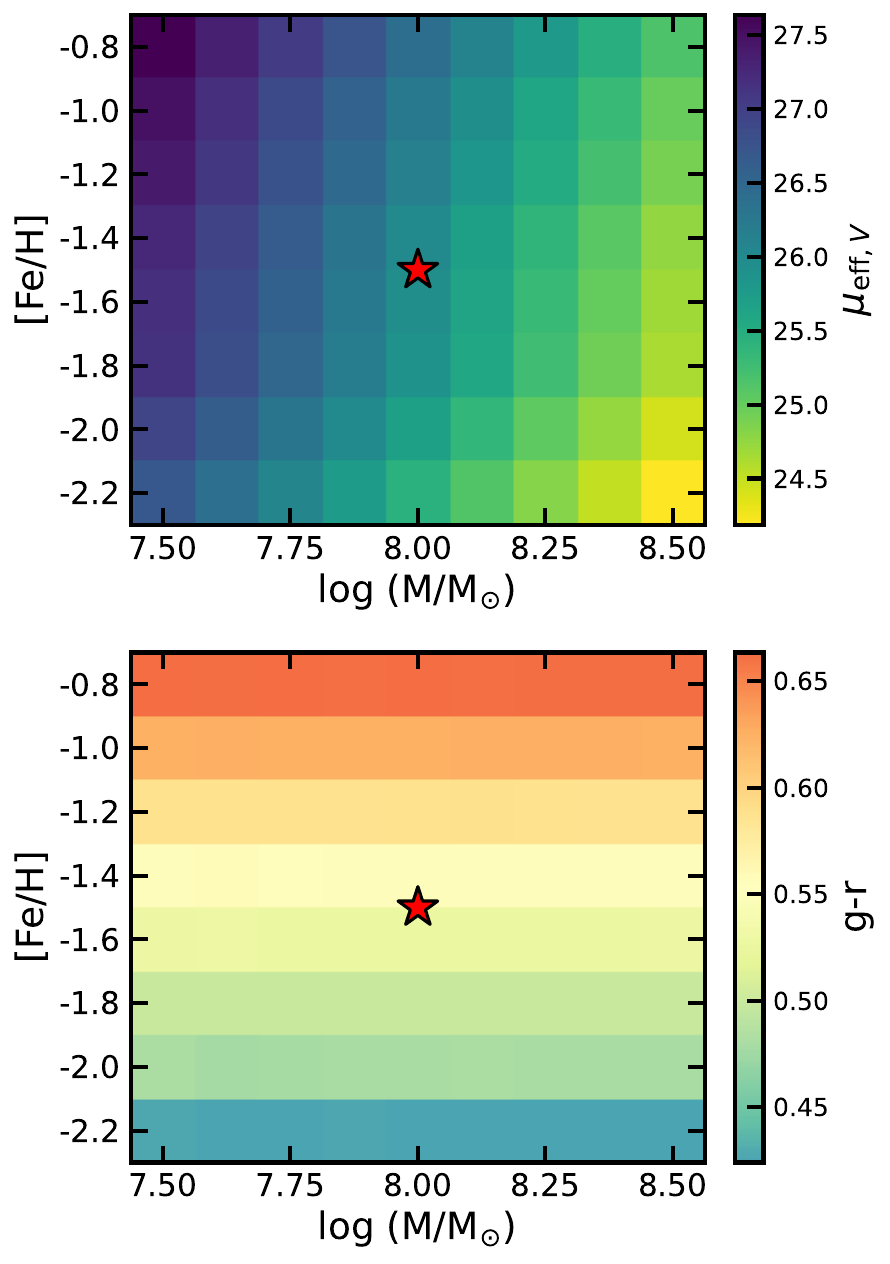}}
  \caption{Mean V-band surface brightness within the effective radius, $\mu_{{\rm eff}, V}$ (upper), and $g-r$ color (lower) of the mass-metallicity grid models.  The red star indicates the fiducial model at log ($M_*/M_{\odot}) = 8$ and [Fe/H]  = -1.5, which is referred to in the main text. The model is fixed at a distance of 20 Mpc with an age of 9 Gyr, a size of $R_{\rm eff}$ = 20{\arcsec} ($\sim$ 2 kpc at 20Mpc), and a S{\'e}rsic index $n_{\rm sersic}=0.8$. Models with lower metallicity have bluer colors, and appear brighter at fixed mass. {Note that the galaxy models do not include Galactic extinction.}}
\label{fig:test_grid}
\end{figure}

\begin{figure*}[!htbp]
\centering
  \resizebox{\hsize}{!}{\includegraphics{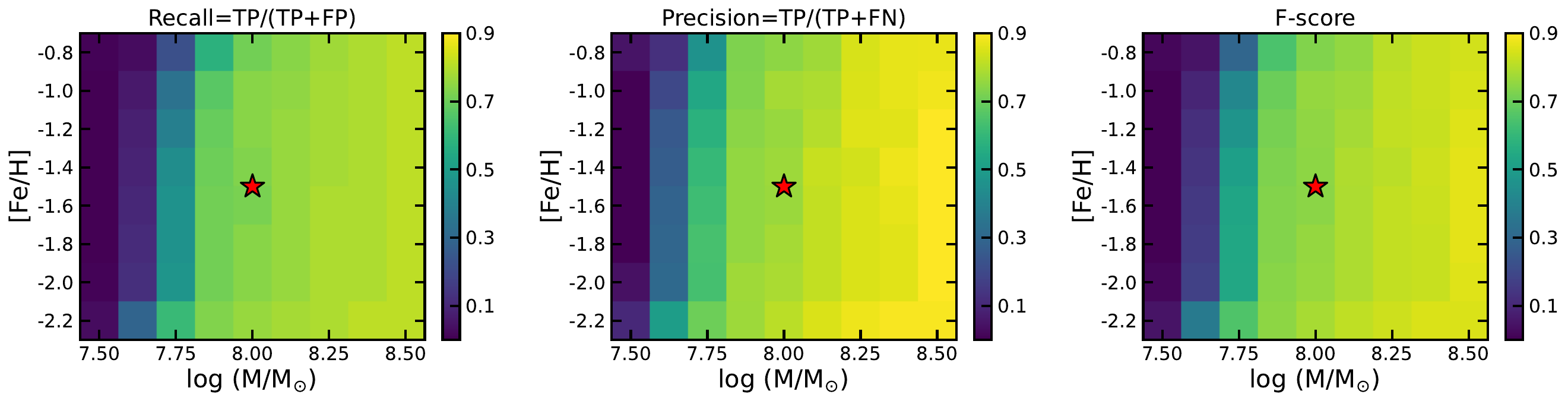}}
  \caption{Performance metrics (recall, precision, and the F-score) for recovering injected galaxies in the cirrus removed image at varying stellar mass and metallicities. The mock galaxies are placed at 20 Mpc with a fixed age of 9 Gyr, a size of $R_{\rm eff}$ = 2 kpc, and an S{\'e}rsic index $n_{\rm sersic}=0.8$. The metrics are reprojected to the grid of apparent parameters ($\mu_{{\rm eff}, V}$ and $g-r$) in Fig.~\ref{fig:test_grid_metrics}. The red star indicates the fiducial model in Sec.~\ref{sec:test_single}.}
\label{fig:test_grid_metrics_original}
\end{figure*}

In Section~\ref{sec:test_grid}, we explored the parameter space of the model galaxy by building a grid of UDG models in $M_*$ and metallicity [Fe/H], and fix the rest of the parameters (D, Age, $R_{\rm eff}$, $n_{\rm sersic}$). The stellar mass varies from log M/M$_{\odot}$ = 7.5 to 8.5 at a step of 0.125. Metallicity is changed from [Fe/H] = -2.2 to -0.8 at a step of 0.2. Note that $M_*$ serves as a normalization that controls the mean surface brightness of the candidates, which can be controlled accordingly by changing the distance, size, and $n_{\rm sersic}$. Likewise, given the age-metallicity degeneracy in stellar population synthesis, a lower metallicity is equivalent to a younger age, which produces a bluer color. Figure \ref{fig:test_grid} shows the variation of the mean surface brightness $\mu_{{\rm eff}, V}$ and $g-r$ of the grid models. {Note that the models do not take Galactic extinction into account, which were injected in the images without additional extinction and reddening correction.} Figure \ref{fig:test_grid_metrics_original} shows the performance metrics (recall, precision, and F-score) evaluated on the original $M_*$-[Fe/H] grid, which was reprojected into $\mu_{\rm eff, V}$ and $g-r$ space in Fig.~\ref{fig:test_grid_metrics}. The UDG model used for the demonstration in Sec.~\ref{sec:test_single} is indicated by the red star in Fig.~\ref{fig:test_grid} and Fig.~\ref{fig:test_grid_metrics_original}. {As noted in the text, galaxies with the same apparent surface brightness and colors will be intrinsically brighter and bluer if overshadowed by optically thicker cirrus. This will be important when connecting the physical parameters to the apparent observables (i.e., from Fig.~\ref{fig:test_grid_metrics_original} to Fig.~\ref{fig:test_grid_metrics}), e.g., when evaluating the completeness of a survey, but would not affect the demonstrative purpose of the experiments for this work.} 

\bibliography{bibtex, Planck_bib, supp_bib}{}

\begin{thebibliography}{}
\expandafter\ifx\csname natexlab\endcsname\relax\def\natexlab#1{#1}\fi
\providecommand{\url}[1]{\href{#1}{#1}}
\providecommand{\dodoi}[1]{doi:~\href{http://doi.org/#1}{\nolinkurl{#1}}}
\providecommand{\doeprint}[1]{\href{http://ascl.net/#1}{\nolinkurl{http://ascl.net/#1}}}
\providecommand{\doarXiv}[1]{\href{https://arxiv.org/abs/#1}{\nolinkurl{https://arxiv.org/abs/#1}}}

\bibitem[{{Abraham} {et~al.}(2003){Abraham}, {van den Bergh}, \& {Nair}}]{2003ApJ...588..218A}
{Abraham}, R.~G., {van den Bergh}, S., \& {Nair}, P. 2003, \apj, 588, 218, \dodoi{10.1086/373919}

\bibitem[{{Abraham} \& {van Dokkum}(2014)}]{2014PASP..126...55A}
{Abraham}, R.~G., \& {van Dokkum}, P.~G. 2014, \pasp, 126, 55, \dodoi{10.1086/674875}

\bibitem[{{Aihara} {et~al.}(2018){Aihara}, {Arimoto}, {Armstrong}, {Arnouts}, {Bahcall}, {Bickerton}, {Bosch}, {Bundy}, {Capak}, {Chan}, {Chiba}, {Coupon}, {Egami}, {Enoki}, {Finet}, {Fujimori}, {Fujimoto}, {Furusawa}, {Furusawa}, {Goto}, {Goulding}, {Greco}, {Greene}, {Gunn}, {Hamana}, {Harikane}, {Hashimoto}, {Hattori}, {Hayashi}, {Hayashi}, {He{\l}miniak}, {Higuchi}, {Hikage}, {Ho}, {Hsieh}, {Huang}, {Huang}, {Ikeda}, {Imanishi}, {Inoue}, {Iwasawa}, {Iwata}, {Jaelani}, {Jian}, {Kamata}, {Karoji}, {Kashikawa}, {Katayama}, {Kawanomoto}, {Kayo}, {Koda}, {Koike}, {Kojima}, {Komiyama}, {Konno}, {Koshida}, {Koyama}, {Kusakabe}, {Leauthaud}, {Lee}, {Lin}, {Lin}, {Lupton}, {Mandelbaum}, {Matsuoka}, {Medezinski}, {Mineo}, {Miyama}, {Miyatake}, {Miyazaki}, {Momose}, {More}, {More}, {Moritani}, {Moriya}, {Morokuma}, {Mukae}, {Murata}, {Murayama}, {Nagao}, {Nakata}, {Niida}, {Niikura}, {Nishizawa}, {Obuchi}, {Oguri}, {Oishi}, {Okabe}, {Okamoto}, {Okura}, {Ono}, {Onodera}, {Onoue}, {Osato}, {Ouchi}, {Price}, {Pyo},
  {Sako}, {Sawicki}, {Shibuya}, {Shimasaku}, {Shimono}, {Shirasaki}, {Silverman}, {Simet}, {Speagle}, {Spergel}, {Strauss}, {Sugahara}, {Sugiyama}, {Suto}, {Suyu}, {Suzuki}, {Tait}, {Takada}, {Takata}, {Tamura}, {Tanaka}, {Tanaka}, {Tanaka}, {Tanaka}, {Terai}, {Terashima}, {Toba}, {Tominaga}, {Toshikawa}, {Turner}, {Uchida}, {Uchiyama}, {Umetsu}, {Uraguchi}, {Urata}, {Usuda}, {Utsumi}, {Wang}, {Wang}, {Wong}, {Yabe}, {Yamada}, {Yamanoi}, {Yasuda}, {Yeh}, {Yonehara}, \& {Yuma}}]{2018PASJ...70S...4A}
{Aihara}, H., {Arimoto}, N., {Armstrong}, R., {et~al.} 2018, \pasj, 70, S4, \dodoi{10.1093/pasj/psx066}

\bibitem[{{Alina} {et~al.}(2022){Alina}, {Shomanov}, \& {Baimukhametova}}]{2022arXiv220500683A}
{Alina}, D., {Shomanov}, A., \& {Baimukhametova}, S. 2022, arXiv e-prints, arXiv:2205.00683, \dodoi{10.48550/arXiv.2205.00683}

\bibitem[{{Appleton} {et~al.}(1993){Appleton}, {Siqueira}, \& {Basart}}]{1993AJ....106.1664A}
{Appleton}, P.~N., {Siqueira}, P.~R., \& {Basart}, J.~P. 1993, \aj, 106, 1664, \dodoi{10.1086/116756}

\bibitem[{{Arai} {et~al.}(2015){Arai}, {Matsuura}, {Bock}, {Cooray}, {Kim}, {Lanz}, {Lee}, {Lee}, {Sano}, {Smidt}, {Matsumoto}, {Nakagawa}, {Onishi}, {Korngut}, {Shirahata}, {Tsumura}, \& {Zemcov}}]{2015ApJ...806...69A}
{Arai}, T., {Matsuura}, S., {Bock}, J., {et~al.} 2015, \apj, 806, 69, \dodoi{10.1088/0004-637X/806/1/69}

\bibitem[{{Arzoumanian} {et~al.}(2011){Arzoumanian}, {Andr{\'e}}, {Didelon}, {K{\"o}nyves}, {Schneider}, {Men'shchikov}, {Sousbie}, {Zavagno}, {Bontemps}, {di Francesco}, {Griffin}, {Hennemann}, {Hill}, {Kirk}, {Martin}, {Minier}, {Molinari}, {Motte}, {Peretto}, {Pezzuto}, {Spinoglio}, {Ward-Thompson}, {White}, \& {Wilson}}]{2011A&A...529L...6A}
{Arzoumanian}, D., {Andr{\'e}}, P., {Didelon}, P., {et~al.} 2011, \aap, 529, L6, \dodoi{10.1051/0004-6361/201116596}

\bibitem[{{Astropy Collaboration} {et~al.}(2013){Astropy Collaboration}, {Robitaille}, {Tollerud}, {Greenfield}, {Droettboom}, {Bray}, {Aldcroft}, {Davis}, {Ginsburg}, {Price-Whelan}, {Kerzendorf}, {Conley}, {Crighton}, {Barbary}, {Muna}, {Ferguson}, {Grollier}, {Parikh}, {Nair}, {Unther}, {Deil}, {Woillez}, {Conseil}, {Kramer}, {Turner}, {Singer}, {Fox}, {Weaver}, {Zabalza}, {Edwards}, {Azalee Bostroem}, {Burke}, {Casey}, {Crawford}, {Dencheva}, {Ely}, {Jenness}, {Labrie}, {Lim}, {Pierfederici}, {Pontzen}, {Ptak}, {Refsdal}, {Servillat}, \& {Streicher}}]{2013A&A...558A..33A}
{Astropy Collaboration}, {Robitaille}, T.~P., {Tollerud}, E.~J., {et~al.} 2013, \aap, 558, A33, \dodoi{10.1051/0004-6361/201322068}

\bibitem[{{Astropy Collaboration} {et~al.}(2018){Astropy Collaboration}, {Price-Whelan}, {Sip{\H{o}}cz}, {G{\"u}nther}, {Lim}, {Crawford}, {Conseil}, {Shupe}, {Craig}, {Dencheva}, {Ginsburg}, {VanderPlas}, {Bradley}, {P{\'e}rez-Su{\'a}rez}, {de Val-Borro}, {Aldcroft}, {Cruz}, {Robitaille}, {Tollerud}, {Ardelean}, {Babej}, {Bach}, {Bachetti}, {Bakanov}, {Bamford}, {Barentsen}, {Barmby}, {Baumbach}, {Berry}, {Biscani}, {Boquien}, {Bostroem}, {Bouma}, {Brammer}, {Bray}, {Breytenbach}, {Buddelmeijer}, {Burke}, {Calderone}, {Cano Rodr{\'\i}guez}, {Cara}, {Cardoso}, {Cheedella}, {Copin}, {Corrales}, {Crichton}, {D'Avella}, {Deil}, {Depagne}, {Dietrich}, {Donath}, {Droettboom}, {Earl}, {Erben}, {Fabbro}, {Ferreira}, {Finethy}, {Fox}, {Garrison}, {Gibbons}, {Goldstein}, {Gommers}, {Greco}, {Greenfield}, {Groener}, {Grollier}, {Hagen}, {Hirst}, {Homeier}, {Horton}, {Hosseinzadeh}, {Hu}, {Hunkeler}, {Ivezi{\'c}}, {Jain}, {Jenness}, {Kanarek}, {Kendrew}, {Kern}, {Kerzendorf}, {Khvalko}, {King}, {Kirkby}, {Kulkarni},
  {Kumar}, {Lee}, {Lenz}, {Littlefair}, {Ma}, {Macleod}, {Mastropietro}, {McCully}, {Montagnac}, {Morris}, {Mueller}, {Mumford}, {Muna}, {Murphy}, {Nelson}, {Nguyen}, {Ninan}, {N{\"o}the}, {Ogaz}, {Oh}, {Parejko}, {Parley}, {Pascual}, {Patil}, {Patil}, {Plunkett}, {Prochaska}, {Rastogi}, {Reddy Janga}, {Sabater}, {Sakurikar}, {Seifert}, {Sherbert}, {Sherwood-Taylor}, {Shih}, {Sick}, {Silbiger}, {Singanamalla}, {Singer}, {Sladen}, {Sooley}, {Sornarajah}, {Streicher}, {Teuben}, {Thomas}, {Tremblay}, {Turner}, {Terr{\'o}n}, {van Kerkwijk}, {de la Vega}, {Watkins}, {Weaver}, {Whitmore}, {Woillez}, {Zabalza}, \& {Astropy Contributors}}]{2018AJ....156..123A}
{Astropy Collaboration}, {Price-Whelan}, A.~M., {Sip{\H{o}}cz}, B.~M., {et~al.} 2018, \aj, 156, 123, \dodoi{10.3847/1538-3881/aabc4f}

\bibitem[{{Astropy Collaboration} {et~al.}(2022){Astropy Collaboration}, {Price-Whelan}, {Lim}, {Earl}, {Starkman}, {Bradley}, {Shupe}, {Patil}, {Corrales}, {Brasseur}, {N{\"o}the}, {Donath}, {Tollerud}, {Morris}, {Ginsburg}, {Vaher}, {Weaver}, {Tocknell}, {Jamieson}, {van Kerkwijk}, {Robitaille}, {Merry}, {Bachetti}, {G{\"u}nther}, {Aldcroft}, {Alvarado-Montes}, {Archibald}, {B{\'o}di}, {Bapat}, {Barentsen}, {Baz{\'a}n}, {Biswas}, {Boquien}, {Burke}, {Cara}, {Cara}, {Conroy}, {Conseil}, {Craig}, {Cross}, {Cruz}, {D'Eugenio}, {Dencheva}, {Devillepoix}, {Dietrich}, {Eigenbrot}, {Erben}, {Ferreira}, {Foreman-Mackey}, {Fox}, {Freij}, {Garg}, {Geda}, {Glattly}, {Gondhalekar}, {Gordon}, {Grant}, {Greenfield}, {Groener}, {Guest}, {Gurovich}, {Handberg}, {Hart}, {Hatfield-Dodds}, {Homeier}, {Hosseinzadeh}, {Jenness}, {Jones}, {Joseph}, {Kalmbach}, {Karamehmetoglu}, {Ka{\l}uszy{\'n}ski}, {Kelley}, {Kern}, {Kerzendorf}, {Koch}, {Kulumani}, {Lee}, {Ly}, {Ma}, {MacBride}, {Maljaars}, {Muna}, {Murphy}, {Norman},
  {O'Steen}, {Oman}, {Pacifici}, {Pascual}, {Pascual-Granado}, {Patil}, {Perren}, {Pickering}, {Rastogi}, {Roulston}, {Ryan}, {Rykoff}, {Sabater}, {Sakurikar}, {Salgado}, {Sanghi}, {Saunders}, {Savchenko}, {Schwardt}, {Seifert-Eckert}, {Shih}, {Jain}, {Shukla}, {Sick}, {Simpson}, {Singanamalla}, {Singer}, {Singhal}, {Sinha}, {Sip{\H{o}}cz}, {Spitler}, {Stansby}, {Streicher}, {{\v{S}}umak}, {Swinbank}, {Taranu}, {Tewary}, {Tremblay}, {de Val-Borro}, {Van Kooten}, {Vasovi{\'c}}, {Verma}, {de Miranda Cardoso}, {Williams}, {Wilson}, {Winkel}, {Wood-Vasey}, {Xue}, {Yoachim}, {Zhang}, {Zonca}, \& {Astropy Project Contributors}}]{2022ApJ...935..167A}
{Astropy Collaboration}, {Price-Whelan}, A.~M., {Lim}, P.~L., {et~al.} 2022, \apj, 935, 167, \dodoi{10.3847/1538-4357/ac7c74}

\bibitem[{{Barriault} {et~al.}(2010){Barriault}, {Joncas}, {Falgarone}, {Marshall}, {Heyer}, {Boulanger}, {Foster}, {Brunt}, {Miville-Desch{\^e}nes}, {Blagrave}, {Kothes}, {Landecker}, {Martin}, {Scott}, {Stil}, \& {Taylor}}]{2010MNRAS.406.2713B}
{Barriault}, L., {Joncas}, G., {Falgarone}, E., {et~al.} 2010, \mnras, 406, 2713, \dodoi{10.1111/j.1365-2966.2010.16871.x}

\bibitem[{{Bensch} {et~al.}(2001){Bensch}, {Stutzki}, \& {Ossenkopf}}]{2001A&A...366..636B}
{Bensch}, F., {Stutzki}, J., \& {Ossenkopf}, V. 2001, \aap, 366, 636, \dodoi{10.1051/0004-6361:20000292}

\bibitem[{{Bertin} \& {Arnouts}(1996)}]{1996A&AS..117..393B}
{Bertin}, E., \& {Arnouts}, S. 1996, \aaps, 117, 393, \dodoi{10.1051/aas:1996164}

\bibitem[{{B{\'\i}lek} {et~al.}(2020){B{\'\i}lek}, {Duc}, {Cuillandre}, {Gwyn}, {Cappellari}, {Bekaert}, {Bonfini}, {Bitsakis}, {Paudel}, {Krajnovi{\'c}}, {Durrell}, \& {Marleau}}]{2020MNRAS.498.2138B}
{B{\'\i}lek}, M., {Duc}, P.-A., {Cuillandre}, J.-C., {et~al.} 2020, \mnras, 498, 2138, \dodoi{10.1093/mnras/staa2248}

\bibitem[{{Boissier} {et~al.}(2015){Boissier}, {Boselli}, {Voyer}, {Bianchi}, {Pappalardo}, {Guhathakurta}, {Heinis}, {Cortese}, {Duc}, {Cuillandre}, {Davies}, \& {Smith}}]{2015A&A...579A..29B}
{Boissier}, S., {Boselli}, A., {Voyer}, E., {et~al.} 2015, \aap, 579, A29, \dodoi{10.1051/0004-6361/201526089}

\bibitem[{{Bowes} \& {Martin}(2023)}]{2023ApJ...959...40B}
{Bowes}, S.~K., \& {Martin}, P.~G. 2023, \apj, 959, 40, \dodoi{10.3847/1538-4357/ad0971}

\bibitem[{{Bracco} {et~al.}(2011){Bracco}, {Cooray}, {Veneziani}, {Amblard}, {Serra}, {Wardlow}, {Thompson}, {White}, {Auld}, {Baes}, {Bertoldi}, {Buttiglione}, {Cava}, {Clements}, {Dariush}, {de Zotti}, {Dunne}, {Dye}, {Eales}, {Fritz}, {Gomez}, {Hopwood}, {Ibar}, {Ivison}, {Jarvis}, {Lagache}, {Lee}, {Leeuw}, {Maddox}, {Micha{\l}owski}, {Pearson}, {Pohlen}, {Rigby}, {Rodighiero}, {Smith}, {Temi}, {Vaccari}, \& {van der Werf}}]{2011MNRAS.412.1151B}
{Bracco}, A., {Cooray}, A., {Veneziani}, M., {et~al.} 2011, \mnras, 412, 1151, \dodoi{10.1111/j.1365-2966.2010.17971.x}

\bibitem[{{Bradley} {et~al.}(2016){Bradley}, {Sipocz}, {Robitaille}, {Tollerud}, {Deil}, {Vin{\'\i}cius}, {Barbary}, {G{\"u}nther}, {Bostroem}, {Droettboom}, {Bray}, {Bratholm}, {Pickering}, {Craig}, {Pascual}, {Greco}, {Donath}, {Kerzendorf}, {Littlefair}, {Barentsen}, {D'Eugenio}, \& {Weaver}}]{2016ascl.soft09011B}
{Bradley}, L., {Sipocz}, B., {Robitaille}, T., {et~al.} 2016, {Photutils: Photometry Tools}.
\newblock \doeprint{1609.011}

\bibitem[{{Brandt} \& {Draine}(2012)}]{2012ApJ...744..129B}
{Brandt}, T.~D., \& {Draine}, B.~T. 2012, \apj, 744, 129, \dodoi{10.1088/0004-637X/744/2/129}

\bibitem[{{Carlsten} {et~al.}(2021){Carlsten}, {Greene}, {Greco}, {Beaton}, \& {Kado-Fong}}]{2021ApJ...922..267C}
{Carlsten}, S.~G., {Greene}, J.~E., {Greco}, J.~P., {Beaton}, R.~L., \& {Kado-Fong}, E. 2021, \apj, 922, 267, \dodoi{10.3847/1538-4357/ac2581}

\bibitem[{{Chapman} {et~al.}(2013){Chapman}, {Widrow}, {Collins}, {Dubinski}, {Ibata}, {Rich}, {Ferguson}, {Irwin}, {Lewis}, {Martin}, {McConnachie}, {Pe{\~n}arrubia}, \& {Tanvir}}]{2013MNRAS.430...37C}
{Chapman}, S.~C., {Widrow}, L., {Collins}, M.~L.~M., {et~al.} 2013, \mnras, 430, 37, \dodoi{10.1093/mnras/sts392}

\bibitem[{{Chellew} {et~al.}(2022){Chellew}, {Brandt}, {Hensley}, {Draine}, \& {Matthaey}}]{2022ApJ...932..112C}
{Chellew}, B., {Brandt}, T.~D., {Hensley}, B.~S., {Draine}, B.~T., \& {Matthaey}, E. 2022, \apj, 932, 112, \dodoi{10.3847/1538-4357/ac6efc}

\bibitem[{{Choi} {et~al.}(2016){Choi}, {Dotter}, {Conroy}, {Cantiello}, {Paxton}, \& {Johnson}}]{2016ApJ...823..102C}
{Choi}, J., {Dotter}, A., {Conroy}, C., {et~al.} 2016, \apj, 823, 102, \dodoi{10.3847/0004-637X/823/2/102}

\bibitem[{{Clark} {et~al.}(2014){Clark}, {Peek}, \& {Putman}}]{2014ApJ...789...82C}
{Clark}, S.~E., {Peek}, J.~E.~G., \& {Putman}, M.~E. 2014, \apj, 789, 82, \dodoi{10.1088/0004-637X/789/1/82}

\bibitem[{{Clarke} {et~al.}(2020){Clarke}, {Williams}, \& {Walch}}]{2020MNRAS.497.4390C}
{Clarke}, S.~D., {Williams}, G.~M., \& {Walch}, S. 2020, \mnras, 497, 4390, \dodoi{10.1093/mnras/staa2298}

\bibitem[{{Compi{\`e}gne} {et~al.}(2011){Compi{\`e}gne}, {Verstraete}, {Jones}, {Bernard}, {Boulanger}, {Flagey}, {Le Bourlot}, {Paradis}, \& {Ysard}}]{2011A&A...525A.103C}
{Compi{\`e}gne}, M., {Verstraete}, L., {Jones}, A., {et~al.} 2011, \aap, 525, A103, \dodoi{10.1051/0004-6361/201015292}

\bibitem[{{Cuillandre} {et~al.}(2024){Cuillandre}, {Bertin}, {Bolzonella}, {Bouy}, {Gwyn}, {Isani}, {Kluge}, {Lai}, {Lan{\c{c}}on}, {Lang}, \& et~al.}]{2024arXiv240513496C}
{Cuillandre}, J.~C., {Bertin}, E., {Bolzonella}, M., {et~al.} 2024, arXiv e-prints, arXiv:2405.13496, \dodoi{10.48550/arXiv.2405.13496}

\bibitem[{{Czekala} {et~al.}(2015){Czekala}, {Andrews}, {Mandel}, {Hogg}, \& {Green}}]{2015ApJ...812..128C}
{Czekala}, I., {Andrews}, S.~M., {Mandel}, K.~S., {Hogg}, D.~W., \& {Green}, G.~M. 2015, \apj, 812, 128, \dodoi{10.1088/0004-637X/812/2/128}

\bibitem[{{Danieli} {et~al.}(2018){Danieli}, {van Dokkum}, \& {Conroy}}]{2018ApJ...856...69D}
{Danieli}, S., {van Dokkum}, P., \& {Conroy}, C. 2018, \apj, 856, 69, \dodoi{10.3847/1538-4357/aaadfb}

\bibitem[{{Danieli} {et~al.}(2020){Danieli}, {Lokhorst}, {Zhang}, {Merritt}, {van Dokkum}, {Abraham}, {Conroy}, {Gilhuly}, {Greco}, {Janssens}, {Li}, {Liu}, {Miller}, \& {Mowla}}]{2020ApJ...894..119D}
{Danieli}, S., {Lokhorst}, D., {Zhang}, J., {et~al.} 2020, \apj, 894, 119, \dodoi{10.3847/1538-4357/ab88a8}

\bibitem[{{DeVore} {et~al.}(2013){DeVore}, {Kristl}, \& {Rappaport}}]{2013JGRD..118.5679D}
{DeVore}, J.~G., {Kristl}, J.~A., \& {Rappaport}, S.~A. 2013, Journal of Geophysical Research (Atmospheres), 118, 5679, \dodoi{10.1002/jgrd.50440}

\bibitem[{{Dey} {et~al.}(2019){Dey}, {Schlegel}, {Lang}, {Blum}, {Burleigh}, {Fan}, {Findlay}, {Finkbeiner}, {Herrera}, {Juneau}, {Landriau}, {Levi}, {McGreer}, {Meisner}, {Myers}, {Moustakas}, {Nugent}, {Patej}, {Schlafly}, {Walker}, {Valdes}, {Weaver}, {Y{\`e}che}, {Zou}, {Zhou}, {Abareshi}, {Abbott}, {Abolfathi}, {Aguilera}, {Alam}, {Allen}, {Alvarez}, {Annis}, {Ansarinejad}, {Aubert}, {Beechert}, {Bell}, {BenZvi}, {Beutler}, {Bielby}, {Bolton}, {Brice{\~n}o}, {Buckley-Geer}, {Butler}, {Calamida}, {Carlberg}, {Carter}, {Casas}, {Castander}, {Choi}, {Comparat}, {Cukanovaite}, {Delubac}, {DeVries}, {Dey}, {Dhungana}, {Dickinson}, {Ding}, {Donaldson}, {Duan}, {Duckworth}, {Eftekharzadeh}, {Eisenstein}, {Etourneau}, {Fagrelius}, {Farihi}, {Fitzpatrick}, {Font-Ribera}, {Fulmer}, {G{\"a}nsicke}, {Gaztanaga}, {George}, {Gerdes}, {Gontcho}, {Gorgoni}, {Green}, {Guy}, {Harmer}, {Hernandez}, {Honscheid}, {Huang}, {James}, {Jannuzi}, {Jiang}, {Joyce}, {Karcher}, {Karkar}, {Kehoe}, {Kneib}, {Kueter-Young}, {Lan},
  {Lauer}, {Le Guillou}, {Le Van Suu}, {Lee}, {Lesser}, {Perreault Levasseur}, {Li}, {Mann}, {Marshall}, {Mart{\'\i}nez-V{\'a}zquez}, {Martini}, {du Mas des Bourboux}, {McManus}, {Meier}, {M{\'e}nard}, {Metcalfe}, {Mu{\~n}oz-Guti{\'e}rrez}, {Najita}, {Napier}, {Narayan}, {Newman}, {Nie}, {Nord}, {Norman}, {Olsen}, {Paat}, {Palanque-Delabrouille}, {Peng}, {Poppett}, {Poremba}, {Prakash}, {Rabinowitz}, {Raichoor}, {Rezaie}, {Robertson}, {Roe}, {Ross}, {Ross}, {Rudnick}, {Safonova}, {Saha}, {S{\'a}nchez}, {Savary}, {Schweiker}, {Scott}, {Seo}, {Shan}, {Silva}, {Slepian}, {Soto}, {Sprayberry}, {Staten}, {Stillman}, {Stupak}, {Summers}, {Sien Tie}, {Tirado}, {Vargas-Maga{\~n}a}, {Vivas}, {Wechsler}, {Williams}, {Yang}, {Yang}, {Yapici}, {Zaritsky}, {Zenteno}, {Zhang}, {Zhang}, {Zhou}, \& {Zhou}}]{2019AJ....157..168D}
{Dey}, A., {Schlegel}, D.~J., {Lang}, D., {et~al.} 2019, \aj, 157, 168, \dodoi{10.3847/1538-3881/ab089d}

\bibitem[{{Dotter}(2016)}]{2016ApJS..222....8D}
{Dotter}, A. 2016, \apjs, 222, 8, \dodoi{10.3847/0067-0049/222/1/8}

\bibitem[{{Draine}(2003)}]{2003ARA&A..41..241D}
{Draine}, B.~T. 2003, \araa, 41, 241, \dodoi{10.1146/annurev.astro.41.011802.094840}

\bibitem[{{Draine}(2011)}]{2011piim.book.....D}
---. 2011, {Physics of the Interstellar and Intergalactic Medium} (Princeton University Press)

\bibitem[{{Draine} \& {Li}(2001)}]{2001ApJ...551..807D}
{Draine}, B.~T., \& {Li}, A. 2001, \apj, 551, 807, \dodoi{10.1086/320227}

\bibitem[{{Draine} \& {Li}(2007)}]{2007ApJ...657..810D}
---. 2007, \apj, 657, 810, \dodoi{10.1086/511055}

\bibitem[{{Duda} \& {Hart}(1972)}]{Duda1972}
{Duda}, R.~O., \& {Hart}, P. 1972, Communications of the Association of Computing Machinery, 15, 11

\bibitem[{{Elmegreen} \& {Scalo}(2004)}]{2004ARA&A..42..211E}
{Elmegreen}, B.~G., \& {Scalo}, J. 2004, \araa, 42, 211, \dodoi{10.1146/annurev.astro.41.011802.094859}

\bibitem[{{Elvey} \& {Roach}(1937)}]{1937ApJ....85..213E}
{Elvey}, C.~T., \& {Roach}, F.~E. 1937, \apj, 85, 213, \dodoi{10.1086/143815}

\bibitem[{{Euclid Collaboration} {et~al.}(2022){Euclid Collaboration}, {Borlaff}, {G{\'o}mez-Alvarez}, {Altieri}, {Marcum}, {Vavrek}, {Laureijs}, {Kohley}, {Buitrago}, {Cuillandre}, {Duc}, {Gaspar Venancio}, {Amara}, {Andreon}, {Auricchio}, {Azzollini}, {Baccigalupi}, {Balaguera-Antol{\'\i}nez}, {Baldi}, {Bardelli}, {Bender}, {Biviano}, {Bodendorf}, {Bonino}, {Bozzo}, {Branchini}, {Brescia}, {Brinchmann}, {Burigana}, {Cabanac}, {Camera}, {Candini}, {Capobianco}, {Cappi}, {Carbone}, {Carretero}, {Carvalho}, {Casas}, {Castander}, {Castellano}, {Castignani}, {Cavuoti}, {Cimatti}, {Cledassou}, {Colodro-Conde}, {Congedo}, {Conselice}, {Conversi}, {Copin}, {Corcione}, {Coupon}, {Courtois}, {Cropper}, {Da Silva}, {Degaudenzi}, {Di Ferdinando}, {Douspis}, {Dubath}, {Duncan}, {Dupac}, {Dusini}, {Ealet}, {Fabricius}, {Farina}, {Farrens}, {Ferreira}, {Ferriol}, {Finelli}, {Flose-Reimberg}, {Fosalba}, {Frailis}, {Franceschi}, {Fumana}, {Galeotta}, {Ganga}, {Garilli}, {Gillis}, {Giocoli}, {Gozaliasl}, {Graci{\'a}-Carpio},
  {Grazian}, {Grupp}, {Haugan}, {Holmes}, {Hormuth}, {Jahnke}, {Keihanen}, {Kermiche}, {Kiessling}, {Kilbinger}, {Kirkpatrick}, {Kitching}, {Knapen}, {Kubik}, {K{\"u}mmel}, {Kunz}, {Kurki-Suonio}, {Liebing}, {Ligori}, {Lilje}, {Lindholm}, {Lloro}, {Mainetti}, {Maino}, {Mansutti}, {Marggraf}, {Markovic}, {Martinelli}, {Martinet}, {Mart{\'\i}nez-Delgado}, {Marulli}, {Massey}, {Maturi}, {Maurogordato}, {Medinaceli}, {Mei}, {Meneghetti}, {Merlin}, {Metcalf}, {Meylan}, {Moresco}, {Morgante}, {Moscardini}, {Munari}, {Nakajima}, {Neissner}, {Niemi}, {Nightingale}, {Nucita}, {Padilla}, {Paltani}, {Pasian}, {Patrizii}, {Pedersen}, {Percival}, {Pettorino}, {Pires}, {Poncet}, {Popa}, {Potter}, {Pozzetti}, {Raison}, {Rebolo}, {Renzi}, {Rhodes}, {Riccio}, {Romelli}, {Roncarelli}, {Rosset}, {Rossetti}, {Saglia}, {S{\'a}nchez}, {Sapone}, {Sauvage}, {Schneider}, {Scottez}, {Secroun}, {Seidel}, {Serrano}, {Sirignano}, {Sirri}, {Skottfelt}, {Stanco}, {Starck}, {Sureau}, {Tallada-Cresp{\'\i}}, {Taylor}, {Tenti}, {Tereno},
  {Teyssier}, {Toledo-Moreo}, {Torradeflot}, {Tutusaus}, {Valentijn}, {Valenziano}, {Valiviita}, {Vassallo}, {Viel}, {Wang}, {Weller}, {Whittaker}, {Zacchei}, {Zamorani}, \& {Zucca}}]{2022A&A...657A..92E}
{Euclid Collaboration}, {Borlaff}, A.~S., {G{\'o}mez-Alvarez}, P., {et~al.} 2022, \aap, 657, A92, \dodoi{10.1051/0004-6361/202141935}

\bibitem[{{Euclid Collaboration} {et~al.}(2024){Euclid Collaboration}, {Mellier}, {Abdurro'uf}, {Acevedo Barroso}, {Ach{\'u}carro}, {Adamek}, {Adam}, {Addison}, {Aghanim}, {Aguena}, \& et~al.}]{2024arXiv240513491E}
{Euclid Collaboration}, {Mellier}, Y., {Abdurro'uf}, {et~al.} 2024, arXiv e-prints, arXiv:2405.13491, \dodoi{10.48550/arXiv.2405.13491}

\bibitem[{{Fliri} \& {Trujillo}(2016)}]{2016MNRAS.456.1359F}
{Fliri}, J., \& {Trujillo}, I. 2016, \mnras, 456, 1359, \dodoi{10.1093/mnras/stv2686}

\bibitem[{{Gautier} {et~al.}(1992){Gautier}, {Boulanger}, {Perault}, \& {Puget}}]{1992AJ....103.1313G}
{Gautier}, T.~N., I., {Boulanger}, F., {Perault}, M., \& {Puget}, J.~L. 1992, \aj, 103, 1313, \dodoi{10.1086/116144}

\bibitem[{{Greco} \& {Danieli}(2022)}]{2022ApJ...941...26G}
{Greco}, J.~P., \& {Danieli}, S. 2022, \apj, 941, 26, \dodoi{10.3847/1538-4357/ac75b7}

\bibitem[{{Greco} {et~al.}(2018){Greco}, {Greene}, {Strauss}, {Macarthur}, {Flowers}, {Goulding}, {Huang}, {Kim}, {Komiyama}, {Leauthaud}, {Leisman}, {Lupton}, {Sif{\'o}n}, \& {Wang}}]{2018ApJ...857..104G}
{Greco}, J.~P., {Greene}, J.~E., {Strauss}, M.~A., {et~al.} 2018, \apj, 857, 104, \dodoi{10.3847/1538-4357/aab842}

\bibitem[{{Guhathakurta} \& {Tyson}(1989)}]{1989ApJ...346..773G}
{Guhathakurta}, P., \& {Tyson}, J.~A. 1989, \apj, 346, 773, \dodoi{10.1086/168058}

\bibitem[{{Hacar} {et~al.}(2023){Hacar}, {Clark}, {Heitsch}, {Kainulainen}, {Panopoulou}, {Seifried}, \& {Smith}}]{2023ASPC..534..153H}
{Hacar}, A., {Clark}, S.~E., {Heitsch}, F., {et~al.} 2023, in Astronomical Society of the Pacific Conference Series, Vol. 534, Protostars and Planets VII, ed. S.~{Inutsuka}, Y.~{Aikawa}, T.~{Muto}, K.~{Tomida}, \& M.~{Tamura}, 153, \dodoi{10.48550/arXiv.2203.09562}

\bibitem[{Harris {et~al.}(2020)Harris, Millman, van~der Walt, Gommers, Virtanen, Cournapeau, Wieser, Taylor, Berg, Smith, Kern, Picus, Hoyer, van Kerkwijk, Brett, Haldane, del R{'{\i}}o, Wiebe, Peterson, G{'{e}}rard-Marchant, Sheppard, Reddy, Weckesser, Abbasi, Gohlke, \& Oliphant}]{harris2020array}
Harris, C.~R., Millman, K.~J., van~der Walt, S.~J., {et~al.} 2020, Nature, 585, 357, \dodoi{10.1038/s41586-020-2649-2}

\bibitem[{{Henyey}(1937)}]{1937ApJ....85..107H}
{Henyey}, L.~G. 1937, \apj, 85, 107, \dodoi{10.1086/143805}

\bibitem[{{Henyey} \& {Greenstein}(1941)}]{1941ApJ....93...70H}
{Henyey}, L.~G., \& {Greenstein}, J.~L. 1941, \apj, 93, 70, \dodoi{10.1086/144246}

\bibitem[{{Hildebrand}(1983)}]{1983QJRAS..24..267H}
{Hildebrand}, R.~H. 1983, \qjras, 24, 267

\bibitem[{{Hogg} {et~al.}(2010){Hogg}, {Bovy}, \& {Lang}}]{2010arXiv1008.4686H}
{Hogg}, D.~W., {Bovy}, J., \& {Lang}, D. 2010, arXiv e-prints, arXiv:1008.4686, \dodoi{10.48550/arXiv.1008.4686}

\bibitem[{Hough(1962)}]{Hough1962}
Hough, P. V.~C. 1962, Method and Means for Recognizing Complex Patterns

\bibitem[{Hunter(2007)}]{Hunter:2007}
Hunter, J.~D. 2007, Computing in Science \& Engineering, 9, 90, \dodoi{10.1109/MCSE.2007.55}

\bibitem[{{Ienaka} {et~al.}(2013){Ienaka}, {Kawara}, {Matsuoka}, {Sameshima}, {Oyabu}, {Tsujimoto}, \& {Peterson}}]{2013ApJ...767...80I}
{Ienaka}, N., {Kawara}, K., {Matsuoka}, Y., {et~al.} 2013, \apj, 767, 80, \dodoi{10.1088/0004-637X/767/1/80}

\bibitem[{{Isobe} {et~al.}(1990){Isobe}, {Feigelson}, {Akritas}, \& {Babu}}]{1990ApJ...364..104I}
{Isobe}, T., {Feigelson}, E.~D., {Akritas}, M.~G., \& {Babu}, G.~J. 1990, \apj, 364, 104, \dodoi{10.1086/169390}

\bibitem[{{Jordi} {et~al.}(2006){Jordi}, {Grebel}, \& {Ammon}}]{2006A&A...460..339J}
{Jordi}, K., {Grebel}, E.~K., \& {Ammon}, K. 2006, \aap, 460, 339, \dodoi{10.1051/0004-6361:20066082}

\bibitem[{{Kawara} {et~al.}(2017){Kawara}, {Matsuoka}, {Sano}, {Brandt}, {Sameshima}, {Tsumura}, {Oyabu}, \& {Ienaka}}]{2017PASJ...69...31K}
{Kawara}, K., {Matsuoka}, Y., {Sano}, K., {et~al.} 2017, \pasj, 69, 31, \dodoi{10.1093/pasj/psx003}

\bibitem[{{Keim} {et~al.}(2022){Keim}, {van Dokkum}, {Danieli}, {Lokhorst}, {Li}, {Shen}, {Abraham}, {Chen}, {Gilhuly}, {Liu}, {Merritt}, {Miller}, {Pasha}, \& {Polzin}}]{2022ApJ...935..160K}
{Keim}, M.~A., {van Dokkum}, P., {Danieli}, S., {et~al.} 2022, \apj, 935, 160, \dodoi{10.3847/1538-4357/ac7dab}

\bibitem[{{Kelvin} {et~al.}(2023){Kelvin}, {Hasan}, \& {Tyson}}]{2023MNRAS.520.2484K}
{Kelvin}, L.~S., {Hasan}, I., \& {Tyson}, J.~A. 2023, \mnras, 520, 2484, \dodoi{10.1093/mnras/stad180}

\bibitem[{{King}(1971)}]{1971PASP...83..199K}
{King}, I.~R. 1971, \pasp, 83, 199, \dodoi{10.1086/129100}

\bibitem[{{Klypin} {et~al.}(1999){Klypin}, {Kravtsov}, {Valenzuela}, \& {Prada}}]{1999ApJ...522...82K}
{Klypin}, A., {Kravtsov}, A.~V., {Valenzuela}, O., \& {Prada}, F. 1999, \apj, 522, 82, \dodoi{10.1086/307643}

\bibitem[{{Koch} \& {Rosolowsky}(2015)}]{2015MNRAS.452.3435K}
{Koch}, E.~W., \& {Rosolowsky}, E.~W. 2015, \mnras, 452, 3435, \dodoi{10.1093/mnras/stv1521}

\bibitem[{{Koch} {et~al.}(2019){Koch}, {Rosolowsky}, {Boyden}, {Burkhart}, {Ginsburg}, {Loeppky}, \& {Offner}}]{2019AJ....158....1K}
{Koch}, E.~W., {Rosolowsky}, E.~W., {Boyden}, R.~D., {et~al.} 2019, \aj, 158, 1, \dodoi{10.3847/1538-3881/ab1cc0}

\bibitem[{{Kroupa}(2001)}]{2001MNRAS.322..231K}
{Kroupa}, P. 2001, \mnras, 322, 231, \dodoi{10.1046/j.1365-8711.2001.04022.x}

\bibitem[{{Lang} {et~al.}(2010){Lang}, {Hogg}, {Mierle}, {Blanton}, \& {Roweis}}]{2010AJ....139.1782L}
{Lang}, D., {Hogg}, D.~W., {Mierle}, K., {Blanton}, M., \& {Roweis}, S. 2010, \aj, 139, 1782, \dodoi{10.1088/0004-6256/139/5/1782}

\bibitem[{{Lanzetta} {et~al.}(2023){Lanzetta}, {Gromoll}, {Shara}, {Berg}, {Valls-Gabaud}, {Walter}, \& {Webb}}]{2023PASP..135a5002L}
{Lanzetta}, K.~M., {Gromoll}, S., {Shara}, M.~M., {et~al.} 2023, \pasp, 135, 015002, \dodoi{10.1088/1538-3873/acaee6}

\bibitem[{{Lauer} {et~al.}(2021){Lauer}, {Postman}, {Weaver}, {Spencer}, {Stern}, {Buie}, {Durda}, {Lisse}, {Poppe}, {Binzel}, {Britt}, {Buratti}, {Cheng}, {Grundy}, {Hor{\'a}nyi}, {Kavelaars}, {Linscott}, {McKinnon}, {Moore}, {N{\'u}{\~n}ez}, {Olkin}, {Parker}, {Porter}, {Reuter}, {Robbins}, {Schenk}, {Showalter}, {Singer}, {Verbiscer}, \& {Young}}]{2021ApJ...906...77L}
{Lauer}, T.~R., {Postman}, M., {Weaver}, H.~A., {et~al.} 2021, \apj, 906, 77, \dodoi{10.3847/1538-4357/abc881}

\bibitem[{{Laureijs} {et~al.}(1987){Laureijs}, {Mattila}, \& {Schnur}}]{1987A&A...184..269L}
{Laureijs}, R.~J., {Mattila}, K., \& {Schnur}, G. 1987, \aap, 184, 269

\bibitem[{{Liu} {et~al.}(2022){Liu}, {Abraham}, {Gilhuly}, {van Dokkum}, {Martin}, {Li}, {Greco}, {Lokhorst}, {Chen}, {Danieli}, {Keim}, {Merritt}, {Miller}, {Pasha}, {Polzin}, {Shen}, \& {Zhang}}]{2022ApJ...925..219L}
{Liu}, Q., {Abraham}, R., {Gilhuly}, C., {et~al.} 2022, \apj, 925, 219, \dodoi{10.3847/1538-4357/ac32c6}

\bibitem[{{Liu} {et~al.}(2023){Liu}, {Abraham}, {Martin}, {Bowman}, {van Dokkum}, {Janssens}, {Chen}, {Keim}, {Lokhorst}, {Pasha}, {Shen}, \& {Zhang}}]{2023ApJ...953....7L}
{Liu}, Q., {Abraham}, R., {Martin}, P.~G., {et~al.} 2023, \apj, 953, 7, \dodoi{10.3847/1538-4357/acdee3}

\bibitem[{{Lotz} {et~al.}(2004){Lotz}, {Primack}, \& {Madau}}]{2004AJ....128..163L}
{Lotz}, J.~M., {Primack}, J., \& {Madau}, P. 2004, \aj, 128, 163, \dodoi{10.1086/421849}

\bibitem[{{Low} {et~al.}(1984){Low}, {Beintema}, {Gautier}, {Gillett}, {Beichman}, {Neugebauer}, {Young}, {Aumann}, {Boggess}, {Emerson}, {Habing}, {Hauser}, {Houck}, {Rowan-Robinson}, {Soifer}, {Walker}, \& {Wesselius}}]{1984ApJ...278L..19L}
{Low}, F.~J., {Beintema}, D.~A., {Gautier}, T.~N., {et~al.} 1984, \apjl, 278, L19, \dodoi{10.1086/184213}

\bibitem[{{Martin} {et~al.}(2022){Martin}, {Bazkiaei}, {Spavone}, {Iodice}, {Mihos}, {Montes}, {Benavides}, {Brough}, {Carlin}, {Collins}, {Duc}, {G{\'o}mez}, {Galaz}, {Hern{\'a}ndez-Toledo}, {Jackson}, {Kaviraj}, {Knapen}, {Mart{\'\i}nez-Lombilla}, {McGee}, {O'Ryan}, {Prole}, {Rich}, {Rom{\'a}n}, {Shah}, {Starkenburg}, {Watkins}, {Zaritsky}, {Pichon}, {Armus}, {Bianconi}, {Buitrago}, {Bus{\'a}}, {Davis}, {Demarco}, {Desmons}, {Garc{\'\i}a}, {Graham}, {Holwerda}, {Hon}, {Khalid}, {Klehammer}, {Klutse}, {Lazar}, {Nair}, {Noakes-Kettel}, {Rutkowski}, {Saha}, {Sahu}, {Sola}, {V{\'a}zquez-Mata}, {Vera-Casanova}, \& {Yoon}}]{2022MNRAS.513.1459M}
{Martin}, G., {Bazkiaei}, A.~E., {Spavone}, M., {et~al.} 2022, \mnras, 513, 1459, \dodoi{10.1093/mnras/stac1003}

\bibitem[{{Martin} {et~al.}(2009){Martin}, {McConnachie}, {Irwin}, {Widrow}, {Ferguson}, {Ibata}, {Dubinski}, {Babul}, {Chapman}, {Fardal}, {Lewis}, {Navarro}, \& {Rich}}]{2009ApJ...705..758M}
{Martin}, N.~F., {McConnachie}, A.~W., {Irwin}, M., {et~al.} 2009, \apj, 705, 758, \dodoi{10.1088/0004-637X/705/1/758}

\bibitem[{{Martin} {et~al.}(2010){Martin}, {Miville-Desch{\^e}nes}, {Roy}, {Bernard}, {Molinari}, {Billot}, {Brunt}, {Calzoletti}, {Digiorgio}, {Elia}, {Faustini}, {Joncas}, {Mottram}, {Natoli}, {Noriega-Crespo}, {Paladini}, {Robitaille}, {Strafella}, {Traficante}, \& {Veneziani}}]{2010A&A...518L.105M}
{Martin}, P.~G., {Miville-Desch{\^e}nes}, M.~A., {Roy}, A., {et~al.} 2010, \aap, 518, L105, \dodoi{10.1051/0004-6361/201014684}

\bibitem[{{Mathis} {et~al.}(1983){Mathis}, {Mezger}, \& {Panagia}}]{1983A&A...128..212M}
{Mathis}, J.~S., {Mezger}, P.~G., \& {Panagia}, N. 1983, \aap, 128, 212

\bibitem[{{Matsuoka} {et~al.}(2011){Matsuoka}, {Ienaka}, {Kawara}, \& {Oyabu}}]{2011ApJ...736..119M}
{Matsuoka}, Y., {Ienaka}, N., {Kawara}, K., \& {Oyabu}, S. 2011, \apj, 736, 119, \dodoi{10.1088/0004-637X/736/2/119}

\bibitem[{{Mattila}(1979)}]{1979A&A....78..253M}
{Mattila}, K. 1979, \aap, 78, 253

\bibitem[{{Mattila} {et~al.}(2023){Mattila}, {V{\"a}is{\"a}nen}, {Lehtinen}, {Haikala}, \& {Haas}}]{2023MNRAS.524.2797M}
{Mattila}, K., {V{\"a}is{\"a}nen}, P., {Lehtinen}, K., {Haikala}, L., \& {Haas}, M. 2023, \mnras, 524, 2797, \dodoi{10.1093/mnras/stad1940}

\bibitem[{{McConnachie} {et~al.}(2009){McConnachie}, {Irwin}, {Ibata}, {Dubinski}, {Widrow}, {Martin}, {C{\^o}t{\'e}}, {Dotter}, {Navarro}, {Ferguson}, {Puzia}, {Lewis}, {Babul}, {Barmby}, {Bienaym{\'e}}, {Chapman}, {Cockcroft}, {Collins}, {Fardal}, {Harris}, {Huxor}, {Mackey}, {Pe{\~n}arrubia}, {Rich}, {Richer}, {Siebert}, {Tanvir}, {Valls-Gabaud}, \& {Venn}}]{2009Natur.461...66M}
{McConnachie}, A.~W., {Irwin}, M.~J., {Ibata}, R.~A., {et~al.} 2009, \nat, 461, 66, \dodoi{10.1038/nature08327}

\bibitem[{{Men'shchikov}(2013)}]{2013A&A...560A..63M}
{Men'shchikov}, A. 2013, \aap, 560, A63, \dodoi{10.1051/0004-6361/201321885}

\bibitem[{{Mihos} {et~al.}(2017){Mihos}, {Harding}, {Feldmeier}, {Rudick}, {Janowiecki}, {Morrison}, {Slater}, \& {Watkins}}]{2017ApJ...834...16M}
{Mihos}, J.~C., {Harding}, P., {Feldmeier}, J.~J., {et~al.} 2017, \apj, 834, 16, \dodoi{10.3847/1538-4357/834/1/16}

\bibitem[{{Miville-Desch{\^e}nes} {et~al.}(2016){Miville-Desch{\^e}nes}, {Duc}, {Marleau}, {Cuillandre}, {Didelon}, {Gwyn}, \& {Karabal}}]{2016A&A...593A...4M}
{Miville-Desch{\^e}nes}, M.~A., {Duc}, P.~A., {Marleau}, F., {et~al.} 2016, \aap, 593, A4, \dodoi{10.1051/0004-6361/201628503}

\bibitem[{{Miville-Desch{\^e}nes} \& {Lagache}(2005)}]{2005ApJS..157..302M}
{Miville-Desch{\^e}nes}, M.-A., \& {Lagache}, G. 2005, \apjs, 157, 302, \dodoi{10.1086/427938}

\bibitem[{{Miville-Desch{\^e}nes} {et~al.}(2007){Miville-Desch{\^e}nes}, {Lagache}, {Boulanger}, \& {Puget}}]{2007A&A...469..595M}
{Miville-Desch{\^e}nes}, M.~A., {Lagache}, G., {Boulanger}, F., \& {Puget}, J.~L. 2007, \aap, 469, 595, \dodoi{10.1051/0004-6361:20066962}

\bibitem[{{Miville-Desch{\^e}nes} {et~al.}(2003){Miville-Desch{\^e}nes}, {Levrier}, \& {Falgarone}}]{2003ApJ...593..831M}
{Miville-Desch{\^e}nes}, M.~A., {Levrier}, F., \& {Falgarone}, E. 2003, \apj, 593, 831, \dodoi{10.1086/376603}

\bibitem[{{Murthy} {et~al.}(2010){Murthy}, {Henry}, \& {Sujatha}}]{2010ApJ...724.1389M}
{Murthy}, J., {Henry}, R.~C., \& {Sujatha}, N.~V. 2010, \apj, 724, 1389, \dodoi{10.1088/0004-637X/724/2/1389}

\bibitem[{{Ossenkopf} {et~al.}(2008){Ossenkopf}, {Krips}, \& {Stutzki}}]{2008A&A...485..917O}
{Ossenkopf}, V., {Krips}, M., \& {Stutzki}, J. 2008, \aap, 485, 917, \dodoi{10.1051/0004-6361:20079106}

\bibitem[{{Paley} {et~al.}(1991){Paley}, {Low}, {McGraw}, {Cutri}, \& {Rix}}]{1991ApJ...376..335P}
{Paley}, E.~S., {Low}, F.~J., {McGraw}, J.~T., {Cutri}, R.~M., \& {Rix}, H.-W. 1991, \apj, 376, 335, \dodoi{10.1086/170283}

\bibitem[{{Pedregosa} {et~al.}(2011){Pedregosa}, {Varoquaux}, {Gramfort}, {Michel}, {Thirion}, {Grisel}, {Blondel}, {M{\"u}ller}, {Nothman}, {Louppe}, {Prettenhofer}, {Weiss}, {Dubourg}, {Vanderplas}, {Passos}, {Cournapeau}, {Brucher}, {Perrot}, \& {Duchesnay}}]{2011JMLR...12.2825P}
{Pedregosa}, F., {Varoquaux}, G., {Gramfort}, A., {et~al.} 2011, Journal of Machine Learning Research, 12, 2825, \dodoi{10.48550/arXiv.1201.0490}

\bibitem[{{\sorthelp{Planck Collaboration 2011S}}{Planck Collaboration XIX}(2011)}]{planck2011-7.0}
{\sorthelp{Planck Collaboration 2011S}}{Planck Collaboration XIX}. 2011, \aap, 536, A19, \dodoi{10.1051/0004-6361/201116479}

\bibitem[{{\sorthelp{Planck Collaboration 2011X}}{Planck Collaboration XXIV}(2011)}]{planck2011-7.12}
{\sorthelp{Planck Collaboration 2011X}}{Planck Collaboration XXIV}. 2011, \aap, 536, A24, \dodoi{10.1051/0004-6361/201116485}

\bibitem[{{\sorthelp{Planck Collaboration 2014K}}{Planck Collaboration XI}(2014)}]{planck2013-p06b}
{\sorthelp{Planck Collaboration 2014K}}{Planck Collaboration XI}. 2014, \aap, 571, A11, \dodoi{10.1051/0004-6361/201323195}

\bibitem[{{\sorthelp{Planck Collaboration IntQ}}{Planck Collaboration Int. XVII}(2014)}]{planck2013-XVII}
{\sorthelp{Planck Collaboration IntQ}}{Planck Collaboration Int. XVII}. 2014, \aap, 566, A55, \dodoi{10.1051/0004-6361/201323270}

\bibitem[{{\sorthelp{Planck Collaboration IntZW}}{Planck Collaboration Int. XLVIII}(2016)}]{planck2016-XLVIII}
{\sorthelp{Planck Collaboration IntZW}}{Planck Collaboration Int. XLVIII}. 2016, \aap, 596, A109, \dodoi{10.1051/0004-6361/201629022}

\bibitem[{{Racine}(1996)}]{1996PASP..108..699R}
{Racine}, R. 1996, \pasp, 108, 699, \dodoi{10.1086/133788}

\bibitem[{{Robitaille} {et~al.}(2020){Robitaille}, {Deil}, \& {Ginsburg}}]{2020ascl.soft11023R}
{Robitaille}, T., {Deil}, C., \& {Ginsburg}, A. 2020, {reproject: Python-based astronomical image reprojection}, Astrophysics Source Code Library, record ascl:2011.023.
\newblock \doeprint{2011.023}

\bibitem[{{Rom{\'a}n} {et~al.}(2020){Rom{\'a}n}, {Trujillo}, \& {Montes}}]{2020A&A...644A..42R}
{Rom{\'a}n}, J., {Trujillo}, I., \& {Montes}, M. 2020, \aap, 644, A42, \dodoi{10.1051/0004-6361/201936111}

\bibitem[{{Salji} {et~al.}(2015){Salji}, {Richer}, {Buckle}, {di Francesco}, {Hatchell}, {Hogerheijde}, {Johnstone}, {Kirk}, {Ward-Thompson}, \& {JCMT GBS Consortium}}]{2015MNRAS.449.1782S}
{Salji}, C.~J., {Richer}, J.~S., {Buckle}, J.~V., {et~al.} 2015, \mnras, 449, 1782, \dodoi{10.1093/mnras/stv369}

\bibitem[{{Sandage}(1976)}]{1976AJ.....81..954S}
{Sandage}, A. 1976, \aj, 81, 954, \dodoi{10.1086/111975}

\bibitem[{{Sandin}(2014)}]{2014A&A...567A..97S}
{Sandin}, C. 2014, \aap, 567, A97, \dodoi{10.1051/0004-6361/201423429}

\bibitem[{{Sano} {et~al.}(2015){Sano}, {Kawara}, {Matsuura}, {Kataza}, {Arai}, \& {Matsuoka}}]{2015ApJ...811...77S}
{Sano}, K., {Kawara}, K., {Matsuura}, S., {et~al.} 2015, \apj, 811, 77, \dodoi{10.1088/0004-637X/811/2/77}

\bibitem[{{Sano} \& {Matsuura}(2017)}]{2017ApJ...849...31S}
{Sano}, K., \& {Matsuura}, S. 2017, \apj, 849, 31, \dodoi{10.3847/1538-4357/aa906c}

\bibitem[{{Sano} {et~al.}(2016){Sano}, {Matsuura}, {Tsumura}, {Arai}, {Shirahata}, \& {Onishi}}]{2016ApJ...821L..11S}
{Sano}, K., {Matsuura}, S., {Tsumura}, K., {et~al.} 2016, \apjl, 821, L11, \dodoi{10.3847/2041-8205/821/1/L11}

\bibitem[{{Saydjari} \& {Finkbeiner}(2022)}]{2022ApJ...933..155S}
{Saydjari}, A.~K., \& {Finkbeiner}, D.~P. 2022, \apj, 933, 155, \dodoi{10.3847/1538-4357/ac6875}

\bibitem[{{Schisano} {et~al.}(2014){Schisano}, {Rygl}, {Molinari}, {Busquet}, {Elia}, {Pestalozzi}, {Polychroni}, {Billot}, {Carey}, {Paladini}, {Noriega-Crespo}, {Moore}, {Plume}, {Glover}, \& {V{\'a}zquez-Semadeni}}]{2014ApJ...791...27S}
{Schisano}, E., {Rygl}, K.~L.~J., {Molinari}, S., {et~al.} 2014, \apj, 791, 27, \dodoi{10.1088/0004-637X/791/1/27}

\bibitem[{{Schisano} {et~al.}(2020){Schisano}, {Molinari}, {Elia}, {Benedettini}, {Olmi}, {Pezzuto}, {Traficante}, {Brescia}, {Cavuoti}, {di Giorgio}, {Liu}, {Moore}, {Noriega-Crespo}, {Riccio}, {Baldeschi}, {Becciani}, {Peretto}, {Merello}, {Vitello}, {Zavagno}, {Beltr{\'a}n}, {Cambr{\'e}sy}, {Eden}, {Li Causi}, {Molinaro}, {Palmeirim}, {Sciacca}, {Testi}, {Umana}, \& {Whitworth}}]{2020MNRAS.492.5420S}
{Schisano}, E., {Molinari}, S., {Elia}, D., {et~al.} 2020, \mnras, 492, 5420, \dodoi{10.1093/mnras/stz3466}

\bibitem[{{Schlegel} {et~al.}(1998){Schlegel}, {Finkbeiner}, \& {Davis}}]{1998ApJ...500..525S}
{Schlegel}, D.~J., {Finkbeiner}, D.~P., \& {Davis}, M. 1998, \apj, 500, 525, \dodoi{10.1086/305772}

\bibitem[{{Schneider} \& {Elmegreen}(1979)}]{1979ApJS...41...87S}
{Schneider}, S., \& {Elmegreen}, B.~G. 1979, \apjs, 41, 87, \dodoi{10.1086/190609}

\bibitem[{{Simon}(2019)}]{2019ARA&A..57..375S}
{Simon}, J.~D. 2019, \araa, 57, 375, \dodoi{10.1146/annurev-astro-091918-104453}

\bibitem[{{Slater} {et~al.}(2009){Slater}, {Harding}, \& {Mihos}}]{2009PASP..121.1267S}
{Slater}, C.~T., {Harding}, P., \& {Mihos}, J.~C. 2009, \pasp, 121, 1267, \dodoi{10.1086/648457}

\bibitem[{{Smirnov} {et~al.}(2023){Smirnov}, {Savchenko}, {Poliakov}, {Marchuk}, {Mosenkov}, {Il'in}, {Gontcharov}, {Rom{\'a}n}, \& {Seguine}}]{2023MNRAS.519.4735S}
{Smirnov}, A.~A., {Savchenko}, S.~S., {Poliakov}, D.~M., {et~al.} 2023, \mnras, 519, 4735, \dodoi{10.1093/mnras/stac3765}

\bibitem[{{Sousbie}(2011)}]{2011MNRAS.414..350S}
{Sousbie}, T. 2011, \mnras, 414, 350, \dodoi{10.1111/j.1365-2966.2011.18394.x}

\bibitem[{{Stutzki} {et~al.}(1998){Stutzki}, {Bensch}, {Heithausen}, {Ossenkopf}, \& {Zielinsky}}]{1998A&A...336..697S}
{Stutzki}, J., {Bensch}, F., {Heithausen}, A., {Ossenkopf}, V., \& {Zielinsky}, M. 1998, \aap, 336, 697

\bibitem[{{Symons} {et~al.}(2023){Symons}, {Zemcov}, {Cooray}, {Lisse}, \& {Poppe}}]{2023ApJ...945...45S}
{Symons}, T., {Zemcov}, M., {Cooray}, A., {Lisse}, C., \& {Poppe}, A.~R. 2023, \apj, 945, 45, \dodoi{10.3847/1538-4357/acaa37}

\bibitem[{{Thilker} {et~al.}(2023){Thilker}, {Lee}, {Deger}, {Barnes}, {Bigiel}, {Boquien}, {Cao}, {Chevance}, {Dale}, {Egorov}, {Glover}, {Grasha}, {Henshaw}, {Klessen}, {Koch}, {Kruijssen}, {Leroy}, {Lessing}, {Meidt}, {Pinna}, {Querejeta}, {Rosolowsky}, {Sandstrom}, {Schinnerer}, {Smith}, {Watkins}, {Williams}, {Anand}, {Belfiore}, {Blanc}, {Chandar}, {Congiu}, {Emsellem}, {Groves}, {Kreckel}, {Larson}, {Liu}, {Pessa}, \& {Whitmore}}]{2023ApJ...944L..13T}
{Thilker}, D.~A., {Lee}, J.~C., {Deger}, S., {et~al.} 2023, \apjl, 944, L13, \dodoi{10.3847/2041-8213/acaeac}

\bibitem[{Van~der Walt {et~al.}(2014)Van~der Walt, Sch{\"o}nberger, Nunez-Iglesias, Boulogne, Warner, Yager, Gouillart, \& Yu}]{van2014scikit}
Van~der Walt, S., Sch{\"o}nberger, J.~L., Nunez-Iglesias, J., {et~al.} 2014, PeerJ, 2, e453

\bibitem[{{van Dokkum} \& {Pasha}(2024)}]{2024PASP..136c4503V}
{van Dokkum}, P., \& {Pasha}, I. 2024, \pasp, 136, 034503, \dodoi{10.1088/1538-3873/ad2866}

\bibitem[{{van Dokkum} {et~al.}(2020){van Dokkum}, {Lokhorst}, {Danieli}, {Li}, {Merritt}, {Abraham}, {Gilhuly}, {Greco}, \& {Liu}}]{2020PASP..132g4503V}
{van Dokkum}, P., {Lokhorst}, D., {Danieli}, S., {et~al.} 2020, \pasp, 132, 074503, \dodoi{10.1088/1538-3873/ab9416}

\bibitem[{{van Dokkum} {et~al.}(2015){van Dokkum}, {Abraham}, {Merritt}, {Zhang}, {Geha}, \& {Conroy}}]{2015ApJ...798L..45V}
{van Dokkum}, P.~G., {Abraham}, R., {Merritt}, A., {et~al.} 2015, \apjl, 798, L45, \dodoi{10.1088/2041-8205/798/2/L45}

\bibitem[{{Virtanen} {et~al.}(2020){Virtanen}, {Gommers}, {Oliphant}, {Haberland}, {Reddy}, {Cournapeau}, {Burovski}, {Peterson}, {Weckesser}, {Bright}, {van der Walt}, {Brett}, {Wilson}, {Millman}, {Mayorov}, {Nelson}, {Jones}, {Kern}, {Larson}, {Carey}, {Polat}, {Feng}, {Moore}, {VanderPlas}, {Laxalde}, {Perktold}, {Cimrman}, {Henriksen}, {Quintero}, {Harris}, {Archibald}, {Ribeiro}, {Pedregosa}, {van Mulbregt}, \& {SciPy 1. 0 Contributors}}]{2020NatMe..17..261V}
{Virtanen}, P., {Gommers}, R., {Oliphant}, T.~E., {et~al.} 2020, Nature Methods, 17, 261, \dodoi{10.1038/s41592-019-0686-2}

\bibitem[{{Watkins} {et~al.}(2024){Watkins}, {Kaviraj}, {Collins}, {Knapen}, {Kelvin}, {Duc}, {Rom{\'a}n}, \& {Mihos}}]{2024MNRAS.528.4289W}
{Watkins}, A.~E., {Kaviraj}, S., {Collins}, C.~C., {et~al.} 2024, \mnras, 528, 4289, \dodoi{10.1093/mnras/stae236}

\bibitem[{{Watkins} {et~al.}(2015){Watkins}, {Mihos}, \& {Harding}}]{2015ApJ...800L...3W}
{Watkins}, A.~E., {Mihos}, J.~C., \& {Harding}, P. 2015, \apjl, 800, L3, \dodoi{10.1088/2041-8205/800/1/L3}

\bibitem[{{Weingartner} \& {Draine}(2001)}]{2001ApJ...548..296W}
{Weingartner}, J.~C., \& {Draine}, B.~T. 2001, \apj, 548, 296, \dodoi{10.1086/318651}

\bibitem[{{Wetzel} {et~al.}(2016){Wetzel}, {Hopkins}, {Kim}, {Faucher-Gigu{\`e}re}, {Kere{\v{s}}}, \& {Quataert}}]{2016ApJ...827L..23W}
{Wetzel}, A.~R., {Hopkins}, P.~F., {Kim}, J.-h., {et~al.} 2016, \apjl, 827, L23, \dodoi{10.3847/2041-8205/827/2/L23}

\bibitem[{{Witt} {et~al.}(2006){Witt}, {Gordon}, {Vijh}, {Sell}, {Smith}, \& {Xie}}]{2006ApJ...636..303W}
{Witt}, A.~N., {Gordon}, K.~D., {Vijh}, U.~P., {et~al.} 2006, \apj, 636, 303, \dodoi{10.1086/498052}

\bibitem[{{Witt} {et~al.}(2008){Witt}, {Mandel}, {Sell}, {Dixon}, \& {Vijh}}]{2008ApJ...679..497W}
{Witt}, A.~N., {Mandel}, S., {Sell}, P.~H., {Dixon}, T., \& {Vijh}, U.~P. 2008, \apj, 679, 497, \dodoi{10.1086/587131}

\bibitem[{{Zagury} {et~al.}(1999){Zagury}, {Boulanger}, \& {Banchet}}]{1999A&A...352..645Z}
{Zagury}, F., {Boulanger}, F., \& {Banchet}, V. 1999, \aap, 352, 645

\bibitem[{{Zaritsky} {et~al.}(2021){Zaritsky}, {Donnerstein}, {Karunakaran}, {Barbosa}, {Dey}, {Kadowaki}, {Spekkens}, \& {Zhang}}]{2021ApJS..257...60Z}
{Zaritsky}, D., {Donnerstein}, R., {Karunakaran}, A., {et~al.} 2021, \apjs, 257, 60, \dodoi{10.3847/1538-4365/ac2607}

\bibitem[{{Zaritsky} {et~al.}(2022){Zaritsky}, {Donnerstein}, {Karunakaran}, {Barbosa}, {Dey}, {Kadowaki}, {Spekkens}, \& {Zhang}}]{2022ApJS..261...11Z}
---. 2022, \apjs, 261, 11, \dodoi{10.3847/1538-4365/ac6ceb}

\bibitem[{{Zavagno} {et~al.}(2023){Zavagno}, {Dup{\'e}}, {Bensaid}, {Schisano}, {Li Causi}, {Gray}, {Molinari}, {Elia}, {Lambert}, {Brescia}, {Arzoumanian}, {Russeil}, {Riccio}, \& {Cavuoti}}]{2023A&A...669A.120Z}
{Zavagno}, A., {Dup{\'e}}, F.~X., {Bensaid}, S., {et~al.} 2023, \aap, 669, A120, \dodoi{10.1051/0004-6361/202244103}

\bibitem[{{Zemcov} {et~al.}(2017){Zemcov}, {Immel}, {Nguyen}, {Cooray}, {Lisse}, \& {Poppe}}]{2017NatCo...815003Z}
{Zemcov}, M., {Immel}, P., {Nguyen}, C., {et~al.} 2017, Nature Communications, 8, 15003, \dodoi{10.1038/ncomms15003}

\bibitem[{{Zhang} {et~al.}(2023){Zhang}, {Martin}, {Cloutier}, {Price-Jones}, {Abraham}, {van Dokkum}, \& {Merritt}}]{2023ApJ...948....4Z}
{Zhang}, J., {Martin}, P.~G., {Cloutier}, R., {et~al.} 2023, \apj, 948, 4, \dodoi{10.3847/1538-4357/acc177}

\bibitem[{{Zhao} {et~al.}(2024){Zhao}, {Zhang}, {Ma}, {Wen}, \& {Wu}}]{2024AJ....168...88Z}
{Zhao}, Y., {Zhang}, W., {Ma}, L., {Wen}, S., \& {Wu}, H. 2024, \aj, 168, 88, \dodoi{10.3847/1538-3881/ad58d5}

\bibitem[{{Zubko} {et~al.}(2004){Zubko}, {Dwek}, \& {Arendt}}]{2004ApJS..152..211Z}
{Zubko}, V., {Dwek}, E., \& {Arendt}, R.~G. 2004, \apjs, 152, 211, \dodoi{10.1086/382351}

\end{thebibliography}
\bibliographystyle{aasjournal}

%% This command is needed to show the entire author+affiliation list when
%% the collaboration and author truncation commands are used.  It has to
%% go at the end of the manuscript.
%\allauthors

%% Include this line if you are using the \added, \replaced, \deleted
%% commands to see a summary list of all changes at the end of the article.
%\listofchanges

\end{document}